\documentclass[preprint,aps,preprintnumbers,amsmath,amssymb]{revtex4}
\usepackage{graphicx}
\usepackage{color}
\usepackage{dcolumn}
\usepackage{bm}

\newcommand{\MINERVA}   {\mbox{\hbox{MINER}$\nu$\hbox{A}}}
\newcommand{\NOVA}      {\mbox{\hbox{NO}$\nu$\hbox{A}}}

\def\bit{\begin{itemize}}
\def\eit{\end{itemize}}

\def\beq{\begin{equation}}
\def\eeq{\end{equation}}
\def\bea{\begin{eqnarray}}
\def\eea{\end{eqnarray}}

\begin{document}

\preprint{Fermilab Proposal P-960}
\title{
\textbf{Proposal to upgrade the MIPP Experiment}}

\author{D.~Isenhower, M.~Sadler, R.~Towell, S.~Watson}
\affiliation{Abilene Christian University$\dag$}
\author{R.~J.~Peterson}
\affiliation{University of Colorado,  Boulder}
\author{W.~Baker, D.~Carey, D.~Christian, M.~Demarteau, D.~Jensen, 
C.~Johnstone, 
H.~Meyer, R.~Raja~\footnote{contactperson}, A.~Ronzhin, N.~Solomey, W.~Wester}
\affiliation{Fermi National Accelerator Laboratory}
\author{H.~Gutbrod, K.~Peters}
\affiliation{GSI,~ Darmstadt, Germany$\dag$}
\author{G.~Feldman}
\affiliation{Harvard~University}
\author{Y.~Torun}
\affiliation{Illinois Institute of Technology}
\author{M.D.~Messier, J.~Paley}
\affiliation{Indiana University}
\author{U.~Akgun, G.~Aydin, F.~Duru, E.~G\"ulmez, Y.~Gunaydin, Y.~Onel, 
A.~Penzo}
\affiliation{University of Iowa}
\author{V.~Avdeichikov, R.~Leitner, J.~Manjavidze, V.~Nikitin, I.~Rufanov, 
A.~Sissakian, T.~Topuria}
\affiliation{Joint Institute of Nuclear Research, Dubna, Russia$\dag$}
\author{D.~M.~Manley}
\affiliation{Kent State University$\dag$}
\author{H.~L\"ohner, J.~Messchendorp}
\affiliation{KVI, Groningen, Netherlands $\dag$}
\author{H.~R.~Gustafson, M.~Longo, T.~Nigmanov, D.~Rajaram}
\affiliation{University of Michigan}
\author{S.~P.~Kruglov, I.~V.~Lopatin, N.~G.~Kozlenko, A.~A.~Kulbardis, 
D.~V.~Nowinsky, A.~K.~Radkov, V.~V.~Sumachev}

\affiliation{Petersburg Nuclear Physics Institute,  Gatchina,  Russia
\footnote{New MIPP Collaborating Institution}}
\author{A.~Bujak, L.~Gutay}
\affiliation{Purdue University}
\author{A.~Godley, S.~R.~Mishra, C.~Rosenfeld}
\affiliation{University of South Carolina}
\author{C.~Dukes, C.~Materniak, K.~Nelson, A.~Norman}
\affiliation{University of Virginia}
\author{P.~Desiati, F.~Halzen, T.~Montaruli}
\affiliation{University of Wisconsin, Madison$\dag$}
\date{\today}

\begin{abstract}
The upgraded MIPP physics results are needed for the support of 
NuMI projects,  
atmospheric cosmic ray and neutrino programs worldwide and will permit 
a systematic study of non-perturbative QCD interctions.
The MIPP TPC is the largest contributor to the MIPP event size by
far. Its readout system and electronics were designed in the 1990's
and limit it to a readout rate of 60~Hz in simple events and $\approx$
20~Hz in complicated events. With the readout chips designed for the
ALICE collaboration at the LHC, we propose a low cost  scheme
of upgrading the MIPP data acquisition speed to 3000 Hz. 
This will also enable us to measure the medium energy numi
target to be used for the \NOVA/\MINERVA{} experiments.  
We outline the 
capabilities of the upgraded MIPP detector to obtain 
high statistics particle production data  on a number of  
nuclei that will help towards the
understanding and simulation of hadronic showers in matter. 
Measurements of nitrogen cross sections will permit a better 
understanding of cosmic ray shower systematics in the atmosphere.
In addition, we
explore the possibilities of providing tagged neutral beams using the
MIPP spectrometer that may be crucial for validating the Particle Flow
Algorithm proposed for calorimeters for the International Linear
Collider detectors. Lastly, we outline the physics potential of such a 
detector in understanding non-perturbative QCD processes.
\end{abstract}
\thispagestyle{empty}
\maketitle
  
\tableofcontents
\section{Current Status of the MIPP Experiment}

We give a brief status report on the MIPP experiment and its
performance to date.  The Main Injector Particle Production Experiment
(FNAL E-907, MIPP)~\cite{mipp} is situated in the Meson Center
beamline at Fermilab. It received  approval~\cite{proposal} in
November 2001 and has installed and operated  both the experiment and a newly
designed secondary beamline in the interim. It received its first
beams in March 2004, had an engineering run to commission the detector
in 2004 and had its physics data-taking run in the period January
2005-March 2006. The experiment is currently busy analyzing its data.

\begin{figure}[htb!]
\includegraphics[width=\textwidth]{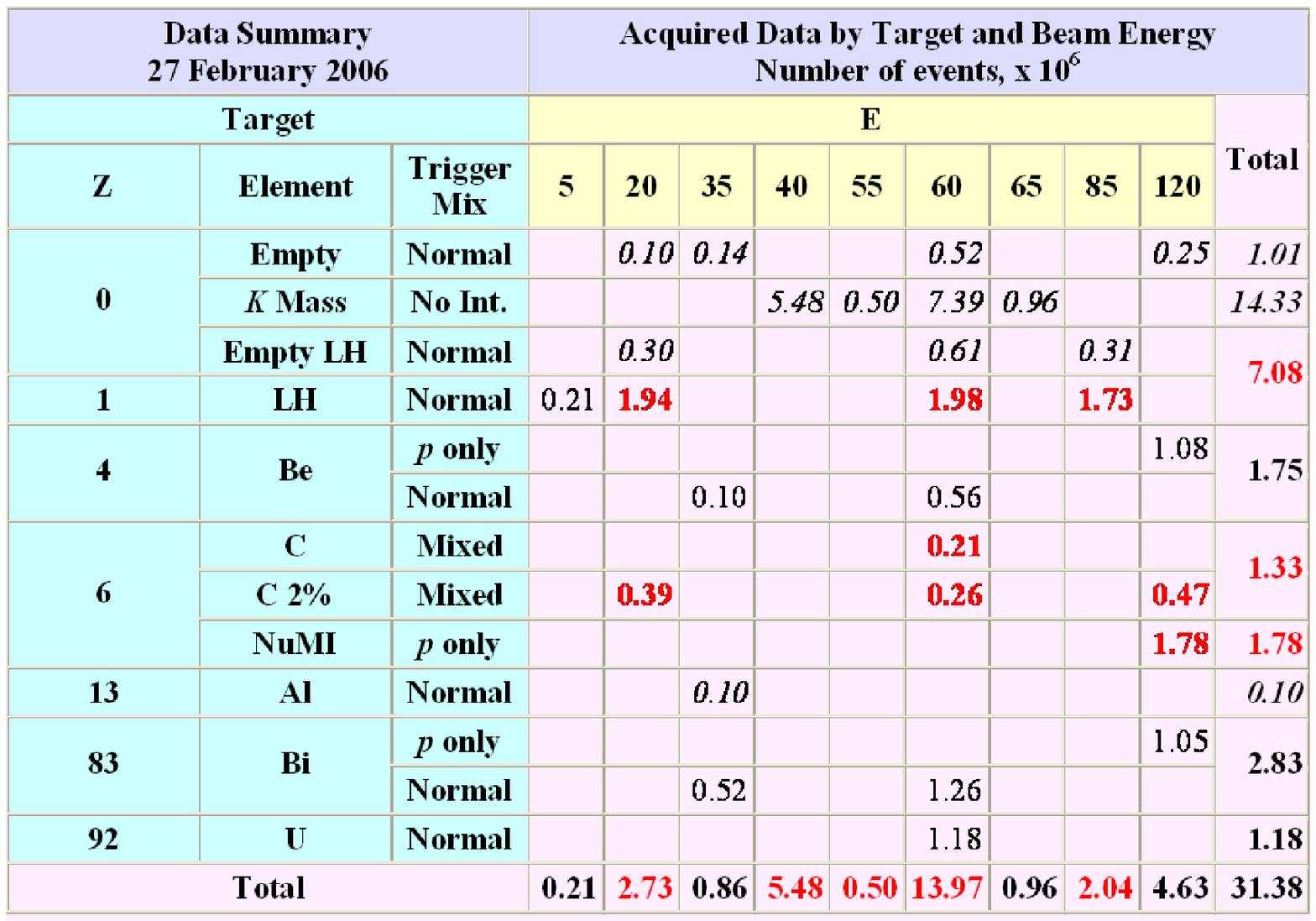}
\caption{The data taken during the first MIPP run as a function of nucleus. 
The numbers are in millions of events. During the last month of the
run, the Jolly Green Giant magnet coils developed shorts. This time
was used to acquire data without the TPC for exploring the feasibility
of measuring the charged kaon mass using the RICH radii.}
\label{tab1}
\end{figure}

MIPP is designed primarily as an experiment to measure and study in
detail the dynamics associated with non-perturbative strong
interactions. It has nearly 100\% acceptance for charged particles and
excellent momentum resolution.  Using particle identification
techniques that encompass $dE/dx$, time-of-flight~\cite{tof},
Multi-Cell \v Cerenkov~\cite{e690} and a Ring Imaging \v Cerenkov
(RICH)detector~\cite{rich}, MIPP is designed to identify charged particles
at the 3$\sigma$ or better level in nearly all of its final state
phase space.  MIPP has acquired data of unparalleled quality and
statistics for beam momenta ranging from 5~GeV/c to 90~GeV/c for 6
beam species ($\pi^\pm, K^\pm~ $and$~ p^\pm$) on a variety of targets
as shown in Figure~\ref{tab1}.

 An important aspect of MIPP data-taking was the measurement of
 particle production off the NuMI~\cite{minos} target in order to
 minimize the systematics in the near/far detector ratio in the
 MINOS~\cite{minos} experiment.  MIPP also made measurements with
 proton beams off various nuclei for the needs of proton
 radiography~\cite{proposal}.

Another physics motivation behind MIPP is to restart the study of
non-perturbative QCD interactions, which constitute over 99\% of the
strong interaction cross section. The available data are of poor
quality, and sparsely populate the beam momentum, $p_T$, and 
atomic Weight phase space
that makes comparisons between different experiments difficult. The Time
Projection Chamber (TPC)~\cite{tpc} that is at the heart of the MIPP
experiment represents the electronic equivalent of the bubble chamber
with vastly superior data acquisition rates. It also digitizes the
charged tracks in three dimensions, obviating the need for track
matching across stereo views.  Coupled with the particle
identification capability of MIPP, the data from MIPP would add
significantly to our knowledge base of non-perturbative QCD. This
would help test inclusive scaling relations and also scaling nuclear
reactions.

\subsection{Experimental Setup}

We designed a secondary beam~\cite{carol} specific to our needs. The
120 GeV/c primary protons are resonantly extracted in a slow spill from
the Fermilab Main Injector and transported down the Meson Center line. 
They impinge on a 20~cm long copper target producing
secondary beam particles. This target is imaged onto an adjustable
momentum selection collimator which controls the momentum spread of
the beam. This collimator is re-imaged on to our interaction target
placed next to the TPC. The beam is tracked using three beam chambers
and identified using two differential \v Cerenkovs~\cite{bckov} filled
with gas, the composition and the pressure of which can be varied
within limits depending on the beam momentum and charge.

\begin{figure}[htb!]
\includegraphics[width=\textwidth]{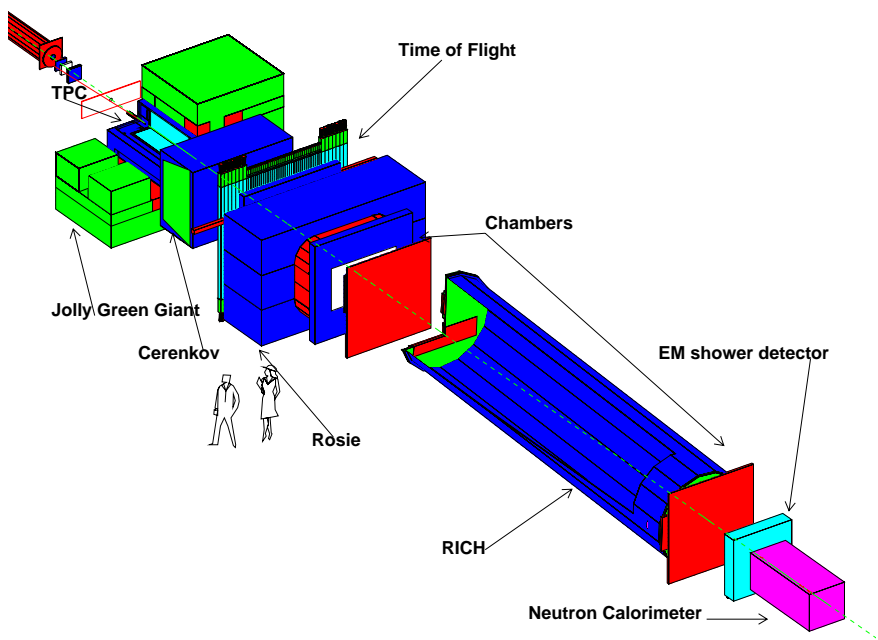}
\caption{The experimental setup. The picture is a rendition in 
Geant3, which is used to simulate the detector.}
\label{mipp}
\end{figure}

Figure~\ref{mipp} shows the layout of the apparatus. The TPC sits in a
wide aperture magnet (the Jolly Green Giant) which has a peak field of
0.7 tesla. Downstream of the TPC are a 96 mirror multi-cell \v
Cerenkov detector filled with $C_4F_{10}$ gas, and a time of flight
system. This is followed by a large aperture magnet (Rosie) which runs
in opposite polarity (at -0.6 tesla) to the Jolly Green Giant to bend
the particles back into the Ring Imaging \v Cerenkov counter. The RICH
has $CO_2$ as the radiator and an array of phototubes of 32 rows and
89 columns~\cite{fire}.
Downstream of the RICH we have an electromagnetic
calorimeter~\cite{ecal}and a hadron
calorimeter~\cite{hcal} to
measure forward-going photons and neutrons. The electromagnetic
calorimeter provides a means of distinguishing forward neutrons 
from photons and will also serve as a device to measure the electron
content of our beam at lower energies, which will be useful for
measuring cross sections.

\begin{figure}[hbtp] 
\begin{minipage}{15pc}
\includegraphics[width=2.5in]{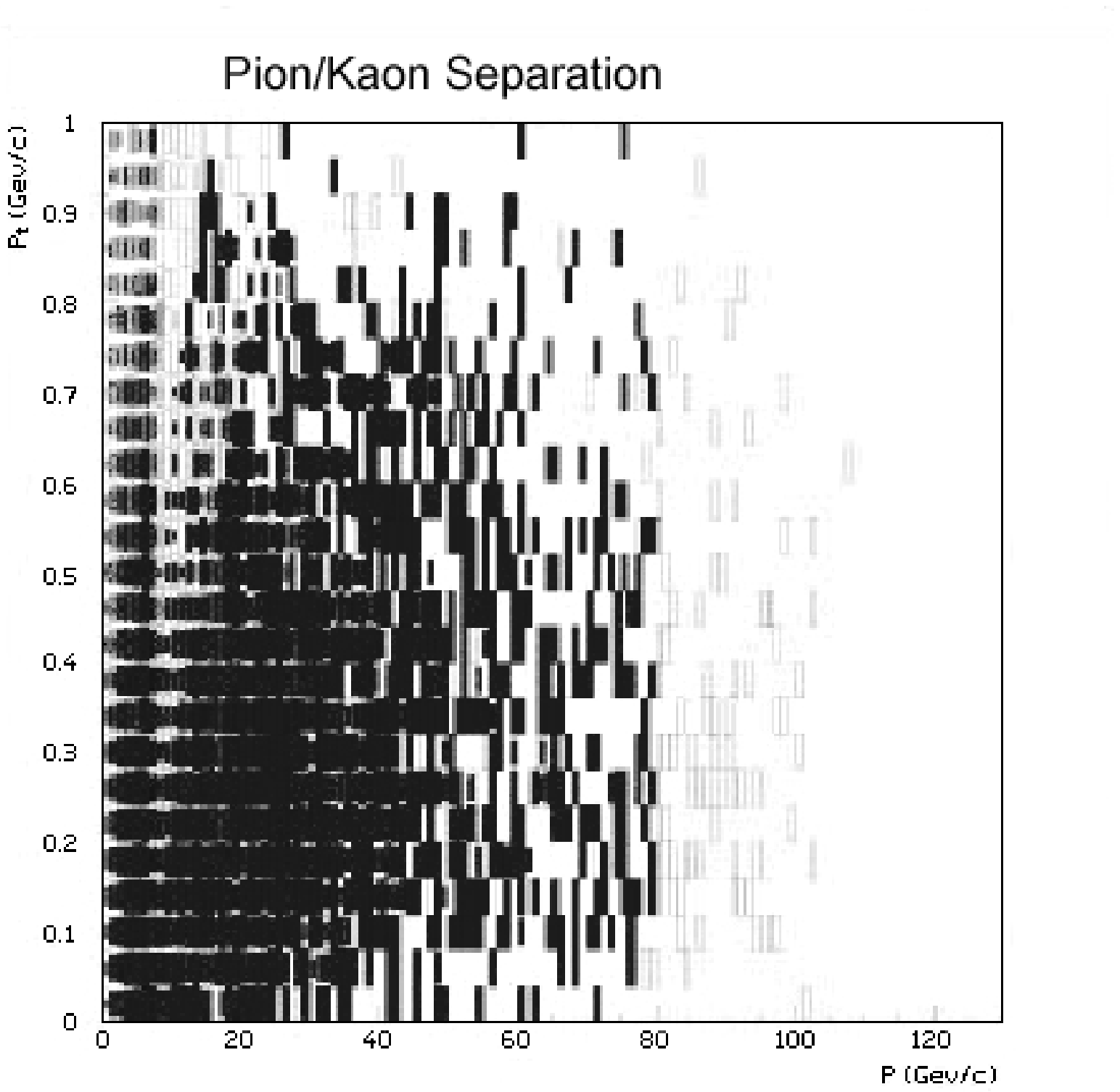}
\end{minipage}
\begin{minipage}{15pc}
\includegraphics[width=2.5in]{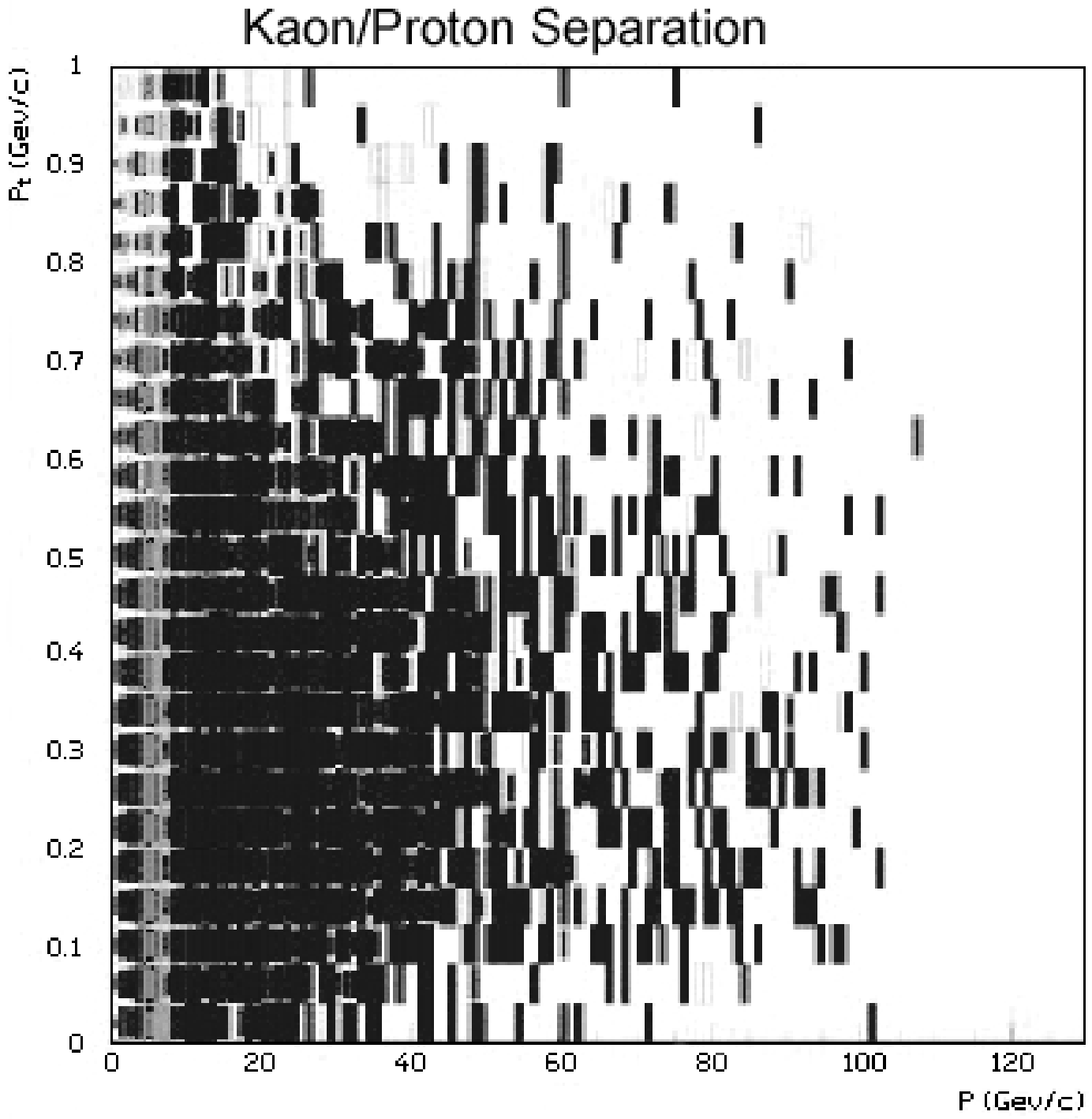}
\end{minipage}
\caption{Particle identification plots for pion/kaon separation and for
kaon/proton separation as a function of the longitudinal and
transverse momentum of the outgoing final state particle. Black
indicates separation at the $3\sigma$ level or better and grey
indicates separation at the $1-3\sigma$ level. The boxes at largest
values of the longitudinal momenta suffer from lack of kaon
statistics.}
\label{pid}
\end{figure}

MIPP uses $dE/dx$ in the TPC to separate pions, kaons and protons for
momenta less than $\approx$ 1~GeV/c  and the time of flight array of
counters to do the particle identification for momenta less than
2~GeV/c. The multi-cell \v Cerenkov detector~\cite{e690} contributes
to particle identification in the momentum range $\approx$ 2.5~GeV/c-14~GeV/c
and the RICH~\cite{rich} for momenta higher than this. By combining
information from all counters, we get the expected particle
identification separation for $K/p$ and $\pi/K$ as shown in
Figure~\ref{pid}. It can be seen that excellent separation at the
$3\sigma$ or higher level exists for both $K/p$ and $\pi/p$ over
almost all of phase space.  Tracking of the beam particles and
secondary beam particles is accomplished by a set of drift
chambers~\cite{chambers1} and proportional chambers~\cite{chambers2}
each of which have 4 stereo layers.
\subsection{Some results from Acquired data}
\begin{figure}[htb!]
\begin{minipage}{30pc}
\includegraphics[width=\textwidth]{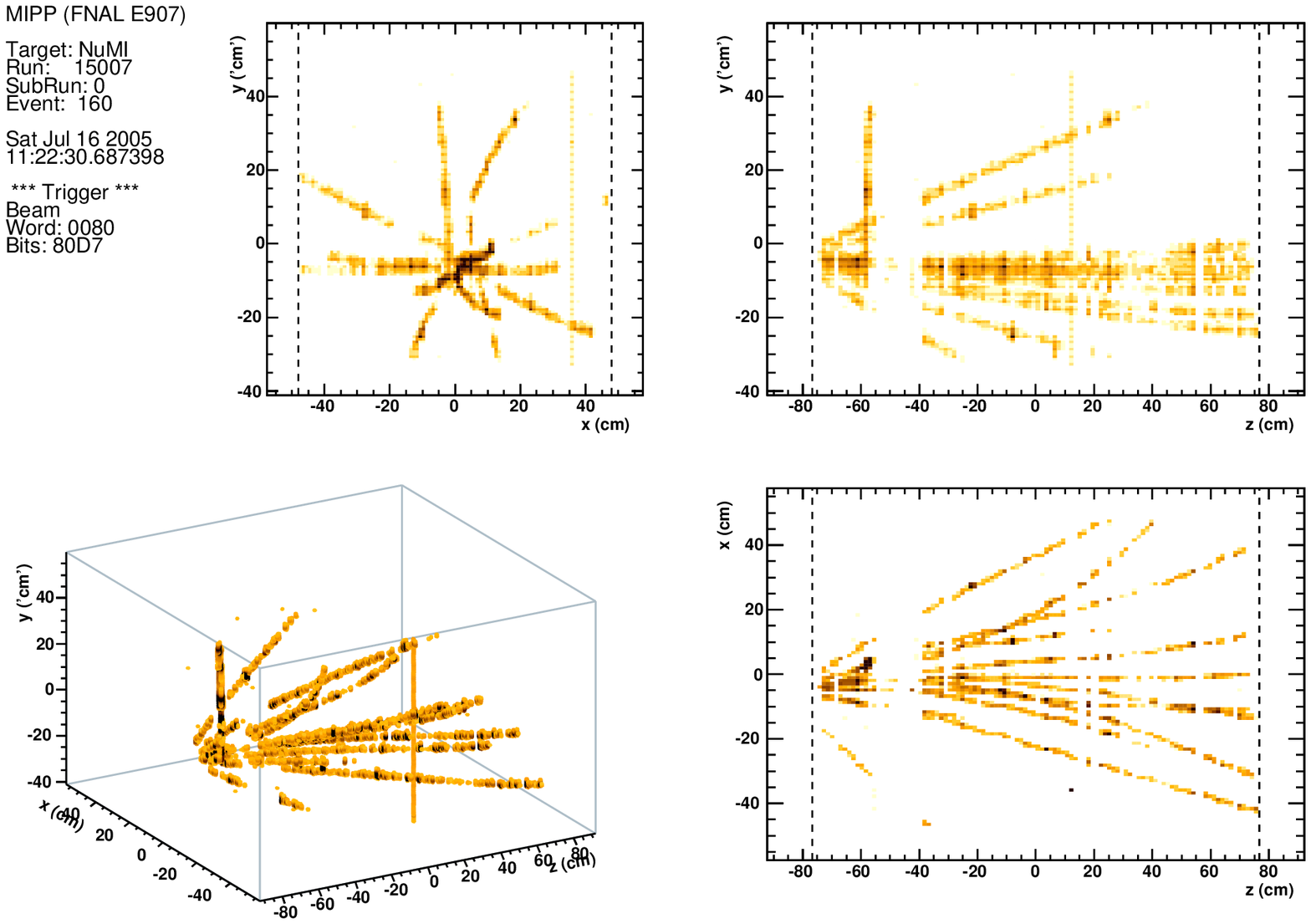}
\end{minipage}
\begin{minipage}{30pc}
\includegraphics[width=\textwidth]{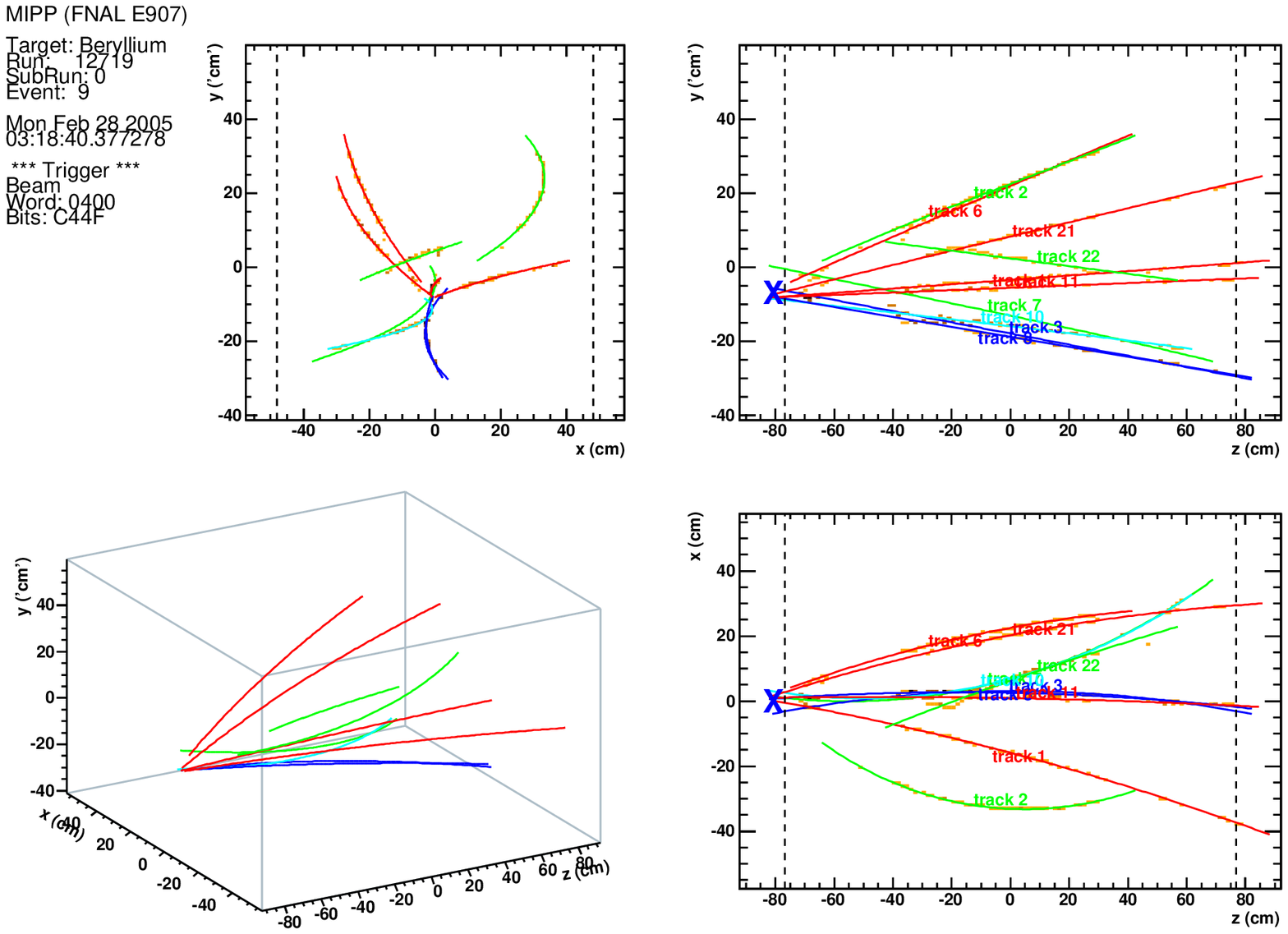}
\end{minipage}
\caption{RAW and Reconstructed TPC tracks  from two different events.}
\label{tpc1}
\end{figure}

Figure~\ref{tpc1} shows the pictures of reconstructed tracks in the
TPC obtained during the data-taking run. The tracks are digitized and
fitted as helices in three dimensions. Extrapolating three dimensional
tracks to the other chambers makes the pattern recognition
particularly easy.

\begin{figure}[htb!]
\begin{minipage}{15pc}
\includegraphics[width=2.5in]{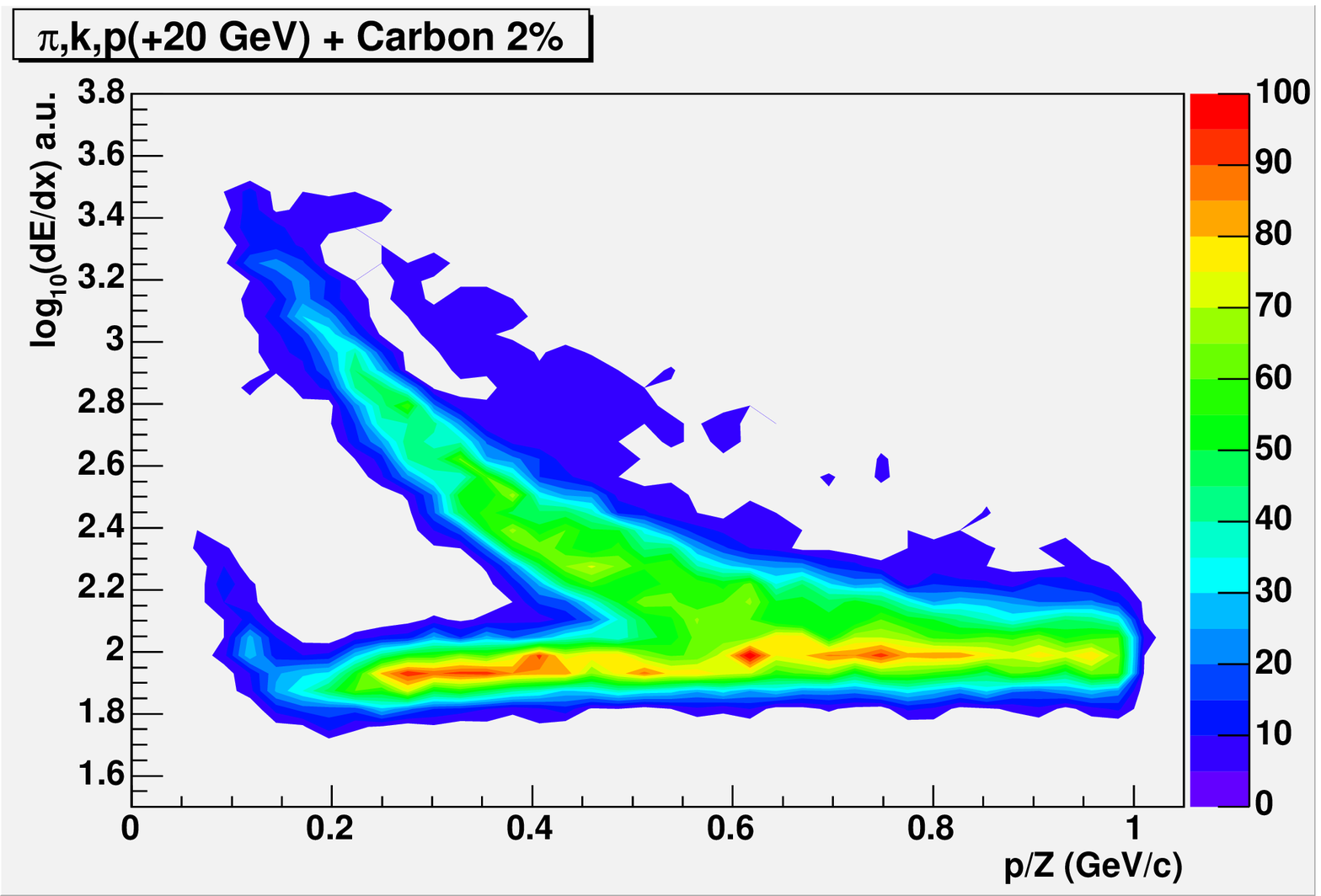}
\end{minipage}
\begin{minipage}{15pc}
\includegraphics[width=2.5in]{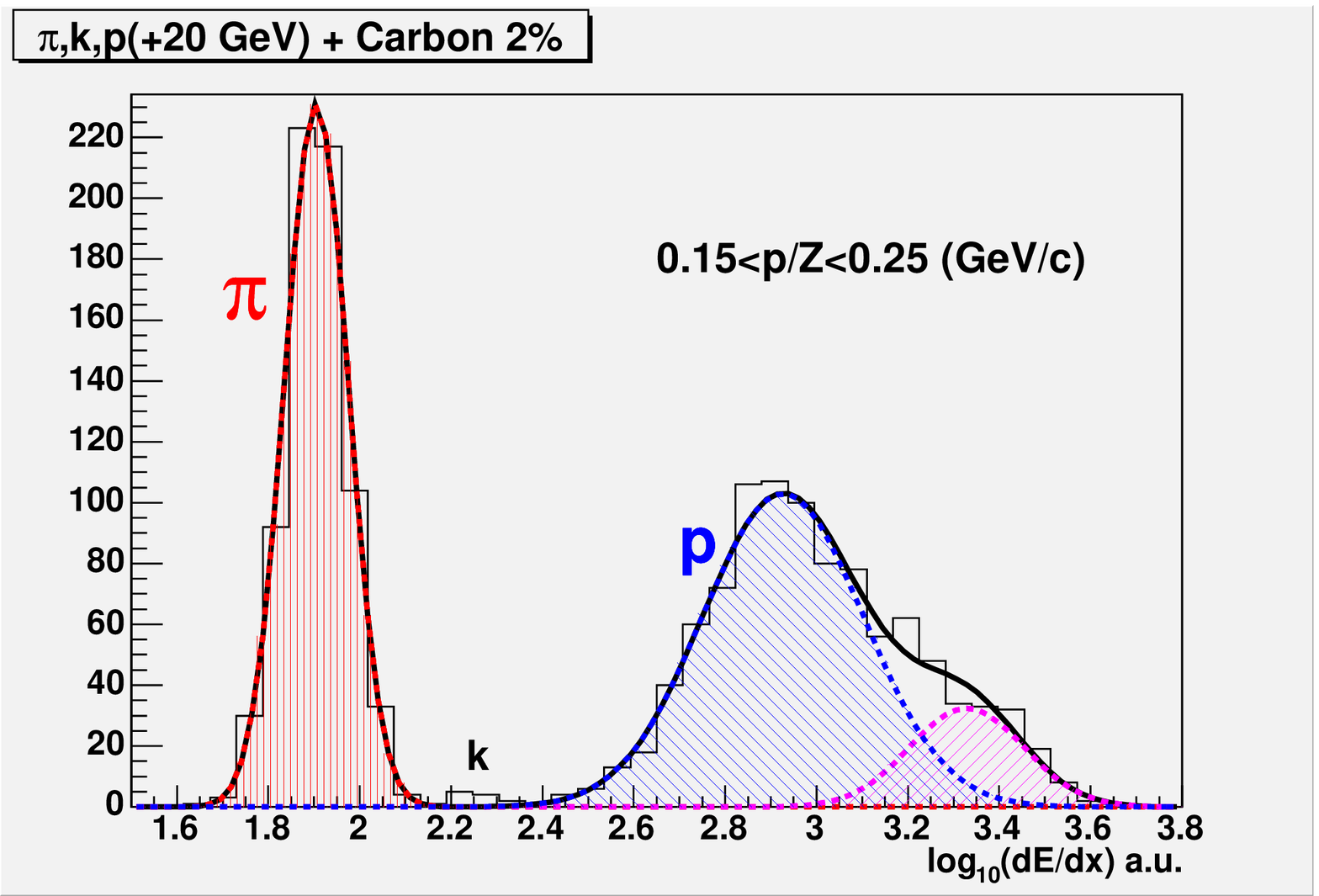}
\end{minipage}
\caption{Preliminary dE/dx distributions in the TPC The scatter plot
shows the pion and proton peaks in the distribution as a function of
the lab momentum. The second plot is the projection on the dE/dx axis
for a momentum slice 150~MeV/c to 250~MeV/c for p-Carbon data.}
\label{dedx}
\end{figure}

\begin{figure}[htb!]
\centerline{
\begin{minipage}{15pc}
\includegraphics[width=2.5in]{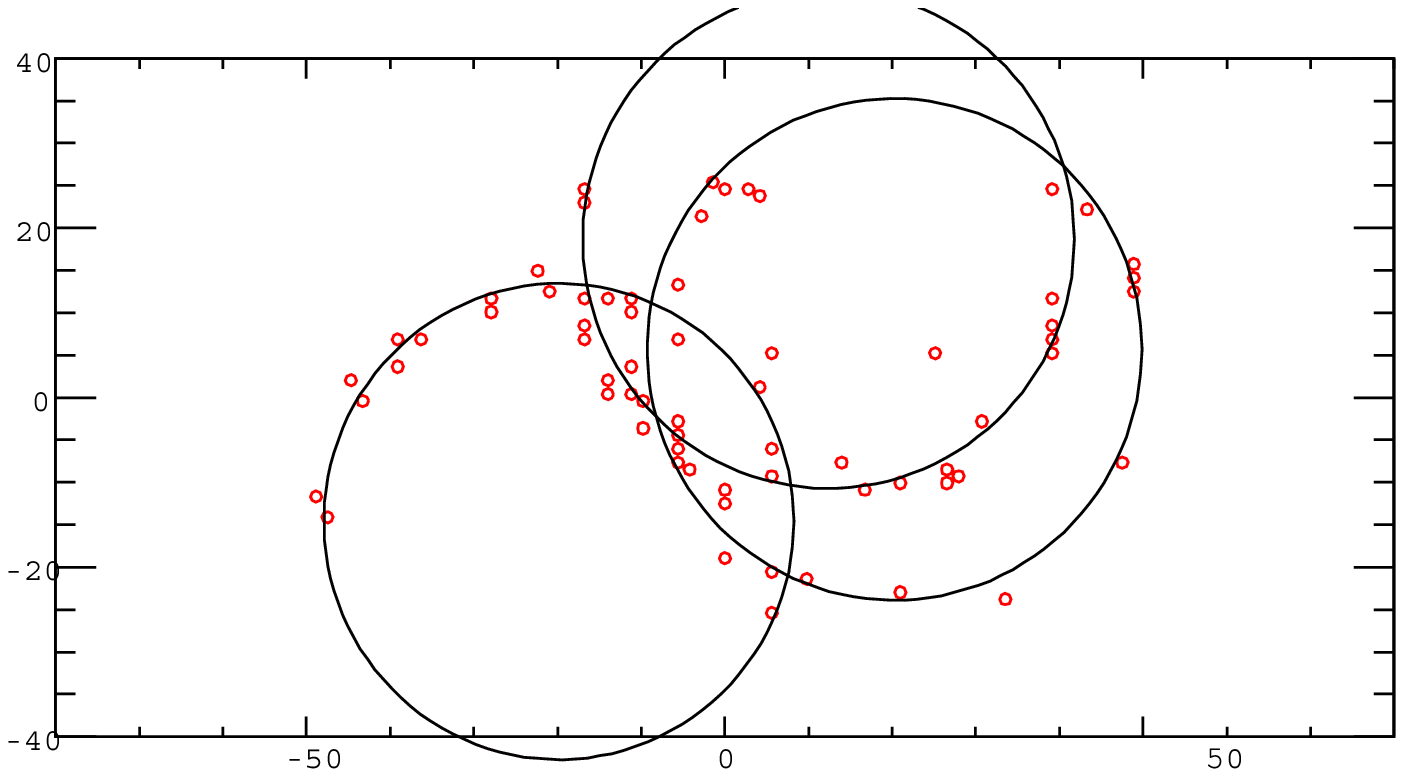}
\end{minipage}
\begin{minipage}{15pc}
\includegraphics[width=2.5in]{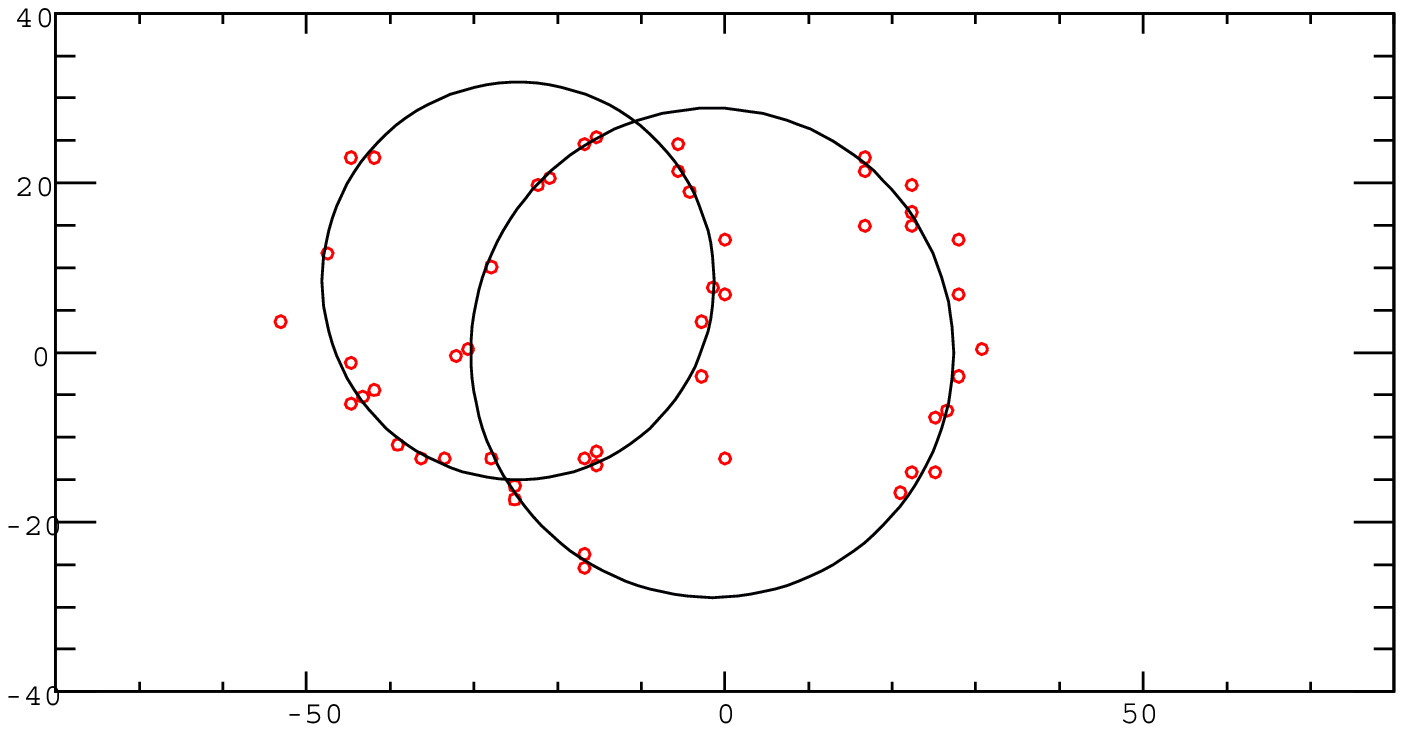}
\end{minipage}
}
\centerline{
\begin{minipage}{15pc}
\includegraphics[width=2.5in]{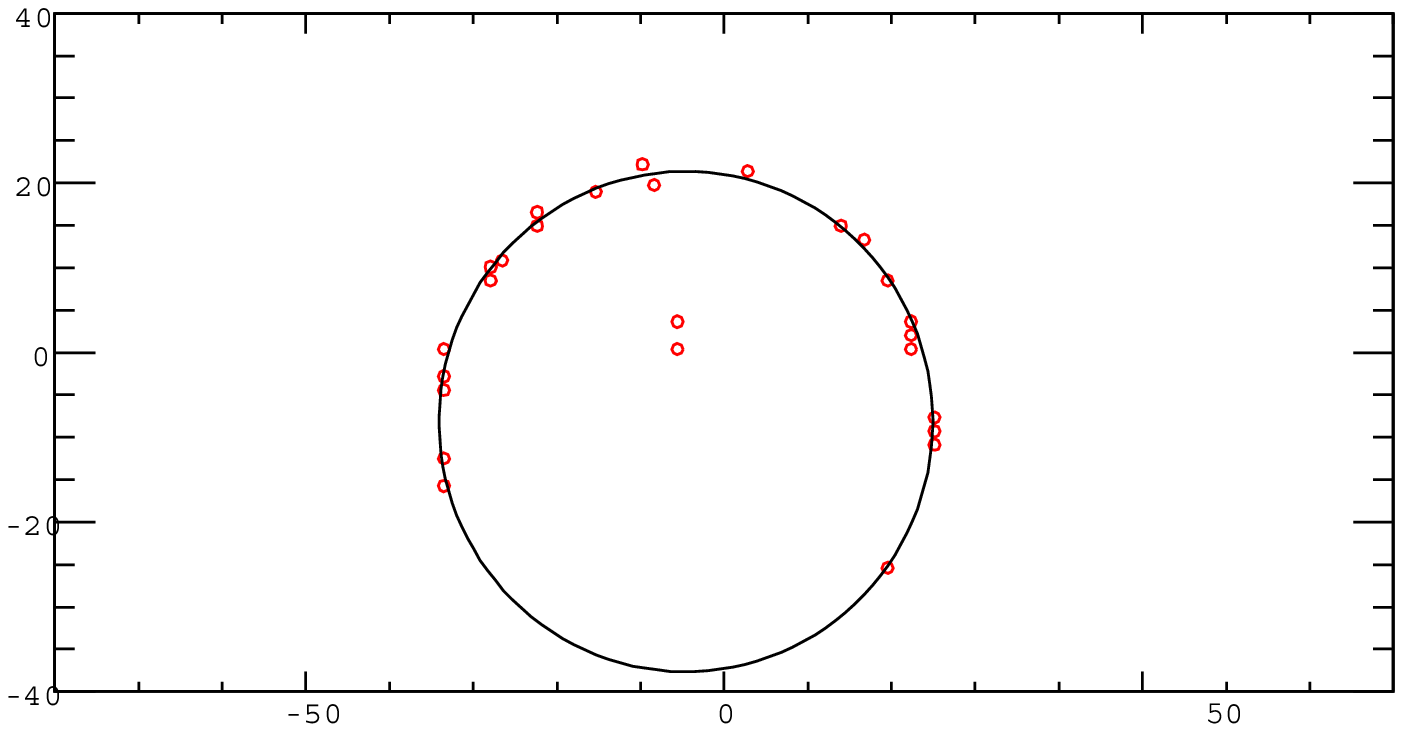}
\end{minipage}
}
\caption{\label{rings}Examples of events with rings in the RICH counter 
for a 40~GeV/c beam. The x and y axes are in cm.}
\end{figure}

\begin{figure}[htb!]
\centerline{
\begin{minipage}{15pc}
\includegraphics[width=2.5in]{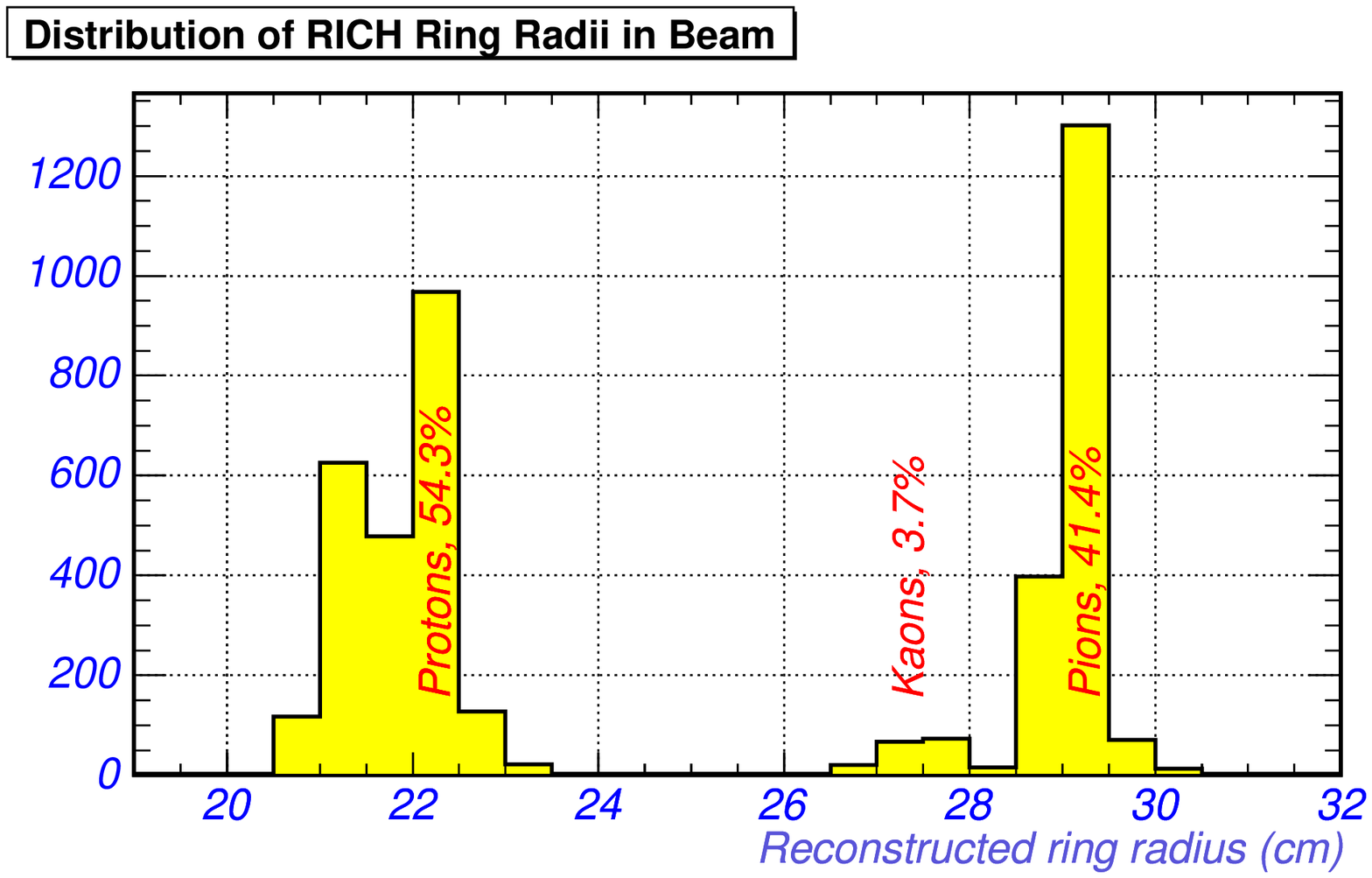}
\end{minipage}
\begin{minipage}{15pc}
\includegraphics[width=2.5in]{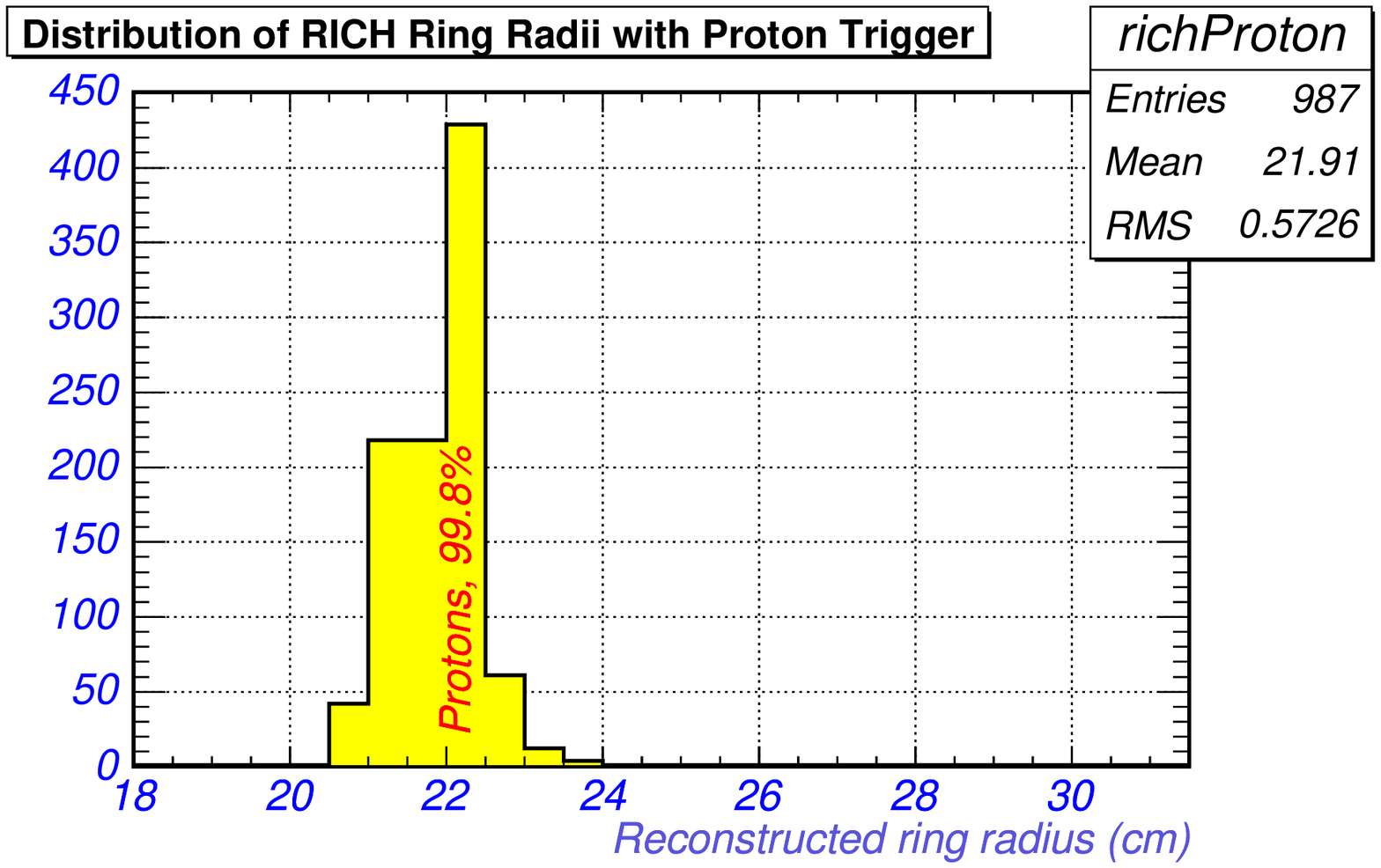}
\end{minipage}
}
\centerline{
\begin{minipage}{15pc}
\includegraphics[width=2.5in]{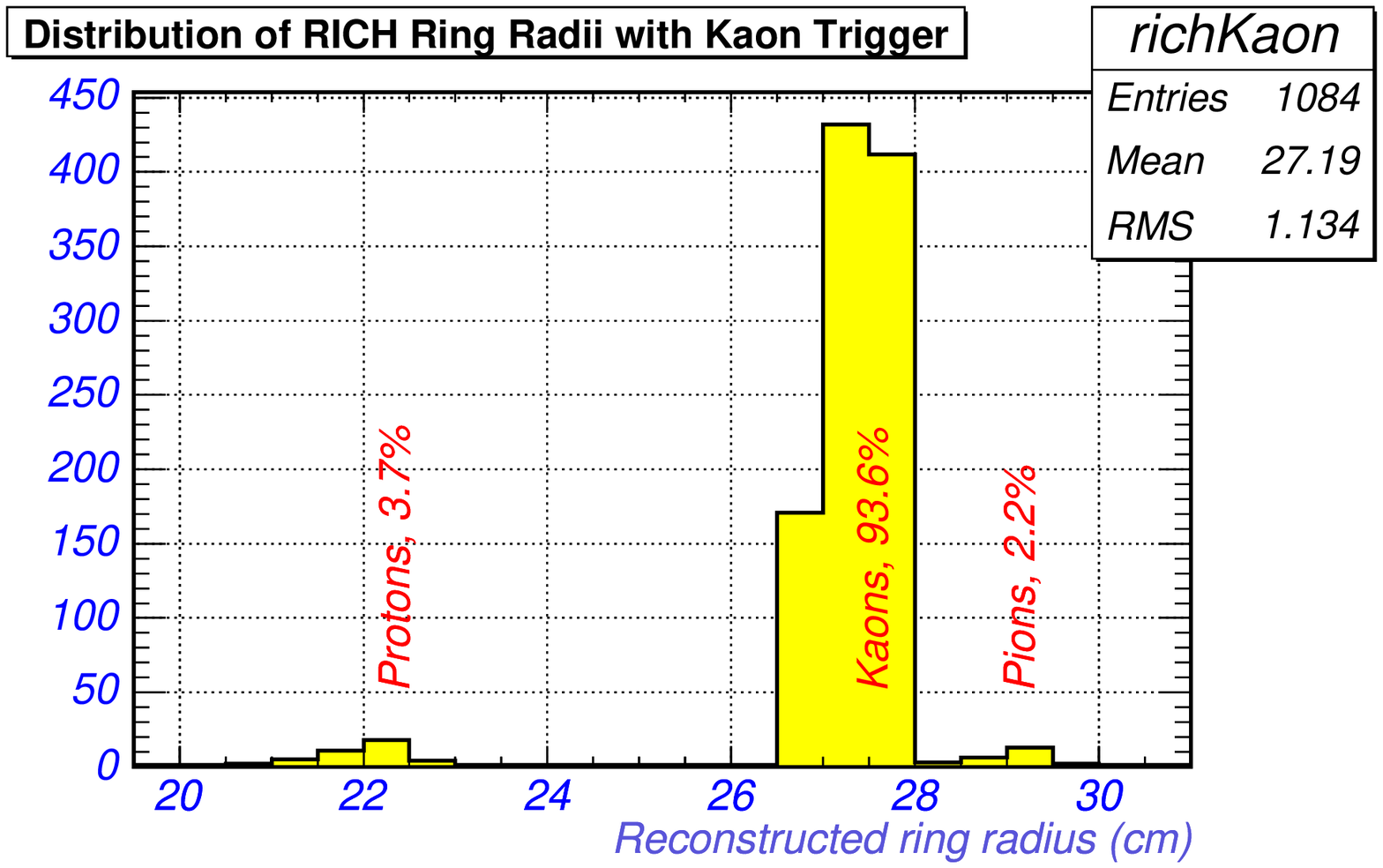}
\end{minipage}
\begin{minipage}{15pc}
\includegraphics[width=2.5in]{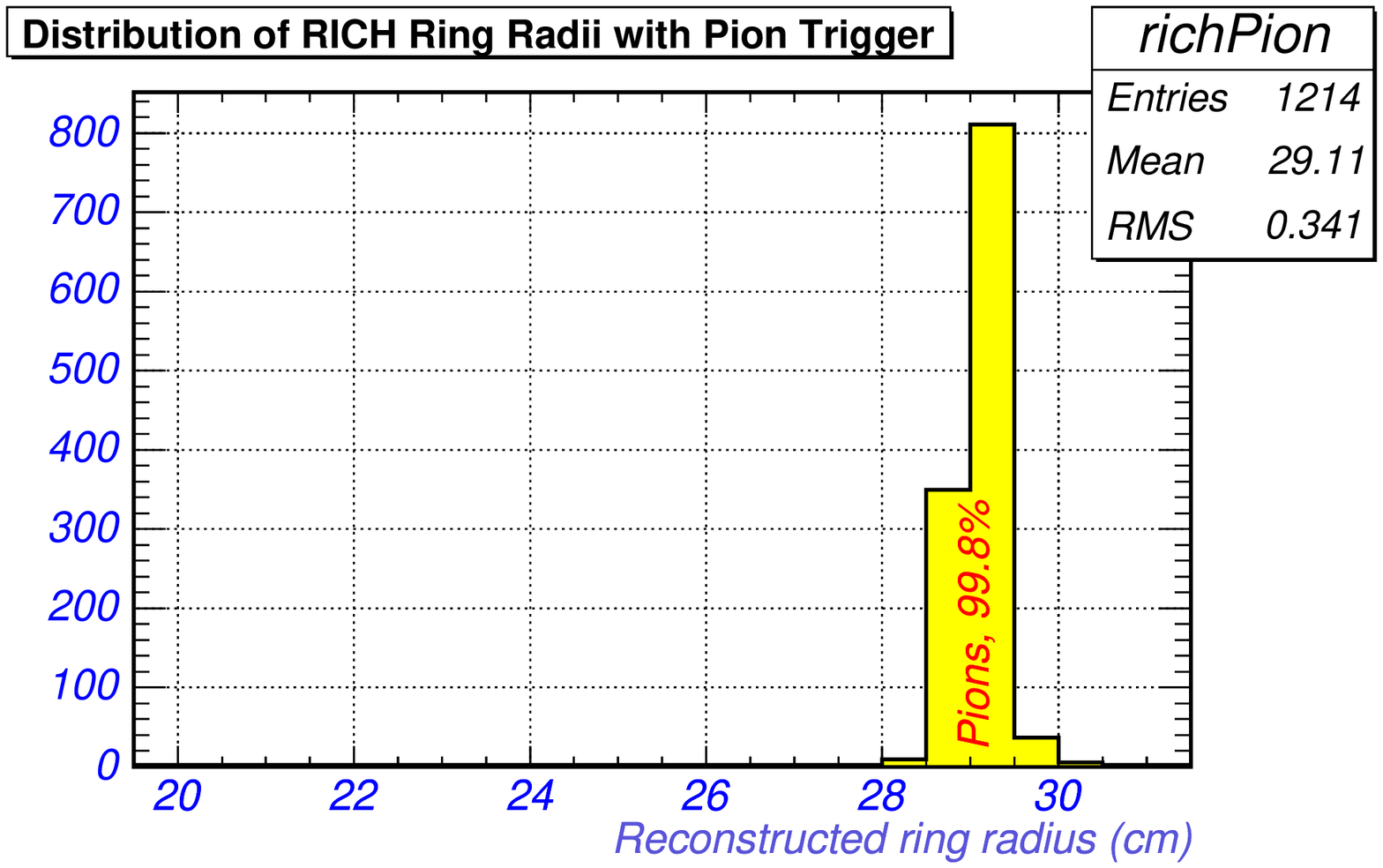}
\end{minipage}
}
\caption{An example of a 40 GeV/c primary beam (non-interacting)
trigger. The RICH identifies protons, kaons and pions by the ring
radii. The beam \v Cerenkov detectors can be used to do the same. When
the beam \v Cerenkov identification is used, one gets a very clean
separation of pions, kaons and protons in the RICH.}
\label{rbck}
\end{figure}

Figure~\ref{dedx} shows the distribution of $dE/dx$ of tracks measured
in the TPC as a function of the track momentum in a preliminary
analysis of p-Carbon data. The TPC is capable of separating pions,
protons and kaons in the momentum range below $\approx$ 1~GeV/c.
Figure~\ref{rings} shows events with rings in the RICH counter. Some
are due to single beam tracks and others are due to tracks from
interactions.
Figure~\ref{rbck} shows the histogram of ring radii for a +40~GeV
secondary beam. There is clean separation between pions, kaons and
protons and their relative abundances~\cite{malensek} match
expectations. Applying the particle identification trigger from the
beam \v Cerenkovs enables us to separate the three particle species
cleanly. The kaons which form ~4\% of the beam are cleanly picked out
by the beam \v Cerenkov with very simple selection criteria. These can
be made much more stringent with offline cuts to produce a very clean
kaon beam. 

The ring radius of the particle contains information on the mass of
the particle. The pion and proton masses are very well known. The
charged kaon mass, however, currently has measurement uncertainties of
the order of 60~keV. Improving the precision of both charged kaon
masses will pay dividends in CP violation experiments involving
charged kaons where the matrix elements depend on the kaon mass raised
to large powers. Towards the end of our physics run, when the Jolly
Green Giant magnet coils failed, we switched off the TPC and acquired
data at the rate of 300~Hz to investigate how well we can measure the
charged kaon mass. These events, whose statistics are indicated in
Figure~\ref{tab1}, are currently being analyzed to evaluate the
systematics involved in such a measurement.
\subsubsection{NuMI target measurements}
\begin{figure}[htb!]
\centerline{
\begin{minipage}{15pc}
\includegraphics[width=\textwidth]{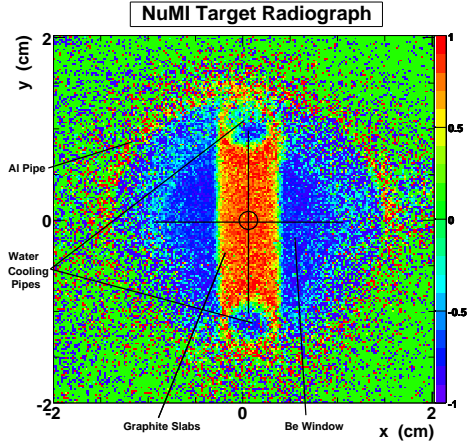}
\end{minipage}
}
\caption{Radiograph of the MINOS target. The beam direction is perpendicular 
to the paper.}
\label{minos}
\end{figure}

\begin{figure}[htb!]
\centerline{
\begin{minipage}{15pc}
\includegraphics[width=\textwidth]{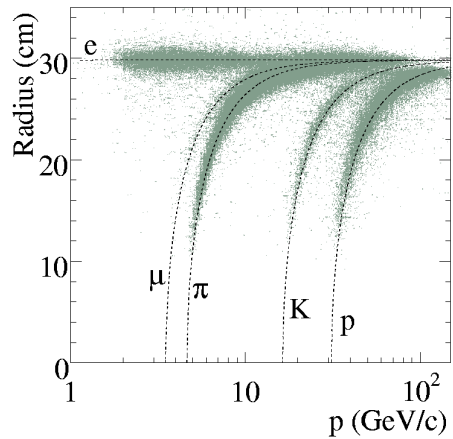}
\end{minipage}
}
\caption{Preliminary data of RICH ring radii of positive 
tracks from the NuMI target vs momentum. Superimposed are the expected curves
for $e,~\mu,~K$ and $p$ particles.}

\label{ringrads}
\end{figure}

\begin{figure}[htb!]
\begin{minipage}{18pc}
\includegraphics[width=0.8\textwidth,angle=-90]{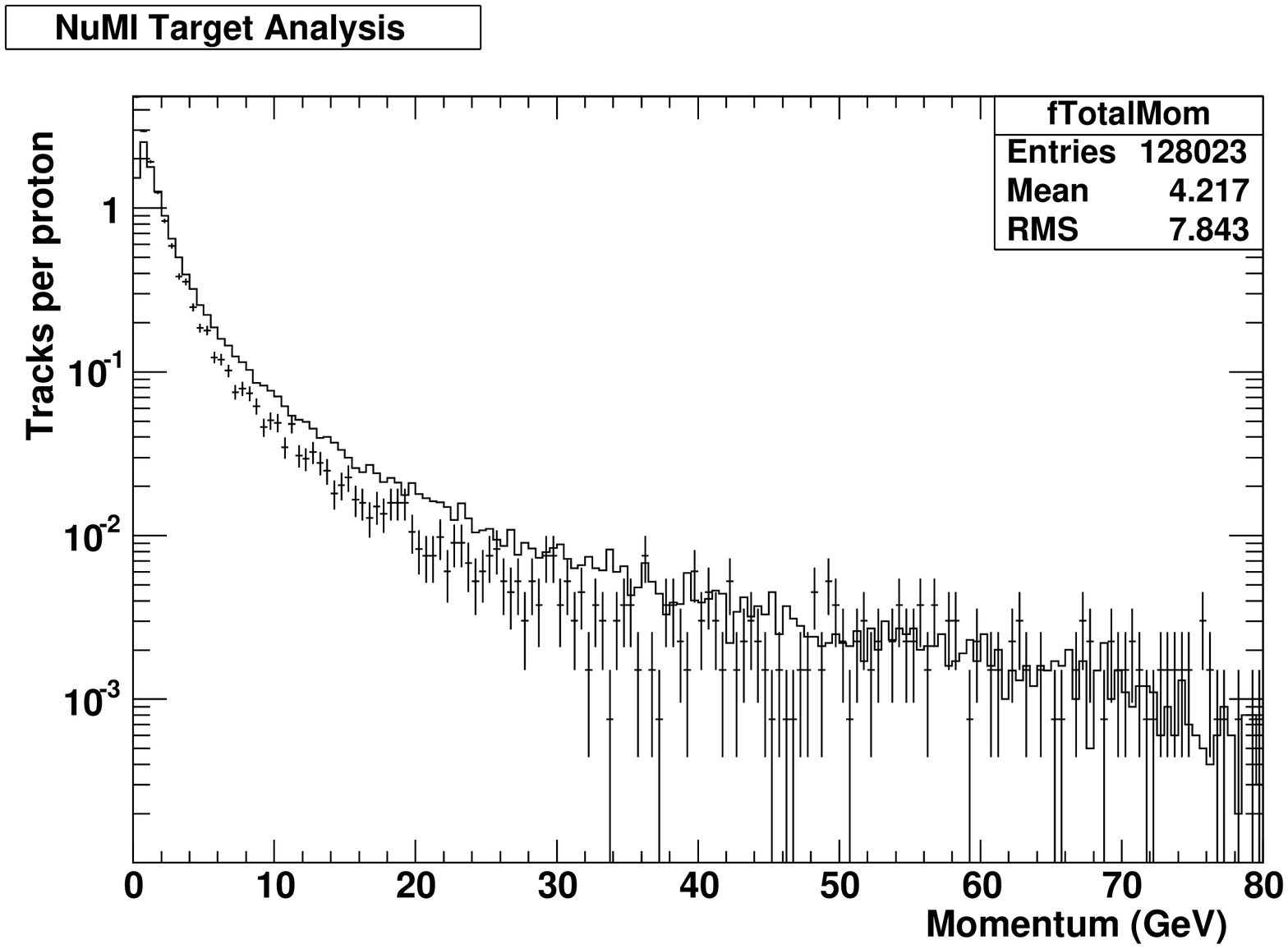}
\end{minipage}
\begin{minipage}{18pc}
\includegraphics[width=0.8\textwidth,angle=-90]{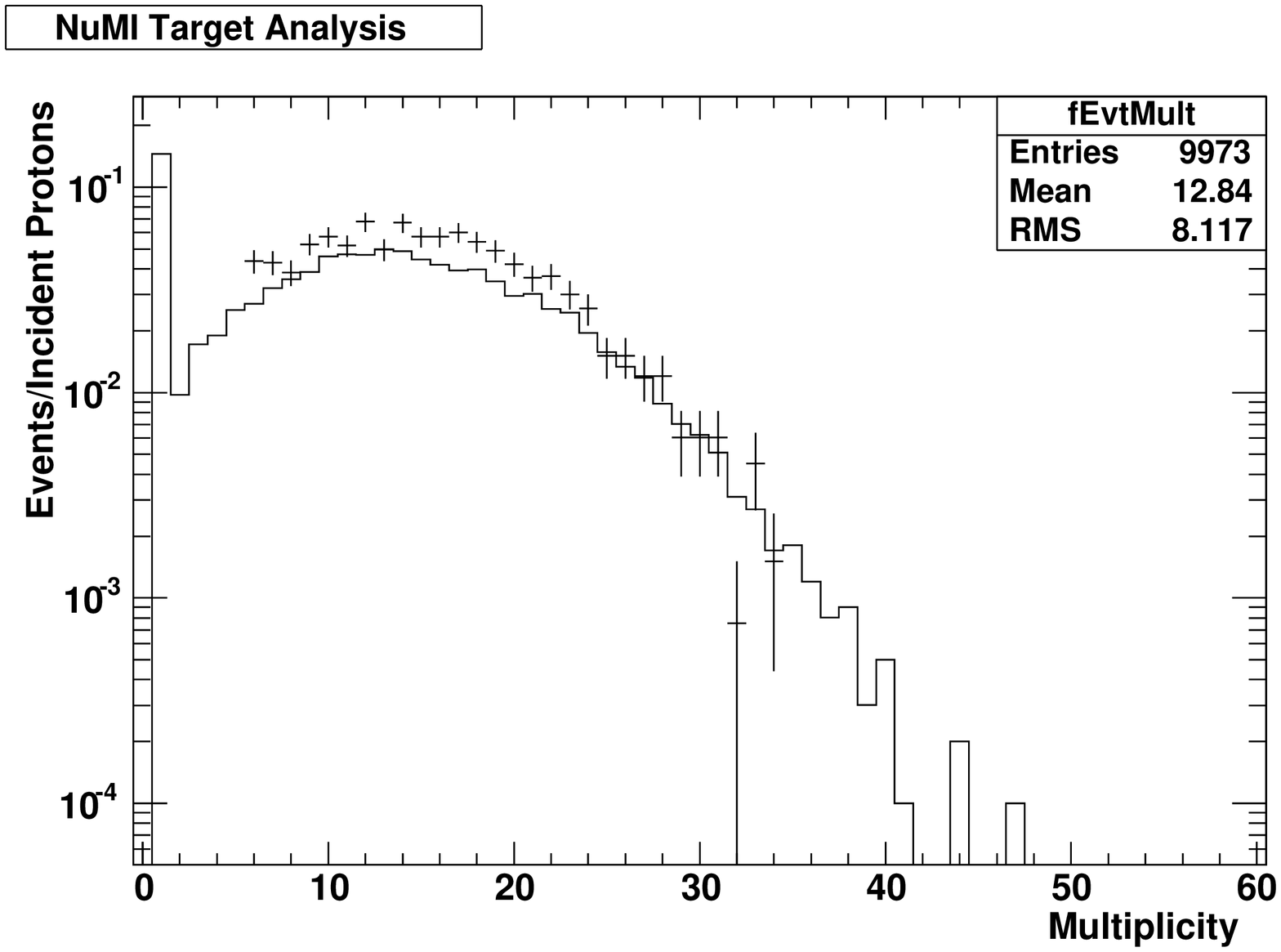}
\end{minipage}
\caption{Comparison (preliminary) of NuMI target data with predictions of the 
FLUKA Monte Carlo in charged particle momentum and multiplicity. 
Data have error bars}
\label{numi-prelim}
\end{figure}

MIPP took 1.75 million events using 120~GeV/c primary beam protons
impinging on the NuMI (spare) target. These events will play a crucial
role in the prediction of neutrino fluxes in the NuMI beamline and
will enable the MINOS experiment to control the systematics in the
near/far detector ratios as well as helping them understand the near
detector performance. Figure~\ref{minos} shows a radiograph of the
MIPP measurements of the MINOS target.  
The graphite slabs and cooling tubes can be seen.
These events were obtained during the commissioning phase of this 
target measurement where the beam was not yet fully focussed and 
aligned on the target. The 1.75~Million events on the NuMI target were 
obtained after the beam was aligned and centered on the target. 
Figure~\ref{ringrads} shows the rich ring radii vs momentum of positive 
tracks originating from the NuMI target. Superimposed are the curves for 
known particles. This shows the excellent particle identification of the
 MIPP detector for forward going particles.
Figure~\ref{numi-prelim} shows the results of a preliminary analysis
of NuMI target data in both energy spectrum of the tracks and track
multiplicity and compares it to the FLUKA Monte Carlo.
\subsubsection{Target fragmentation multiplicities as a function of
Atomic Number} 
\begin{figure}[htb!]
\begin{center}
\includegraphics[width=0.6\textwidth]{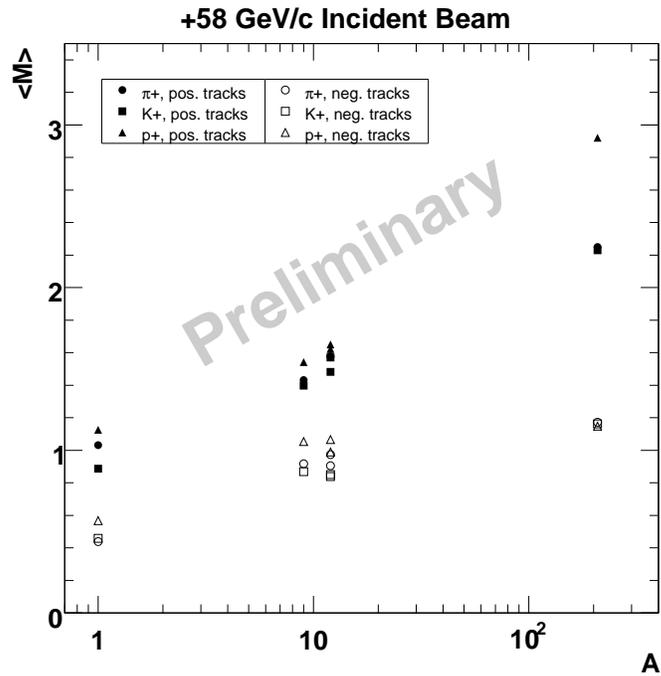}
\end{center}
\caption{Mean multiplicity $<M>$ 
of positive and negative tracks in the momentum range 
0.1~GeV/c- 1.0~GeV/c (target fragmentation region) as a function of
atomic number A of the target ($H_2$, Be, C and Bi) at 58~GeV/c beam
momentum for the positive beam species $\pi^+$, $K^+$ and $p$.}
\label{cmult}
\end{figure}

We have analyzed the multiplicities in the TPC as a
function of Atomic number A of the target. Here we show
(Figure~\ref{cmult}) a preliminary analysis of the multiplicity of
both positive and negative tracks in the momentum range
0.1~GeV/c-1.0~GeV/c for the nuclear targets $H_2$, Be, C, and Bi for 3
positive beam species $\pi^+, K^+$ and p at 58~GeV/c incident momentum. 
The data show a rise in target fragmentation multiplicity as a function of A.
The positive multiplicities are higher, reflecting the charge of the target. 
%

\section{The proposal in a nutshell}

\subsection{Beam Delivery rate assumed}
In what follows, we will assume that the Main Injector delivers one 
4~second slow spill
every two minutes to MIPP in the upgraded mode, with a
 machine downtime of 42\%, 
and that the MIPP DAQ has been upgraded to run at 3~kHz. These are 
conservative estimates of machine delivery rate and downtime.
At this rate, we are capable of acquiring 5~Million events per day.

\subsection{Replacing the Jolly Green Giant Coils}
One month before the end of our run in March 2006, the Jolly Green
Giant magnet failed. Two of its 4 coils became inoperative due to
shorts. The Jolly Green Giant coils were fabricated in the 1960's and
have seen a lot of power cycles. Even if we fix the broken coils,
there is no guarantee as to how long the remaining coils will last. We
have decided to replace all four coils with two aluminum coils with
the same field strength as before. The aluminum conductor is cheaper
than copper. In the process, we have made the coils longer along the
beam direction by $\pm$ 9 inches so as to provide a more uniform field
for the electron drift in the TPC.

\subsection{TPC DAQ upgrade}

The MIPP sub-detector with the largest data output is our TPC. It is also
the slowest in outputing this data, since its data acquisition
electronics were designed and built~\cite{Rai90} in the early
1990's. The TPC  runs at $\approx$ 60~Hz for very simple events (single beam
tracks). For complicated events this rate currently falls to
$\approx$ 20~Hz. 
With modern electronics, it is possible to increase the DAQ rate
to 3000~Hz, resulting in an over-all increase of 150 in our data
acquisition capability. We propose to use the ALTRO/PASA chips
designed and  tested for the ALICE collaboration at the
LHC~\cite{Musa03}. This technology is also being used for the STAR ,
BONUS and TOTEM experiments. A production run for these experiments 
is being negotiated now. If MIPP were to get in on this order, the 
total cost to MIPP to procure these chips (1100 of them) would 
be $\approx$ \$80,000.

We propose to acquire these chips and design and fabricate new front 
end electronics cards for the TPC. With one 4-second slow spill
every two minutes and allowing for a machine down-time of 42\%, we
can accumulate 5~Million events per day.

\subsection{Upgrade of the Rest of the DAQ}

We propose to improve the way MIPP is triggered, the DAQ
electronics of the drift and wire chambers, the time of flight
 and threshold \v Cerenkov counters and the calorimeters.
The only system that remains unaltered
is the RICH for which we built front end electronics for the first
run. We briefly outline the changes to the detector DAQ here.

\subsubsection{Triggering}

We will trigger MIPP using silicon pixel systems that were developed for the
BTeV experiment. We will have one pixel plane before the target and
two after the target. We will have a trigger scheme that will project
on the downstream pixels planes the expected un-interacted beam
position (the bulls-eye) and trigger the experiment if pixels outside
this bull's eye are hit. This scheme will enable us to trigger on low
multiplicity events (including elastics) in an unbiased fashion. This
would improve the existing MIPP trigger system that utilized a
combination of a scintillator counter in conjunction with the first
drift chamber multiplicity to provide the interaction trigger. While
this system performed satisfactorily in our first run, it suffered
from Landau fluctuations in the scintillator and periodic
over-efficiency in the wire chamber. The digital nature of the pixel
signal eliminates Landau tails completely.

\subsubsection{Chamber electronics}

We propose to replace the aging drift chamber electronics 
(remnants of the E690 system) with a more modern electronics system designed 
and built in-house at Fermilab. We will use the same electronics for the
Proportional chambers to replace RMH electronics.

The current drift chamber electronics with a large power consumption 
stresses the air conditioning in MC7 to the limit. The new electronics 
will cause significantly less heating.

The Chamber electronics will utilize the same VME readout cards as the 
new TPC electronics. This reduces initial design costs and also 
simplifies the detector readout.

\subsubsection{Time of flight system and threshold \v Cerenkov detector}

We propose to replace the ToF and \v Cerenkov detector readout
electronics with electronics designed and built in-house. This design
will also utilize the same VME readout cards as used for the TPC
readout.

\subsubsection{Calorimeter readout}

The calorimeter readout electronics is currently too slow to obtain
the desired 3000Hz readout rate. We propose to replace this
electronics with a FERA adc system read out through new commercial
CAMAC branch controllers.

\subsection{Upgrading the beamline to run at lower momenta}
\begin{figure}[tbh!] 
\begin{center}
\includegraphics[width=\textwidth,angle=-90]{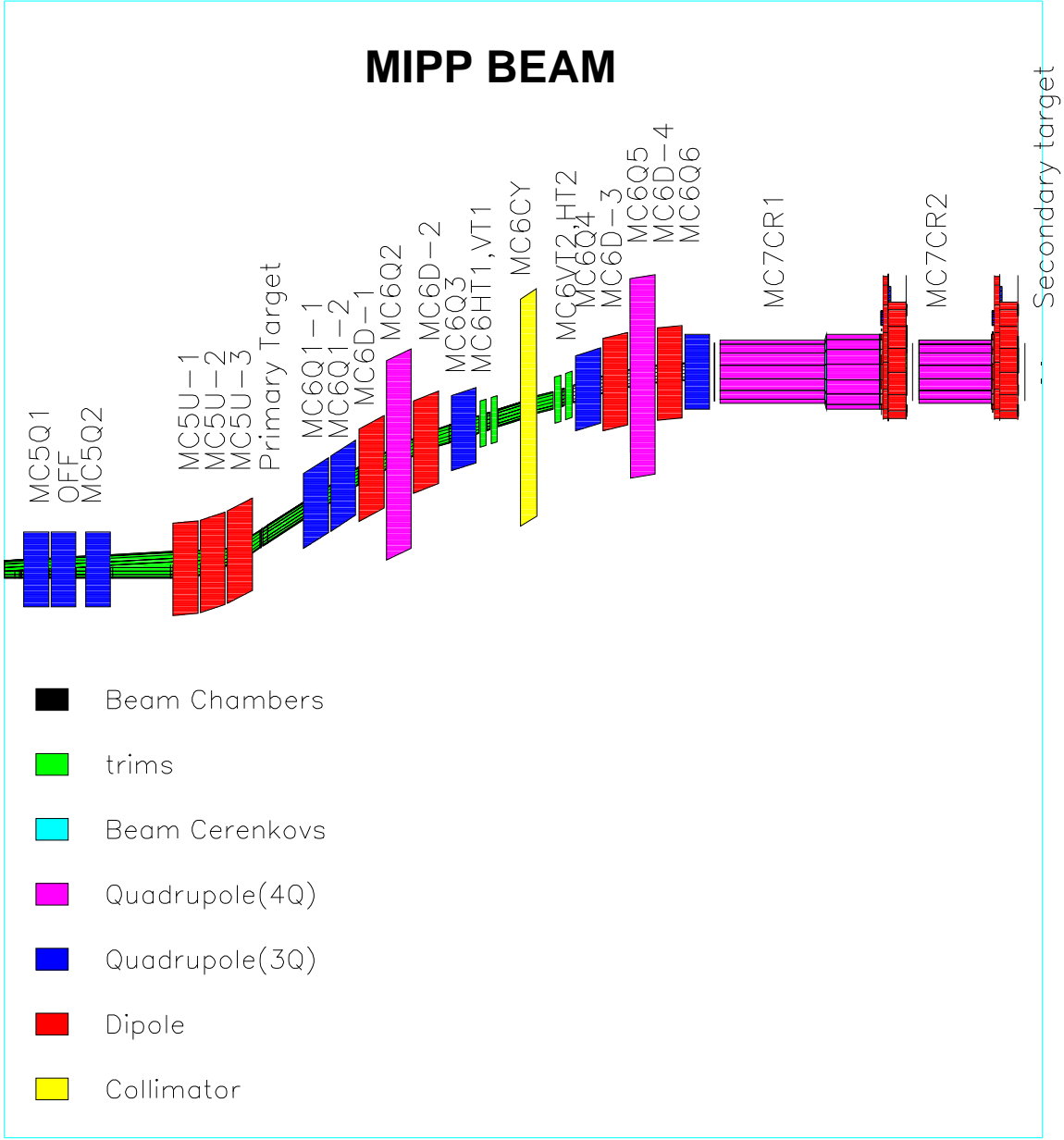} 
\caption{The MIPP secondary beamline. ~\label{beam}
}
\end{center}
\end{figure}

The MIPP secondary beamline performed very well during our physics run which 
concluded in March 2006. 
Figure~\ref{beam} shows a cut view of the beamline elements.
During our run, we managed to operate the beamline as low as 5~GeV/c and
as high as 120~GeV/c for the NuMI target run (The secondary beam
production target was removed and a pinhole collimator inserted to
reduce the Main Injector intensity). The beamline has been operated as low
as 1~GeV/c to measure the flux available, which was adequate. What is
lacking is the ability of the magnet power supplies to regulate the
very small currents. We propose to add low-current magnet power supplies that
are capable of this. We will switch to these power supplies during our
low momentum running. The residual field of the iron yokes becomes
significant at these low currents. Hall probes will be added to the
beamline magnets to monitor their field so that hysteresis effects can
be compensated for.

\subsection{Other tasks}
We propose several small improvements to the MIPP experiment to increase
reliability and maintainability of the experiment. These include several
changes to the gas systems, the cryogenic target, and slow monitoring,
as well as maintenance on Drift chambers and photomultiplier tubes
for the RICH and CKOV detectors.

In the present experiment, 
we operated a single scinitillator veto counter (which had a hole in it to 
let the beam through) to guard against beam spray. We plan to add to this 
counter and build a veto wall for the upgrade. 
In addition, 
a recoil detector (the Plastic Ball) will be added to detect low energy 
particles (both charged and neutral) 
that are produced at wide angles ($>80\deg$ to the beam) and miss the TPC.

\section{Summary of the proposed physics for the Upgraded MIPP run}

\subsection{Particle Production  on neutrino targets in the NuMI beam}

We outline here the need to measure the particle production on the
NuMI targets. In a disappearance experiment such as MINOS, the
evidence of neutrino oscillations is obtained from the difference in
shapes of the energy spectrum of the neutrino charged current events
in the far detector and the near detector. Because of the finite size
of the NuMI target and decay region, the angles of the decaying pions
that produce neutrinos reaching the near detector have a different
distribution than those reaching the far detector.  Put another way, the
neutrinos that interact in the near detector come from the decay of a
different kinematic mix of pions than those that interact in the far
detector. It is thus important to measure the dynamics of pion
production off the NuMI target.

\begin{figure}[htb!]
\centering
\includegraphics[width=\textwidth]{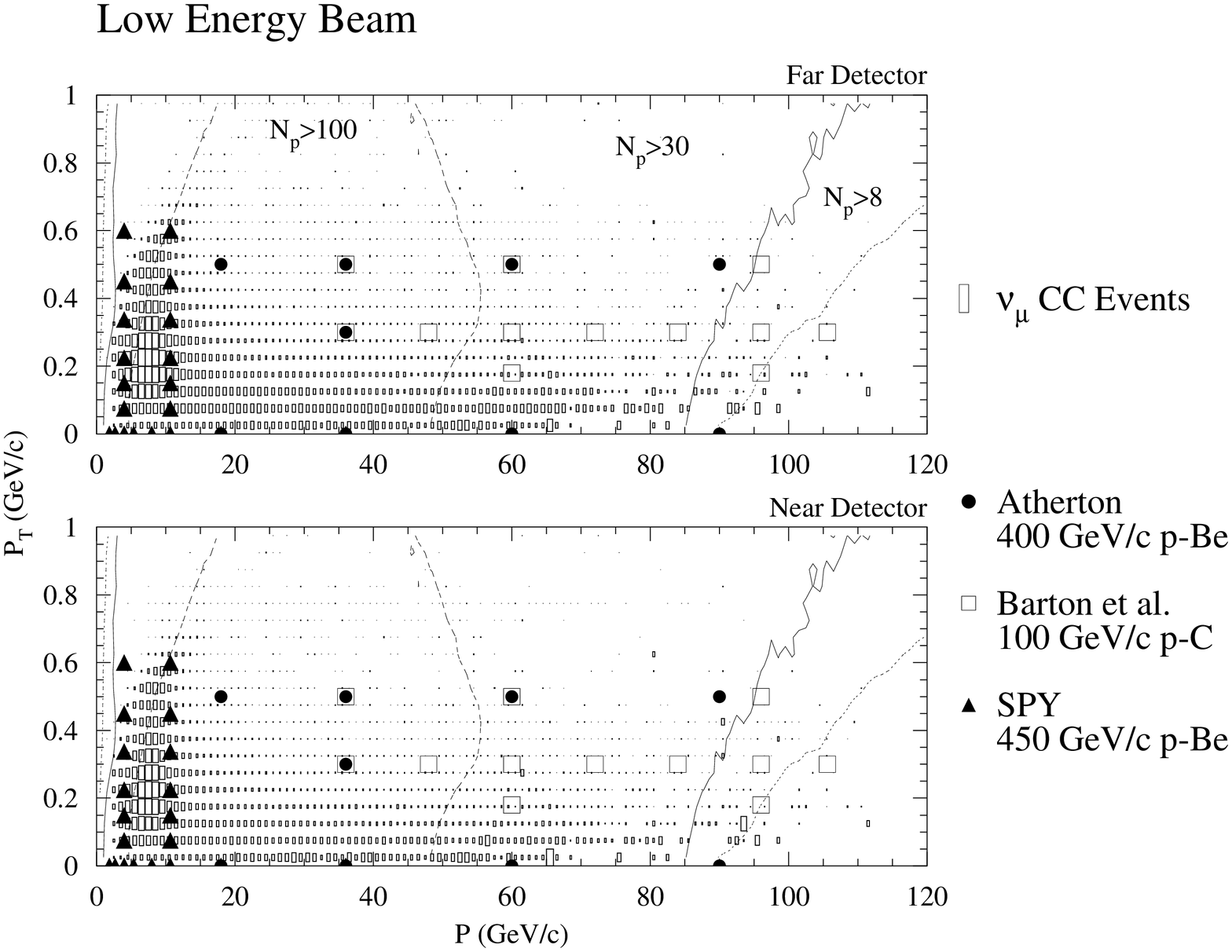}
 \caption{The distribution in longitudinal and transverse momentum of
 secondary pions produced on the NuMI target. Secondaries have been
 weighted by their contribution to the neutrino event rate at the far
 (top) and the near (bottom) detectors. Overlaid are the locations of
 existing hadron production measurements.}
\label{numippt}
\end{figure}

Figure~\ref{numippt} shows the distributions in longitudinal and
transverse momentum of pions weighted by their 
contribution to the neutrino event
rate in the far and near MINOS detectors. These weightings are
different in detail. Superimposed on this plot are  existing data on hadron
production obtained from mainly single arm spectrometer measurements~\cite{hadrprod},
which explains their discreteness in $p,p_T$ space.

\begin{figure}[htb!]
\centering
\includegraphics[width=\textwidth]{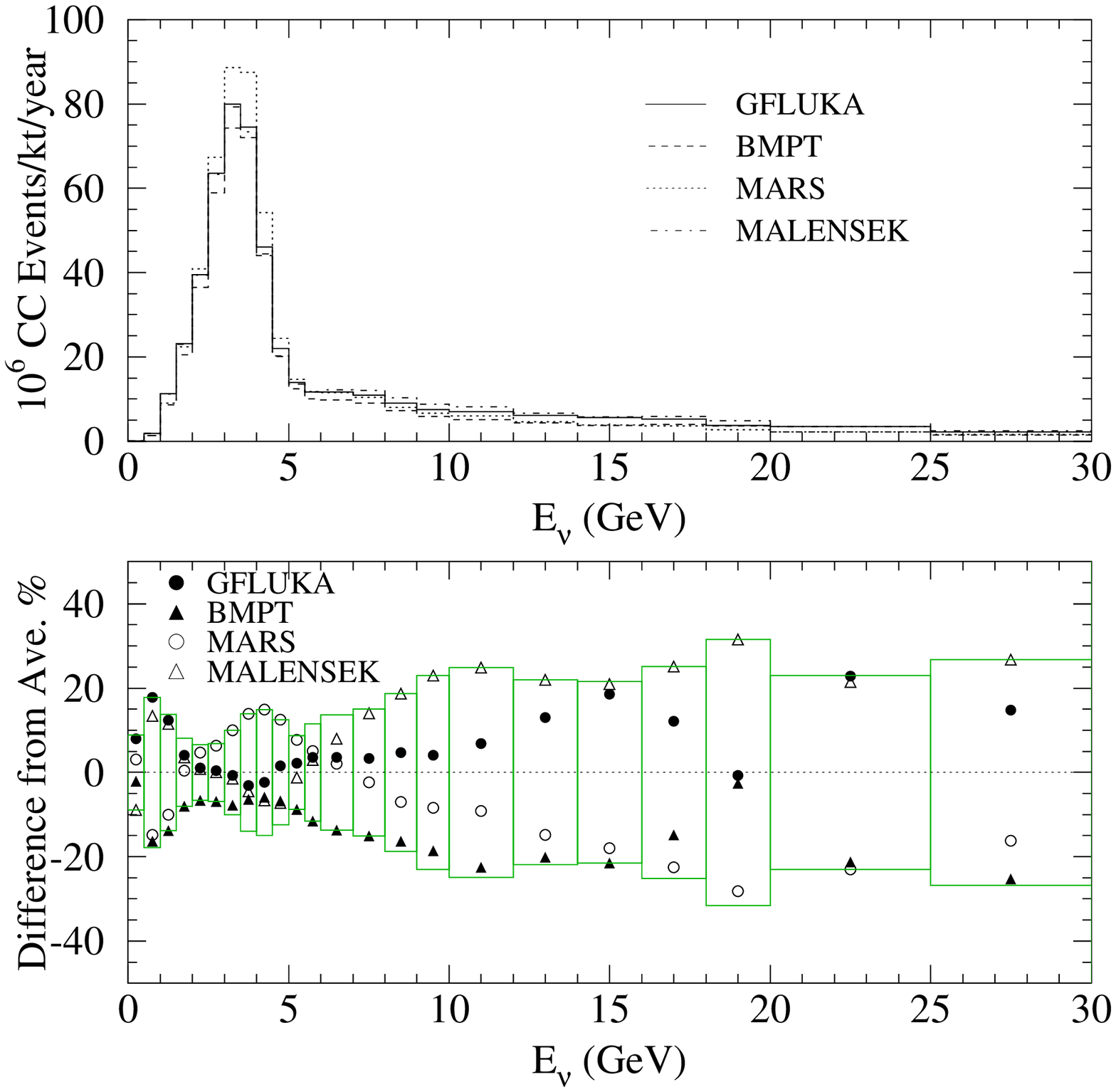}
 \caption{The predictions of the absolute neutrino rates at the MINOS near 
detector using various hadron production models.}
\label{nurates}
\end{figure}

\begin{figure}[htb!]
\centering
\includegraphics[width=\textwidth]{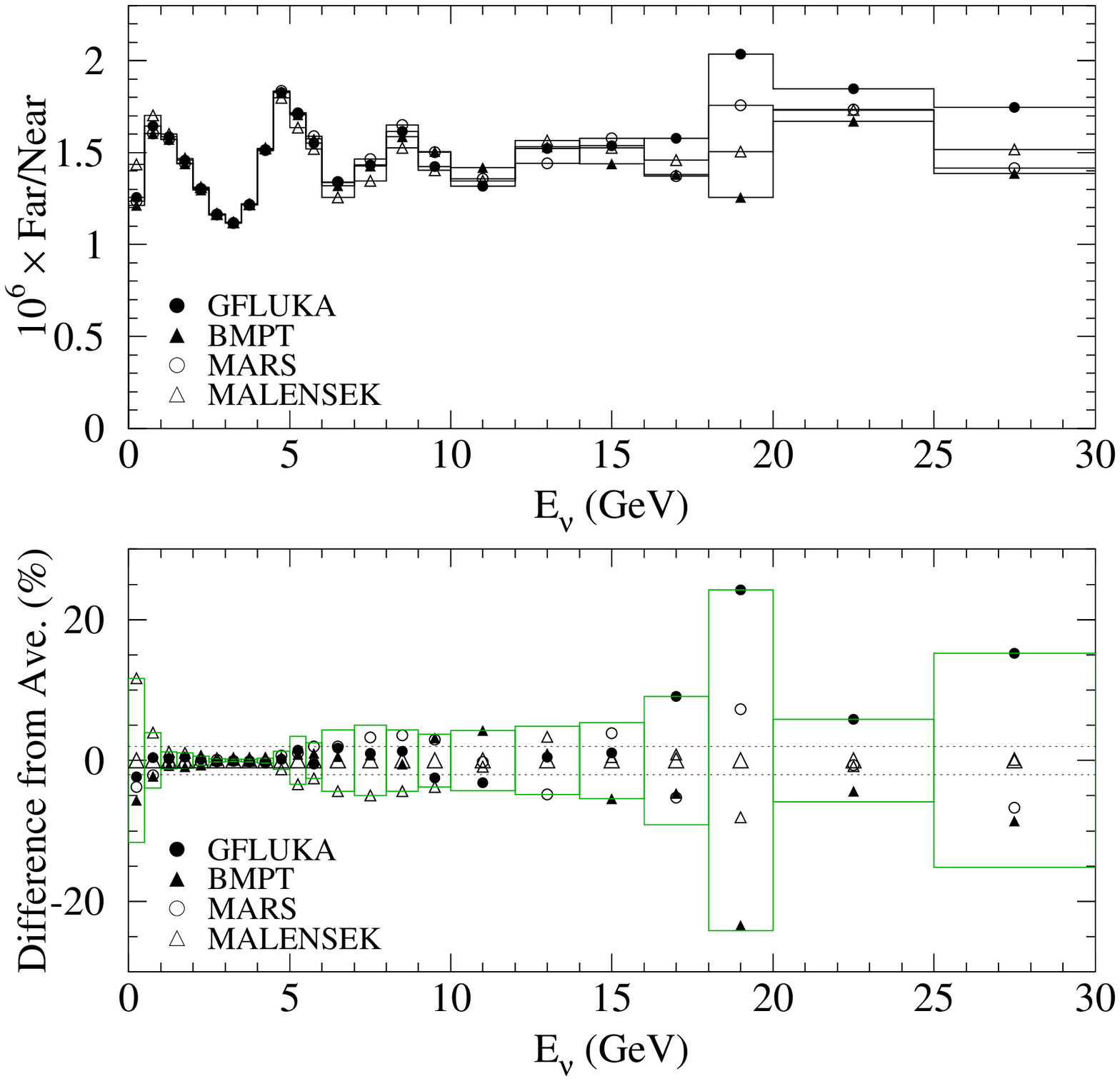}
 \caption{The predictions for the ratio of the far neutrino flux to the 
near neutrino flux for various hadron production models.}
\label{farnear}
\end{figure}

\begin{figure}[htb!]
\centering
\includegraphics[width=\textwidth]{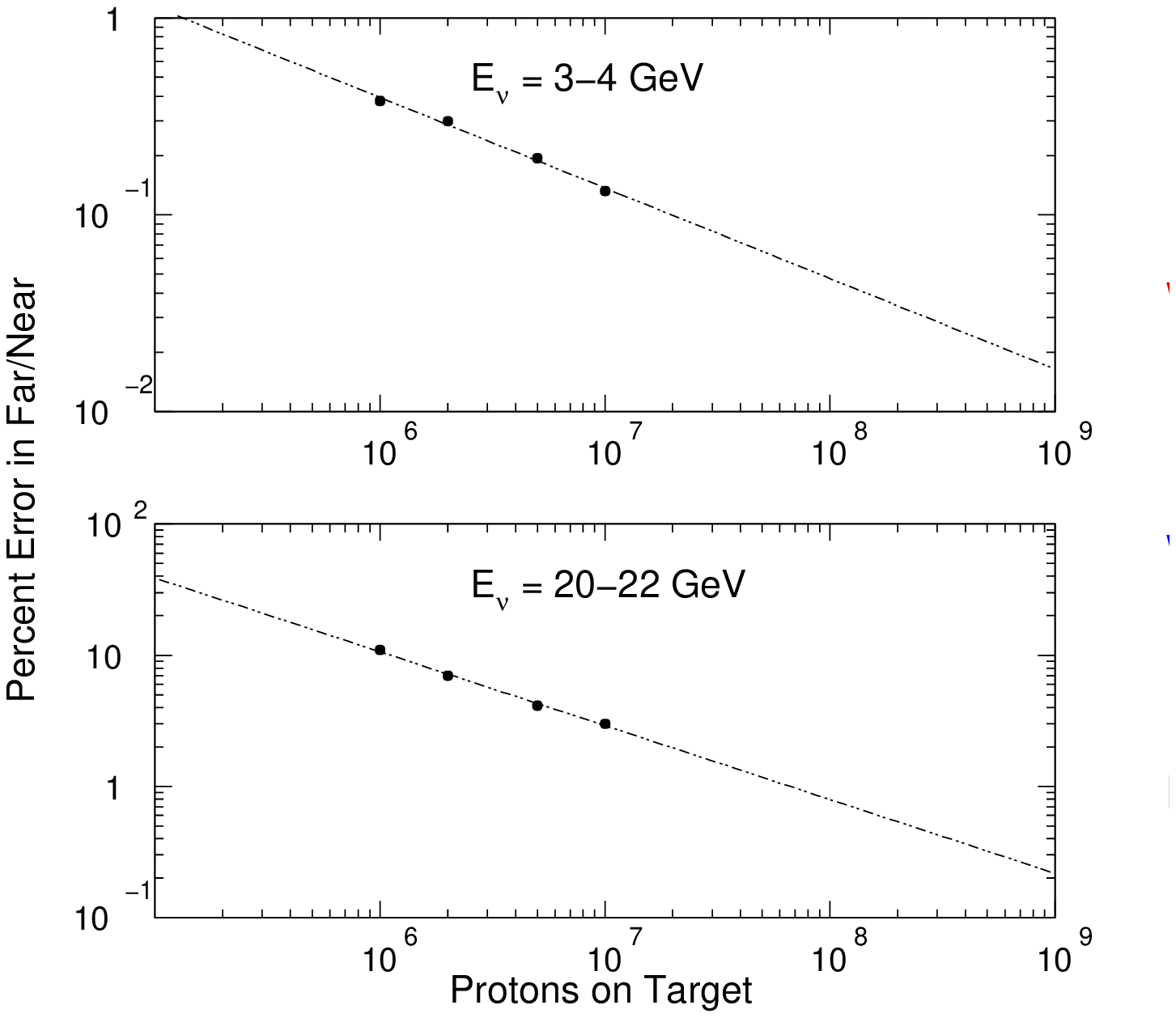}
 \caption{The percentage error in the far/near ratio as a function of
 the number of events in MIPP obtained for the NuMI target (labeled
 protons on target) measurement as a function of the neutrino
 energy. }
\label{mippnumistat}
\end{figure}

Figure~\ref{nurates} shows the predictions of the absolute neutrino
rates in the MINOS near detector using four existing hadron production
models~\cite{hadmod}. 
The model predictions differ from the average by as much as
20\% as a function of neutrino energy.
Figure~\ref{farnear} shows the predictions of the ratio of the far to
the near neutrino flux using the same four models. Again, there is
considerable uncertainty in the predictions,
which increases in the high energy tail of
the spectrum. The evidence for oscillations is obtained by normalizing
the far detector spectrum to the near detector spectrum ($\approx 10^6$ more
events in the near detector). The shapes of the two spectra have to
agree in the high energy tail (no oscillations), before one can take
seriously the expected deficit due to oscillations (in the low energy
part of the spectrum).
 Figure~\ref{mippnumistat} shows the variation of the percentage
error in the far/near detector ratio as a function of the number of
events obtained in MIPP off the NuMI target, for neutrino energies
(3-4 GeV, low energy part, oscillation deficit) and for neutrino
energies (20-22 GeV, high energy tail). It can be seen that one needs
$\approx 10^7$ events in MIPP on the NuMI target for this percentage error to
drop below 3\% in the high energy tail.
MIPP has so far obtained 1.75 million events on the NuMI target currently 
installed in the MINOS experiment using 120~GeV/c protons from the 
Main Injector.

\subsubsection{MINOS analysis}
\begin{figure}[htb!]
\centering
\includegraphics[width=\textwidth]{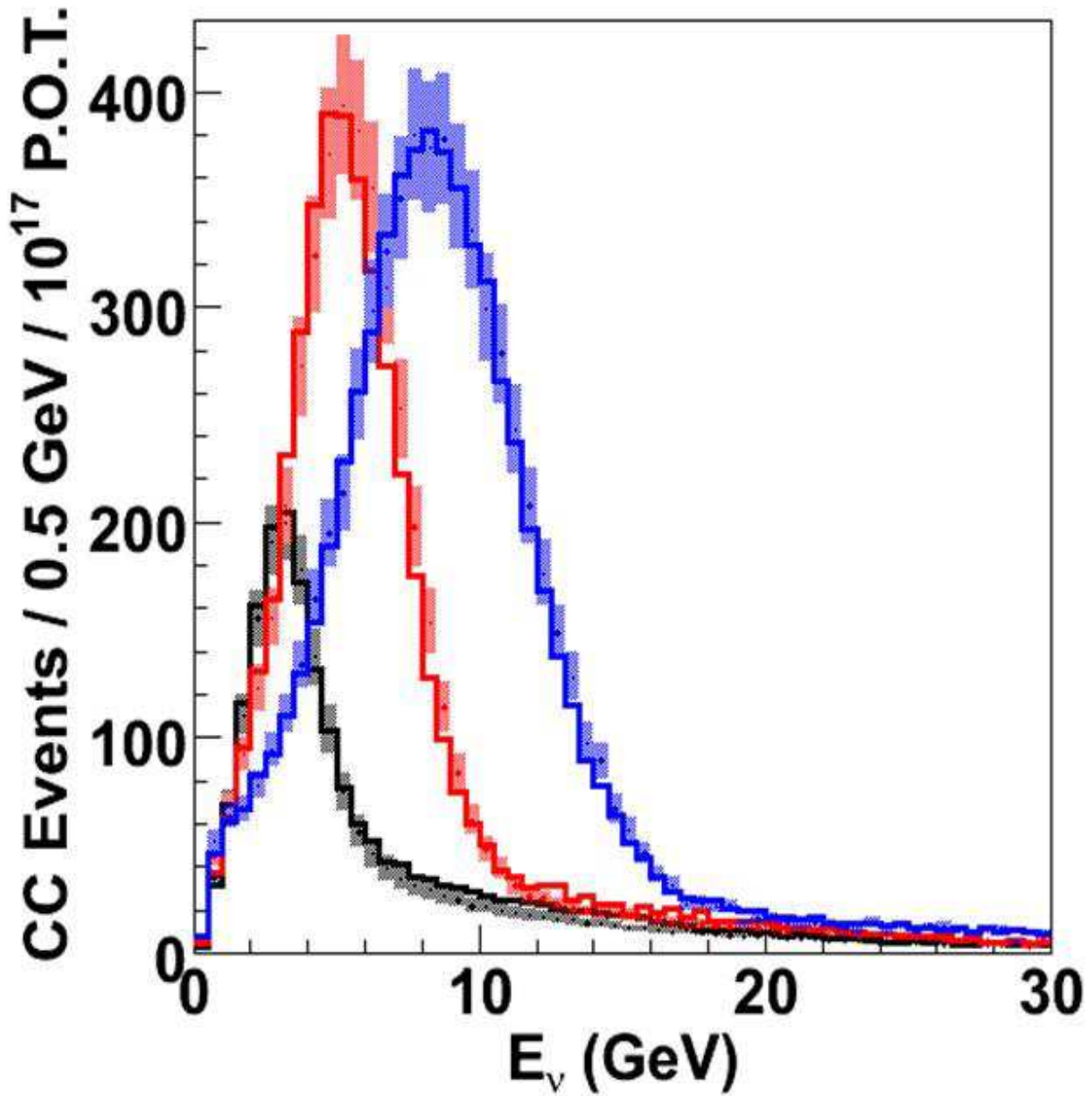}
 \caption{MINOS predictions of the near detector charged current event
 spectra compared with data for a variety of Monte Carlos compared
 with data for the low, medium and high energy NuMI beam settings. The
 Monte Carlo spread is indicated by the shaded error bars and the data
 represented by the solid lines.}
\label{kopp1}
\end{figure}

\begin{figure}[htb!]
\centering
\includegraphics[width=\textwidth]{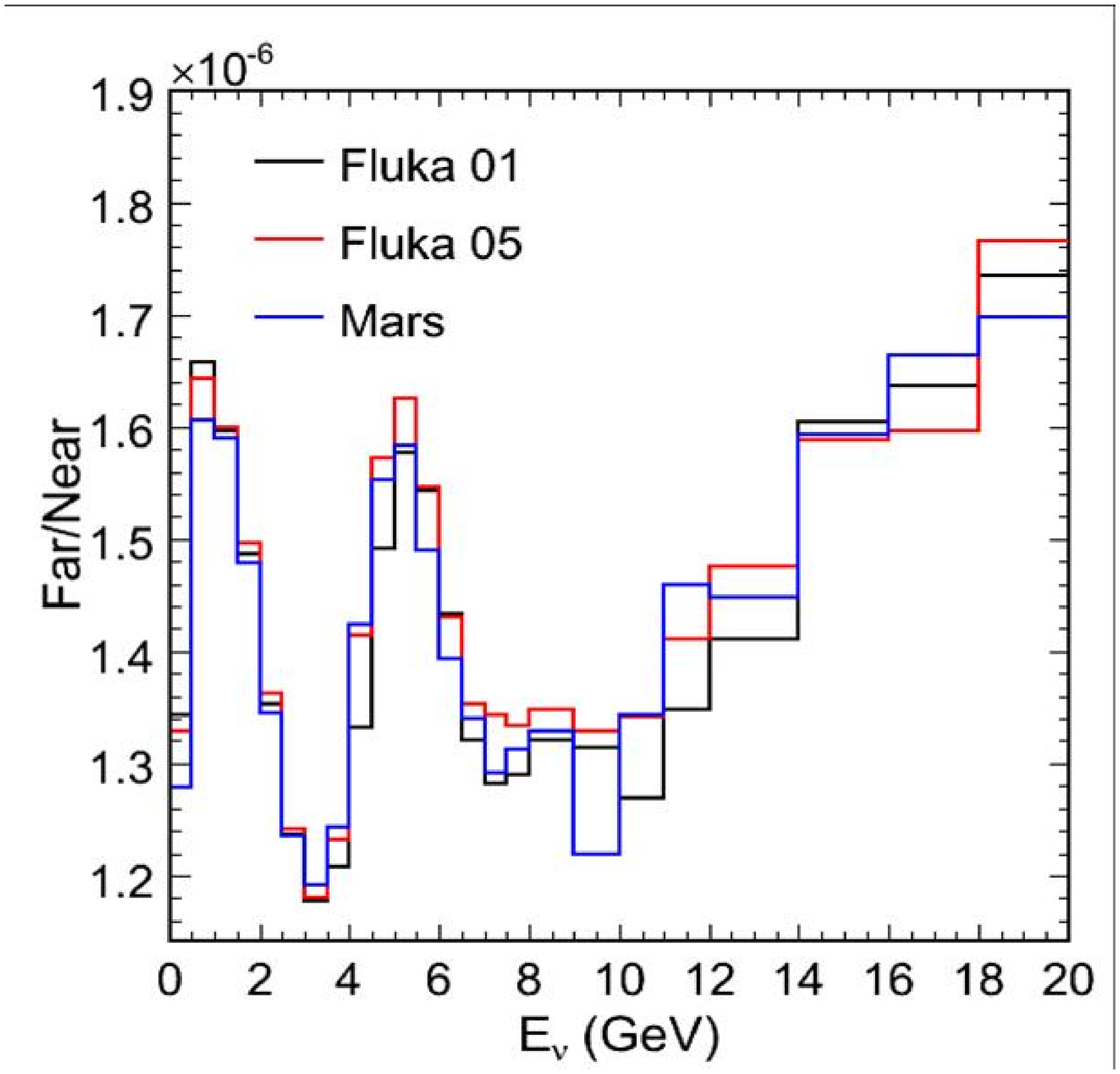}
 \caption{MINOS predictions of the near/far detector ratio using the
 Monte Carlos, FLUKA01, FLUKA05 and MARS. The predictions differ
 between the FLUKA01 and FLUKA05 purely due to algorithmic development
 and not due to introduction of new particle production data. }
\label{kopp2}
\end{figure}

\begin{figure}[htb!]
\centering
\includegraphics[width=\textwidth]{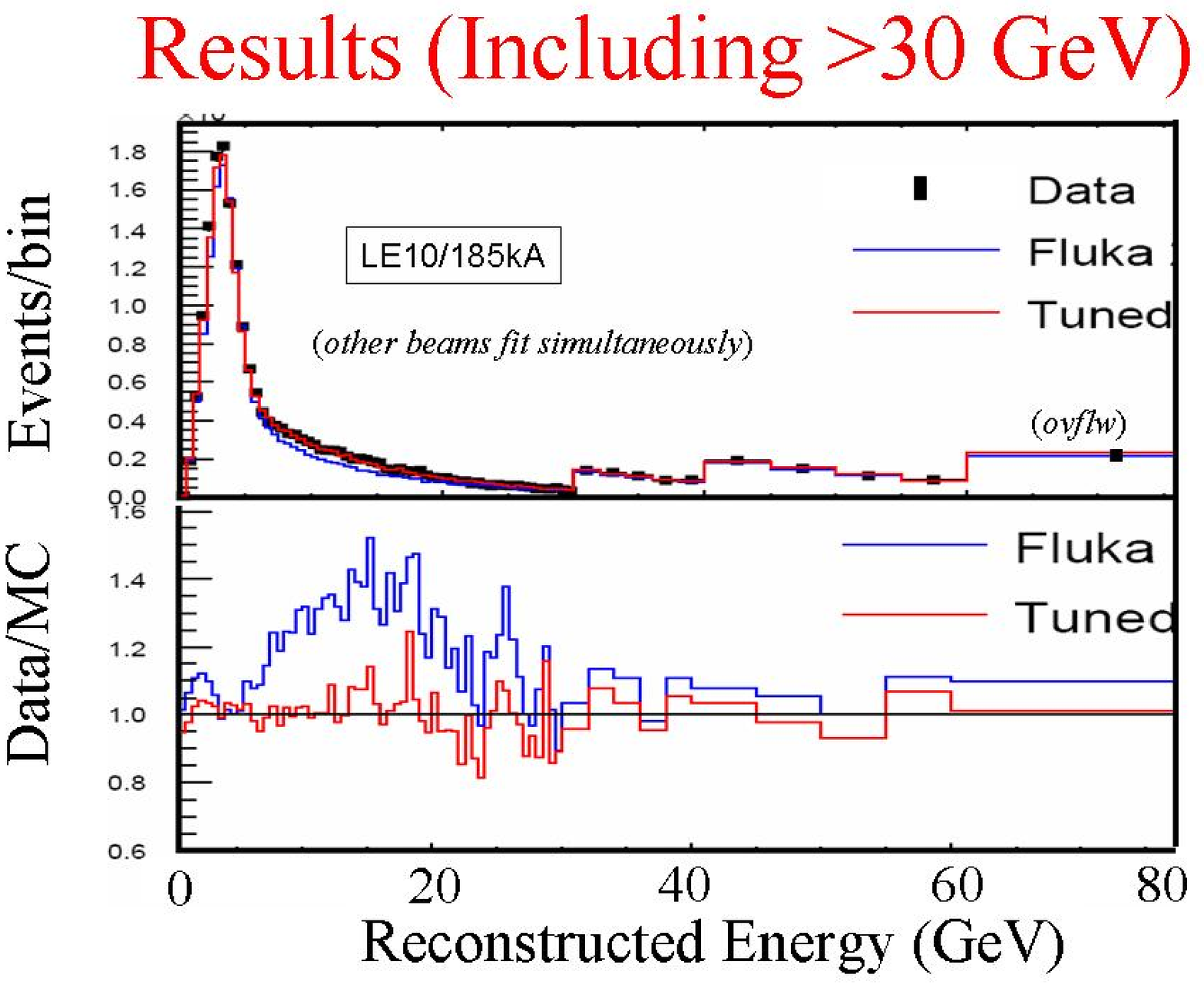}
 \caption{MINOS predictions of the near detector spectrum for the
 target at the LE10 setting and the horn current at 185kA. The monte
 carlo prediction and the weight factor needed to make the Monte Carlo
 agree with data are shown.}
\label{kopp4}
\end{figure}

MINOS has analyzed its near and far detector data and published confirmatory
evidence of neutrino oscillations~\cite{minobs} and the best 
estimates for the oscillation parameters $sin^2 2\theta_{23}$ and
 $\delta m^2_{32}$. Details of their near
and far detector analysis have been reported at
conferences~\cite{sacha}. We reproduce some relevant plots from the
analysis done so far to highlight the uncertainties associated with
hadron production Monte Carlos and the need to measure the particle
production first hand. Figure~\ref{kopp1} shows the prediction of the
near detector spectrum using a number of Monte Carlos for the low, 
medium and high energy NuMI beam settings. The spread
in the Monte Carlos is indicated by the shaded error bar. The
predictions of the near detector spectra utilize 
the Monte Carlo fluxes, the neutrino
cross section and the detector resolution and trigger efficiencies and
acceptances.  The Monte Carlos systematically underestimate the low
energy spectrum in the 6-18~GeV range. They also overestimate the the
medium energy spectrum in the same energy range. This leads to the
conclusion that the predictions of the hadron spectra are to blame and not
the neutrino cross section for the mismatches. MINOS then proceeds to 
weight the Monte Carlo spectrum to match the near detector events and 
then uses the re-weighted spectrum to predict the far detector response 
as a function of the oscillation parameters.  
Figure~\ref{kopp2} shows the predictions of the near/far ratio for
three Monte Carlos, FLUKA01, FLUKA05 and MARS. The predictions between
FLUKA01 and FLUKA05 differ, even though there has been no new particle
production data introduced into the code. The change is purely due to
model development.
MINOS has done a more thorough analysis of the near detector predictions 
as a function of their target position and also the horn current.
Figure~\ref{kopp4} shows the predictions for the low energy target setting with the target at the ``10~cm'' position LE10 and the horn current at 185~kA. At the point of maximum disagreement between the Monte Carlo predictions and data, the weight factor is $\approx$ 1.4.
These re-weighting uncertainties in the Monte Carlo are ameliorated in
MINOS's case by the excellent performance of their near detector. It
would however be still desirable to remove all uncertainties by
obtaining the needed $10^7$ events, which the MIPP upgrade can do in 2
calendar days of running.

All of MINOS's largest systematic uncertainties are related in some fashion to
neutrino cross section uncertainties and event shape modeling. These will also
improve with improved neutrino beam predictions.

\subsubsection {Measuring the \NOVA/\MINERVA{} target}

\MINERVA{} has requested sufficient running in the NuMI LE beam, as 
used by MINOS, to measure the low-energy cross sections and nuclear
effects so important for  MINOS systematics. Neutrino cross
sections need a first principles measurement of the particle
spectrum. The MIPP sample of $10^7$ events off the LE target can be used
directly by \MINERVA{} to estimate their neutrino spectrum without
recourse to any particle production Monte Carlos.

The \NOVA{} medium energy target has still to be designed. When it 
becomes available, MIPP can measure the particle spectrum from 
this target, again obtaining a sample of $10^7$ events. This will help the 
\MINERVA{} experiment obtain cross sections using the medium energy target 
and the \NOVA{} experiment with its backgrounds and systematics in 
its search for electron neutrino appearance.

\subsubsection{Benchmark test of Monte Carlos at the Hadronic Shower Simulation Workshop}

\begin{figure}[htb!]
\begin{minipage}{18pc}
\includegraphics[width=3in]{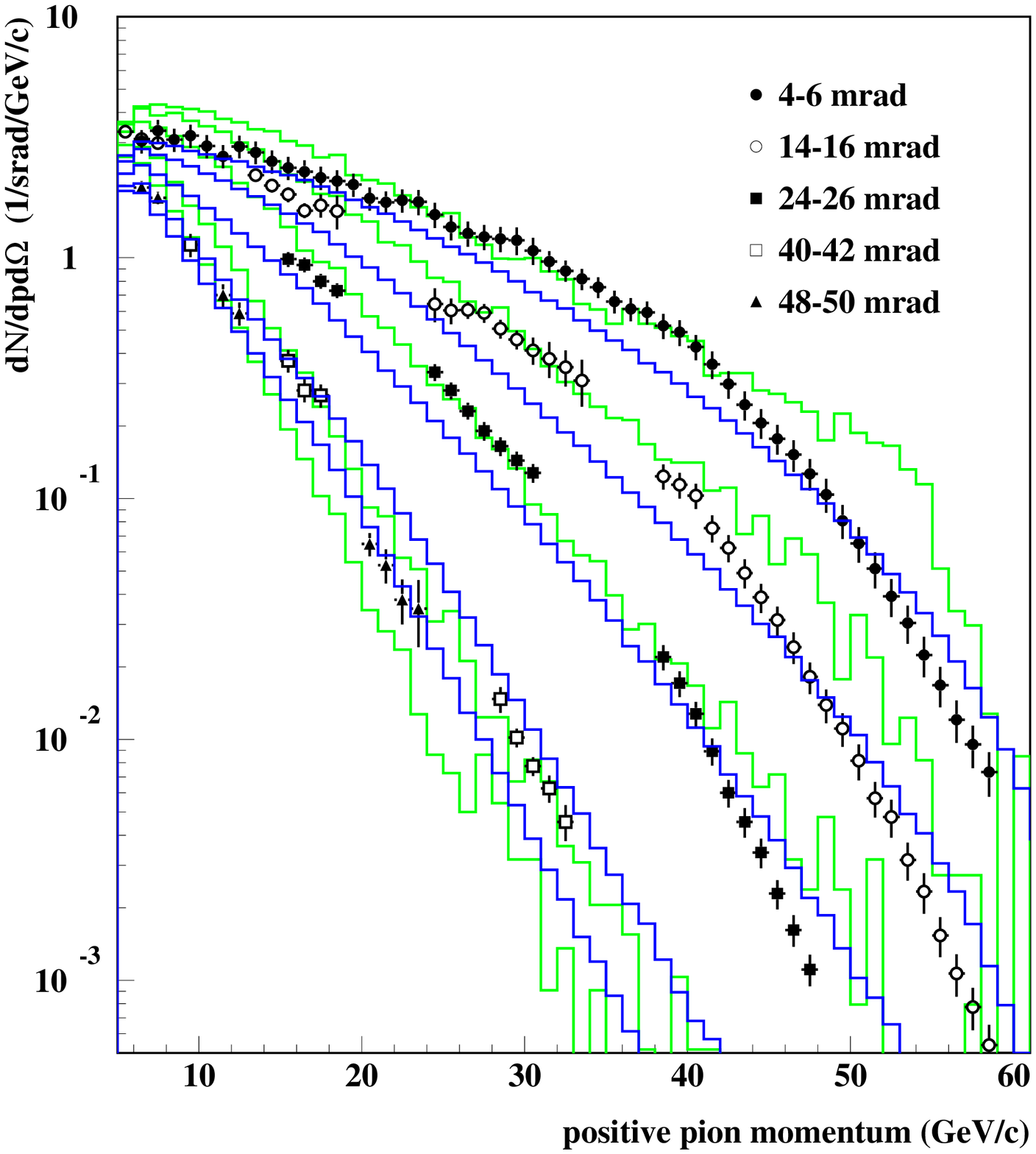}
\end{minipage}
\begin{minipage}{18pc}
\includegraphics[width=3in]{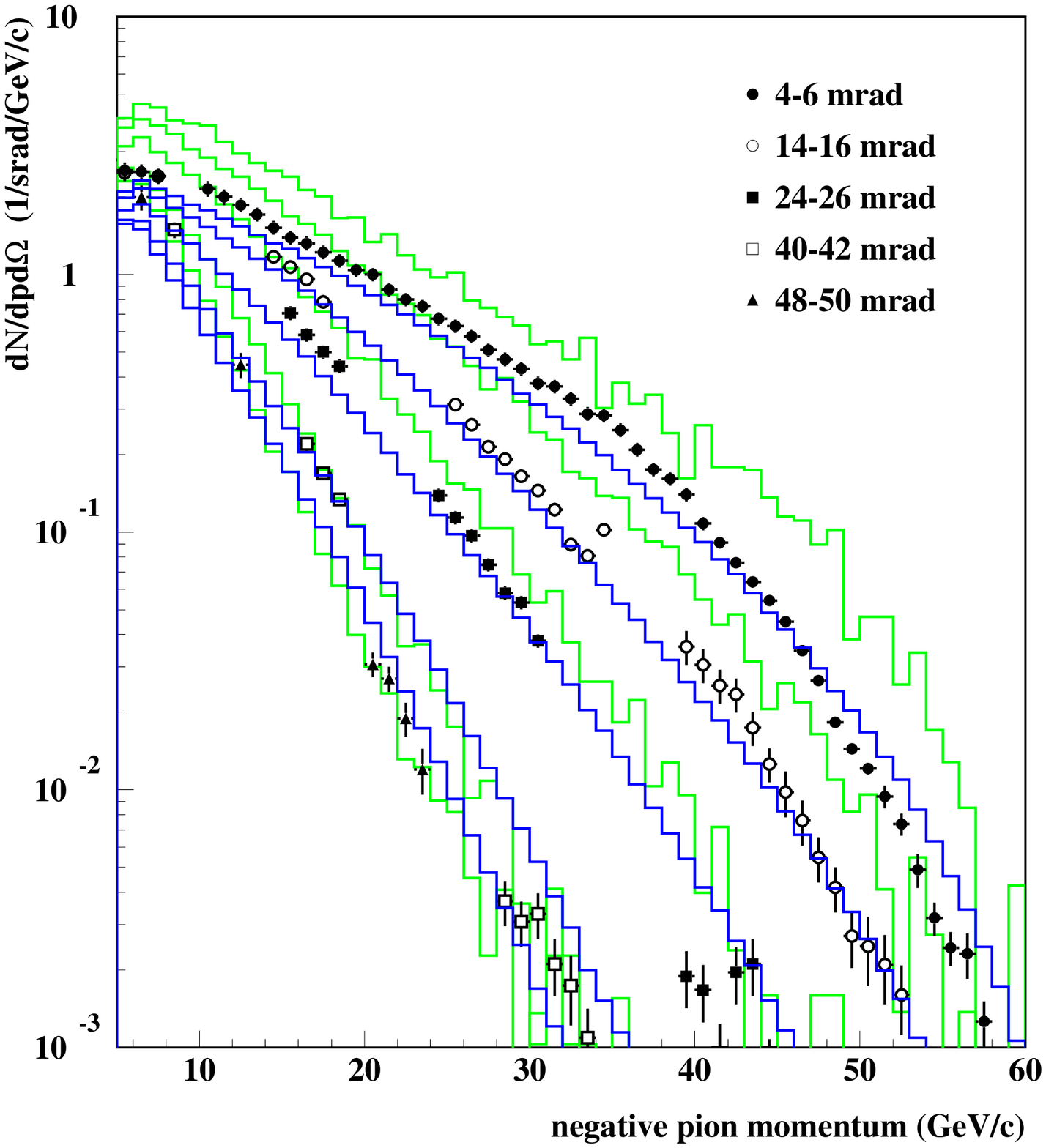}
\end{minipage}
\caption{Comparison of the predictions of the MARS15 (blue) and PHITS (green) Monte Carlos with the production  $\pi^+$ and $\pi^-$ mesons produced by 67~GeV/c protons on a 60~cm long aluminum target.}
\label{pival}
\end{figure}

\begin{figure}[htb!]
\begin{minipage}{18pc}
\includegraphics[width=3in]{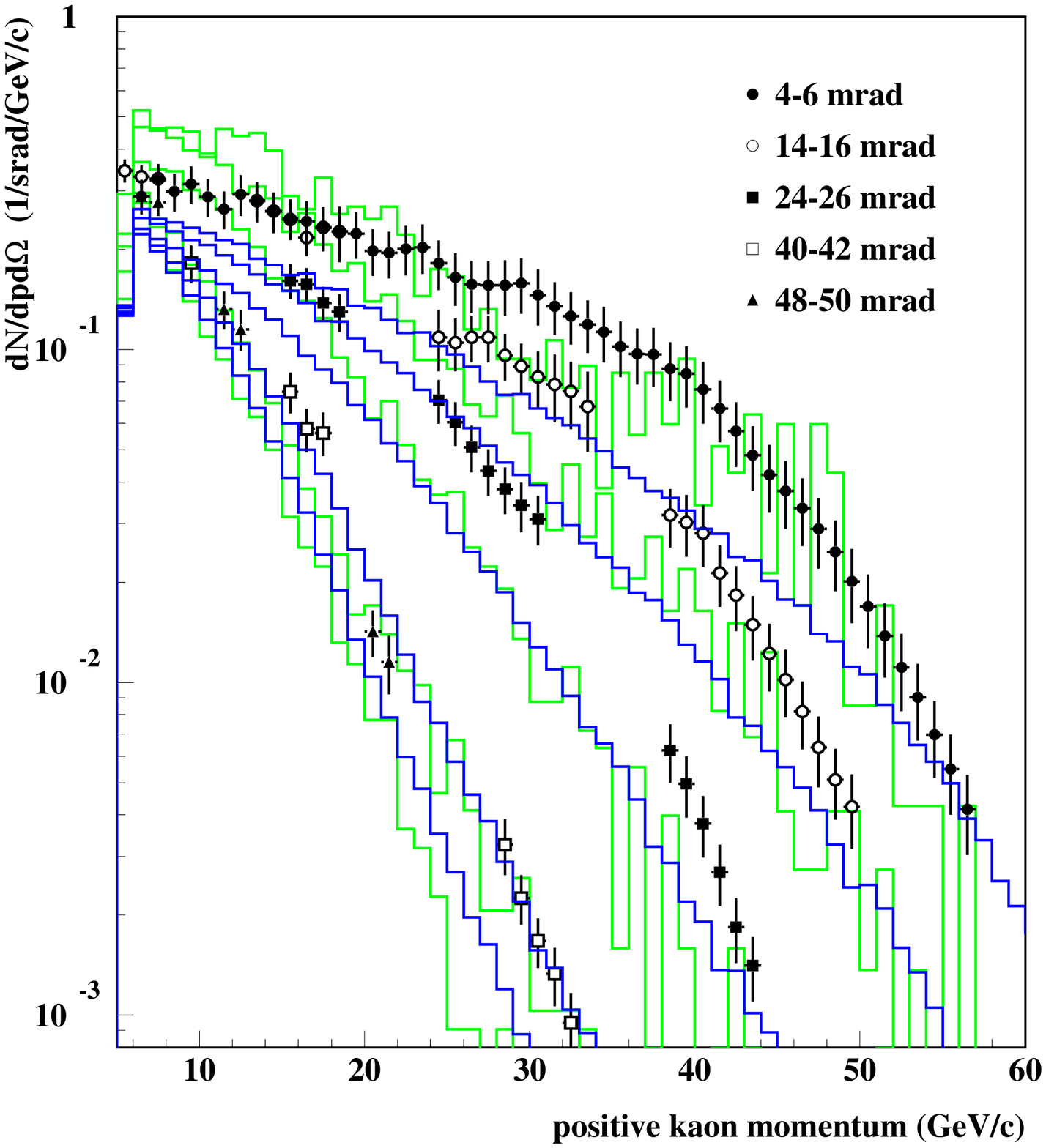}
\end{minipage}
\begin{minipage}{18pc}
\includegraphics[width=3in]{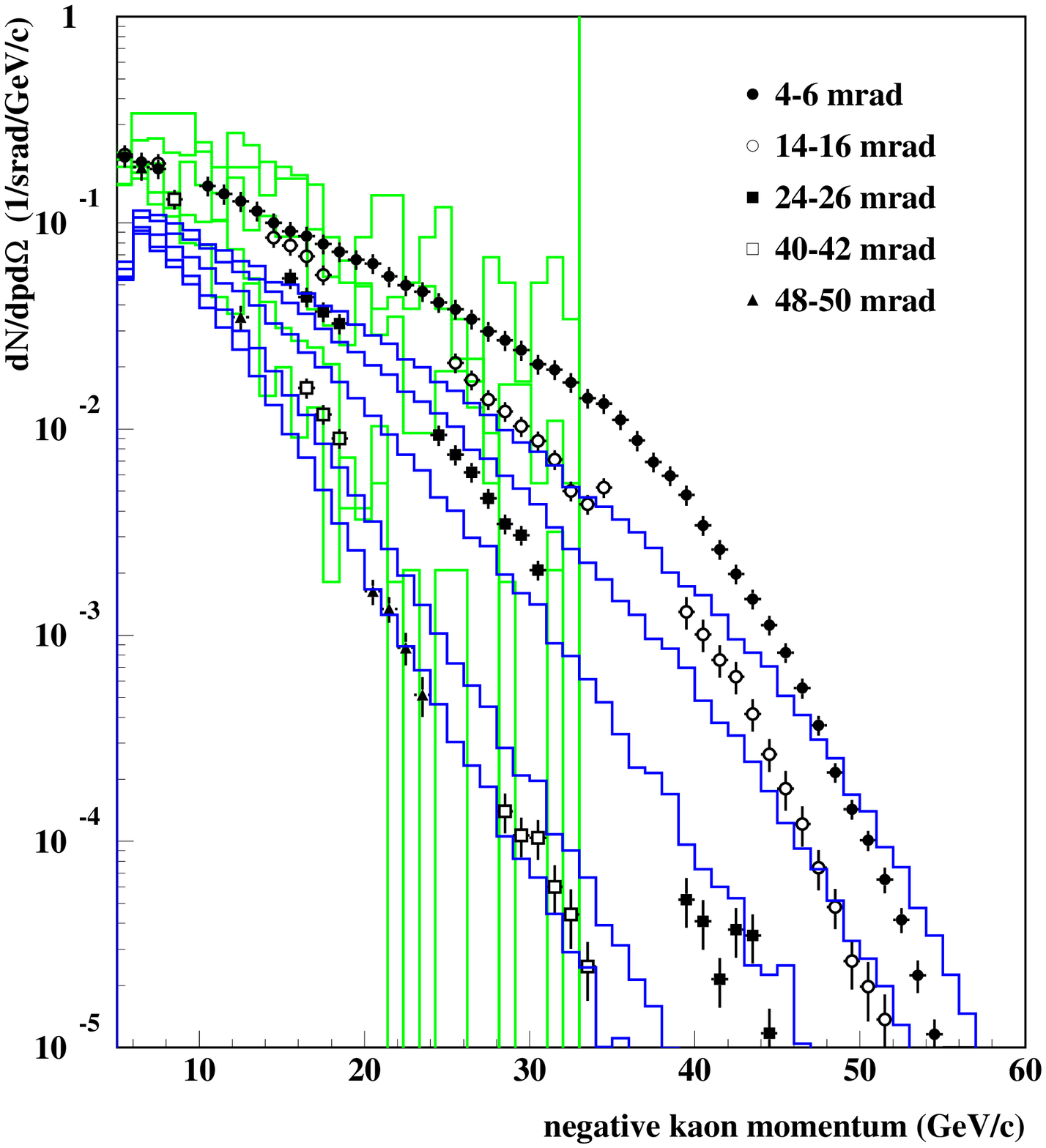}
\end{minipage}
\caption{Comparison of the predictions of the MARS15 (blue) and PHITS (green) Monte Carlos with the production  $K^+$ and $K^-$ mesons produced by 67~GeV/c protons on a 60~cm long aluminum target.}
\label{kval}
\end{figure}

At the recently concluded Hadronic Shower Simulation Workshop at
Fermilab~\cite{hssw06}, a series of benchmark tests were performed to
test various Monte Carlo codes. We show results from one such
benchmark that is relevant to the prediction of neutrino fluxes. Data
on particle production by 67~GeV/c protons on an aluminum target
(60~cm long and 3~cm in radius) obtained in Protvino were compared to
the predictions of the MARS and PHITS monte
carlos. Figure~\ref{pival} shows the comparison of the $\pi^\pm$
spectra with the Monte Carlo predictions and Figure~\ref{kval} shows
the comparison of the $K^\pm$ spectra with the Monte Carlo predictions
as a function of production angle and energy. It is clear that the
Monte Carlos disagree with each other and data underscoring the need
for a first principles measurement of particle production of neutrino
targets.

\subsection{Particle production on Nitrogen and the question of Cosmic Ray  Showers in the Atmosphere}

We propose to measure particle production on a cryogenic nitrogen
target using positive and negative beams, which is needed by experiments
measuring cosmic ray air showers (Pierre Auger, HiRes etc) and also
atmospheric neutrinos (Amanda, Ice Cube, HyperK etc). The problem is
illustrated in a recent paper~\cite{meurer} that simulates
extensive air showers to illustrate the problem. They simulate the air
showers produced by protons of $10^6$~GeV energy in the
atmosphere. The shower goes through several generations of
interactions and produce pions and kaons that decay to produce muons
and neutrinos. The muons and neutrinos are observed in the detectors
and are termed the daughter particles. The mesons that produced the
muons and neutrinos are termed the mother particles and the particles
that interacted in the atmosphere to produce the mother particles are
termed the grandmother particles in their jargon.

\begin{figure}[htb!]
\centering
\includegraphics[width=\textwidth]{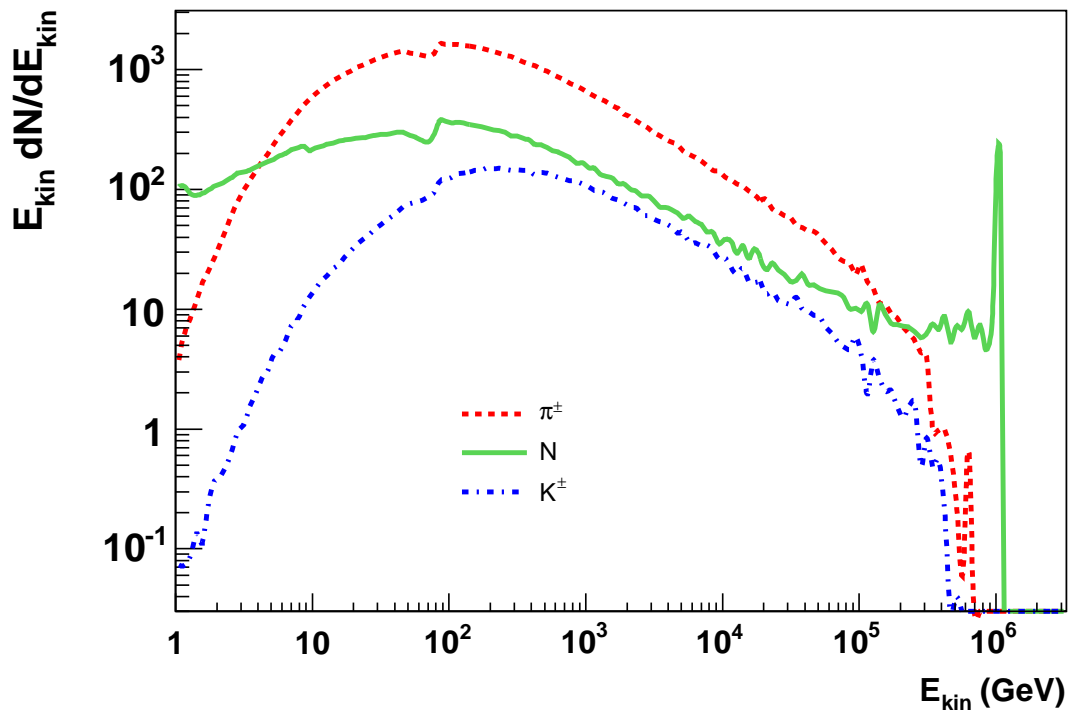}
 \caption{The energy distribution of the grandmother particles in a
 vertical air shower produced by $10^6$~GeV proton interacting in the
 atmosphere as a function of particle type. These particles interact
 further in the atmosphere to produce more particles which then decay
 into muons and neutrinos. The muons are detected at a distance of 
 0-500 meters from the shower center at 
 ground level. It can be seen that these spectra peak at
  energies relevant to the MIPP energy scale. }
\label{meurer1}
\end{figure}

\begin{figure}[htb!]
\centering
\includegraphics[width=\textwidth]{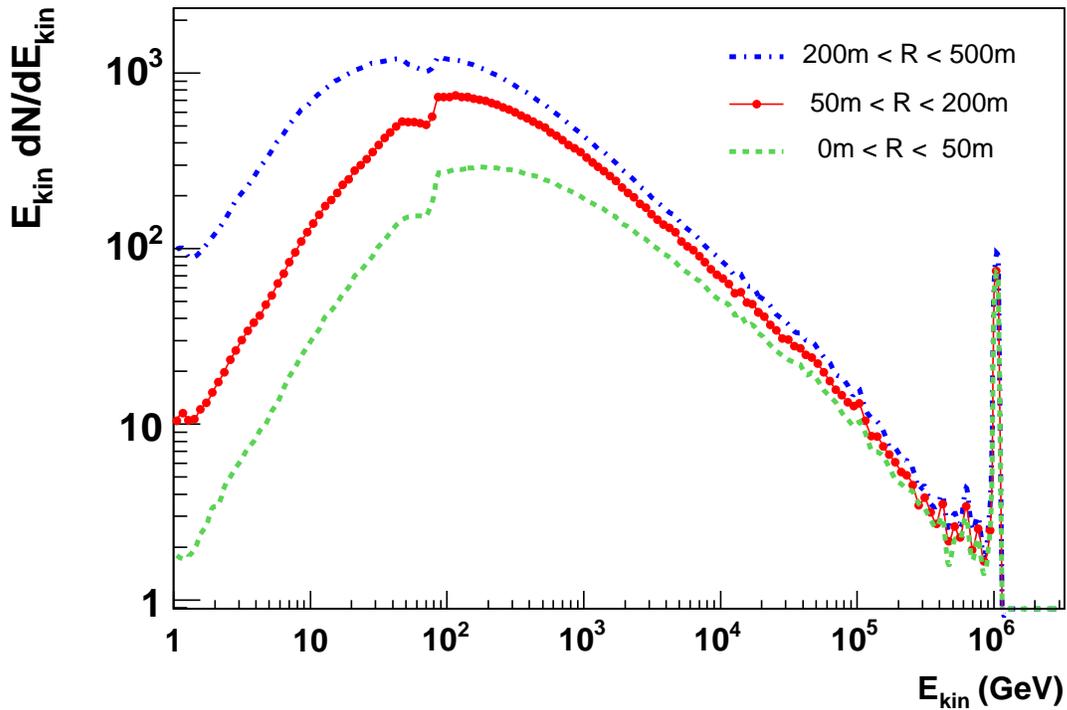}
 \caption{The energy distribution of the grandmother particles in a
 vertical air shower produced by $10^6$~GeV proton interacting in the
 atmosphere as a function of distance from the shower center. 
The spectrum peaks 
at energies relevant to the MIPP energy scale.}

\label{meurer2}
\end{figure}

Figure~\ref{meurer1} shows the energy spectrum of the grandmother
particles ($\pi, K$ and $p$ in an air shower that are produced
by a primary proton of $10^6$~GeV. The spectrum for the pions peaks at
100~GeV and the kaons and protons at somewhat higher energies. These
particles interact with the nitrogen (and oxygen) in the atmosphere to
produce the atmospheric neutrinos and muons. In other words, the beam
energies available at MIPP are relevant to the simulation of the
cosmic ray air showers. The muon flux measurement is a critical component 
of estimating the energy scale of the cosmic ray shower. MIPP measurements thus
will help reduce the systematics in the cosmic ray energy scale measurements. 
As the primary cosmic ray energy increases, the peaks in this plot do not 
shift to higher energies. 
Understanding the shower systematics at the peak of this 
spectrum (i.e MIPP energies) will help the energy systematics of 
cosmic rays of all energies.
Figure~\ref{meurer2} shows the distribution of grandmother particles
at different lateral distances from the shower center for all particle
types.  For the lower energy interactions, the simulation code Gheisha
is used to simulate the interactions of the particles with the
atmosphere. For higher energy interactions, the simulation code QGSJET 01
is used. The sharp break in the spectra at 100~GeV is where the two
codes meet and disagree at places by a factor of two. This illustrates
the problem. These codes at present are ``tuned'' on single arm spectrometer
data and disagree with each other.

\begin{figure}[htb!]
\centering
\includegraphics[width=\textwidth]{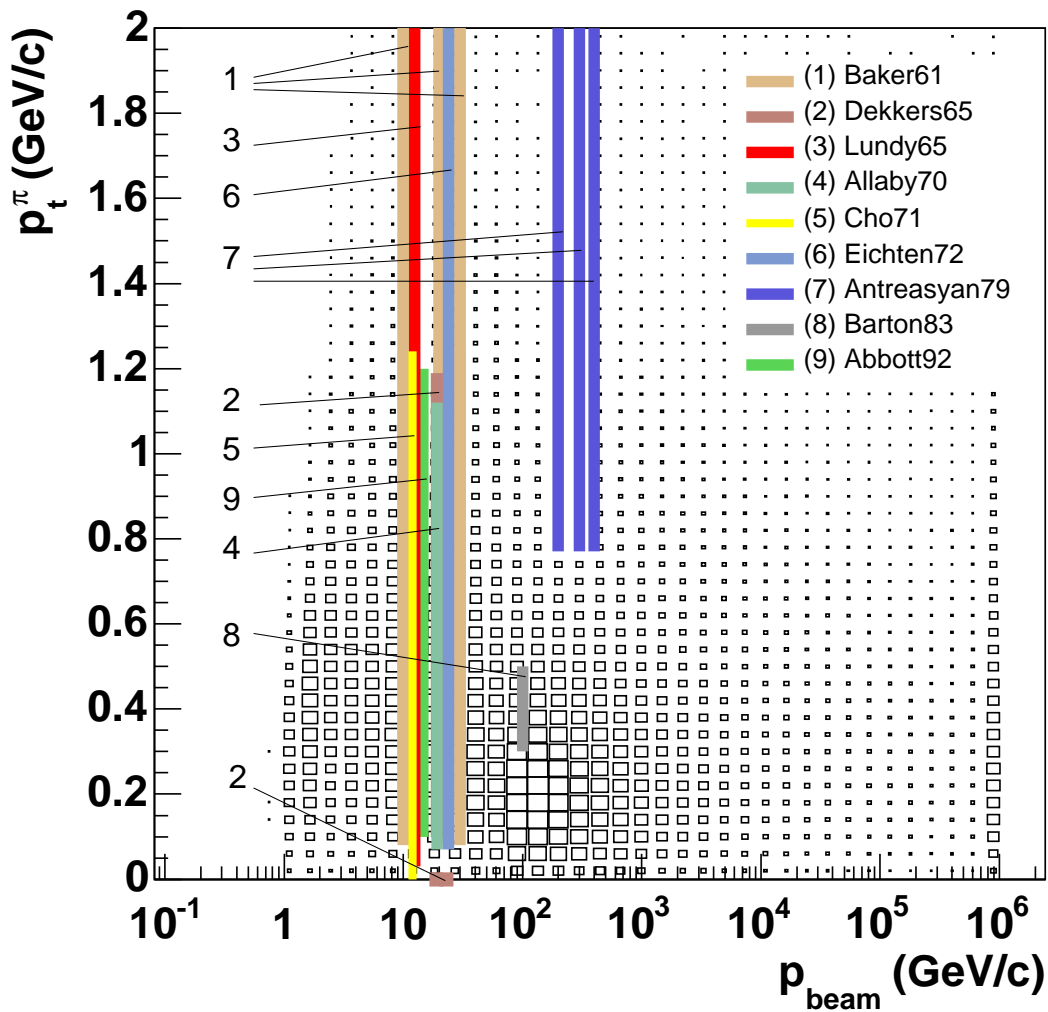}
 \caption{The momentum of the grandmother particle (abscissa) that
 interacts to produce the mother pion that decays to produce the muon
 and atmospheric neutrino. The relevant existing experimental data are
 plotted. The data are over 25 years old and are often from a
 Beryllium target and obtained in single arm spectrometers at discrete
 transverse momenta.} 

\label{meurer3}
\end{figure}

Figure~\ref{meurer3} shows the momentum of grandmother particles
versus the momentum of the mother pions in the air shower and plots
the existing data relevant to simulating the process. Most of the data
are over 25 years old and were obtained using Beryllium targets and
single arm spectrometers resulting in a discrete angular
coverage. MIPP will measure the outgoing pion and kaon spectrum for
proton and pion beams in its full secondary beam momentum range. These
cross sections are the most important in simulating cosmic ray showers
in the atmosphere.  In addition, it will also have kaon and antiproton
interactions on nitrogen of which virtually nothing is known. Being an
open geometry experiment, the MIPP angular coverage will be
continuous, not discrete.
The need for MIPP data is recognized by the cosmic ray community, 
some of whom have joined this proposal.
\subsubsection{MIPP Measurement of $\pi/K$ ratios}

Because of its excellent particle identification capabilities, MIPP
upgrade will measure the ratio of charged kaons to pions as a function
of $p_L,p_T$ of the final state particle. This measurement is of
importance to both the NuMI target measurements and the atmospheric
neutrino measurements, since the charged K's produce $\nu_e's$ which
are a background to the oscillation search $\nu_\mu\rightarrow \nu_e$.

\subsection
{Hadronic production on Nuclei and the Hadronic Shower Simulation Problem}

At the recently concluded workshop on Hadronic Shower
Simulations~\cite{hssw06}, experts in shower simulation codes from
five major Monte Carlos (GEANT4, FLUKA, MARS, MCNPX, and PHITS) (and several 
less well-known ones)  met
and reviewed their code status and what needs to be done further to
improve codes. It was  acknowledged that more particle production data would 
help improve algorithms a great deal and there were calls for a rapid 
publication of existing MIPP data and enthusiastic support for the 
MIPP upgrade.

The problem of hadronic shower simulations stems from our lack of
understanding of the strong interaction. Though a theory of strong
interaction exists (QCD), it cannot be used to calculate fundamental
non-perturbative processes such as elastic cross sections, diffractive
cross sections or any of the particle production cross sections that
comprise 99\% of the total cross section. This is in stark contrast
to the simulation of electromagnetic showers where Monte Carlos such
as EGS regularly make predictions that can be verified by observation.

The minimum bias cross section is modeled in most of the above
mentioned Monte Carlos using the supercritical pomeron (which in
itself violates unitarity).  By application of the optical theorem,
cut pomerons produce the total cross section. The cut pomeron is
approximated by a quark gluon string which is then hadronized. This is
the basis of the particle production models such as DPMJET and QGSJET
which are used in Monte Carlos such as GEANT4 and MARS. The soft part of the
 scattering and the hard part of the scattering cross section are joined 
very carefully, but arbitrarily.

Nuclear break-up is handled using a plethora of models that go by the
name of binary cascade, Bertini cascade, CHIPS (Chiral Invariant Phase
Space), CEM03 and GEM2, each of which have different assumptions on
the nuclear break-up mechanism.

Each of these Monte Carlos are ``validated'' using existing
data. Inclusive particle spectra from single-arm spectrometer
experiments that are over 30 years old are used. These data are
discrete in the transverse momentum variable and have systematics that
are significantly greater than open geometry experiments such as HARP,
NA49 and MIPP. These models are made to agree with inclusive particle
spectra. Predictions of correlations between particles are not tested
against, since such data do not exist in readily testable
form. However, calorimeter designers are currently asking questions
such as how wide a hadronic shower is in a calorimeter, which depends
on particle correlations.

Another important part of calorimetric simulation of hadronic showers
is the nuclear break-up and the number of spallation neutrons and
protons emitted, as emphasized by Wigmans~\cite{wigmans}. 
The linearity of the
calorimeter and the resolution of the calorimeter depend critically on
compensating for the ``invisible energy'' in a hadronic shower carried
away by neutrons and nuclear binding energy. It is important to model
these processes correctly. It is not at all clear as to how well the
above mentioned nuclear break-up models simulate these processes.

To illustrate this further, table~\ref{wigm} 
shows~\cite{wigmans} how the energy is deposited
by a 1.3~GeV pion in lead. The energy is deposited as ionization
$dE/dx$, as binding energy required to split the nucleus, as cascade
nucleons and evaporation nucleons (isotropic emission).  
Please note that on average only 478~MeV of a 
1.3~GeV pion ends up as
ionization energy and the rest is carried away as neutrons and also
absorbed as binding energy. The whole question of compensating
calorimetry hinges on using the neutrons to produce knock-on protons
to compensate for this invisible energy, since 
energy from the neutrons can be made to be visible by introducing
hydrogenous materials in the calorimeter in appropriate proportions,
resulting in knock-on protons caused by neutron elastic scattering
that deposit visible energy. 
\begin{table}
\begin{center}
\begin{tabular}{|c|c|c|c|c|c|}
\hline
\hline
& Binding & Evaporation n & Cascade n & Ionization & Target \\
& Energy & (\# neutrons) & (\# neutrons) & (\#cascade p) & recoil \\
\hline
Before first reaction & & & & (250)($\pi_{in}$) & \\
First reaction & 126 & 27(9) & 519 (4.2) & 350(2.8) & 28 \\
Generation 2& 187 & 63(21) & 161(1.7) & 105(1.1) & 3 \\
Generation 3 & 77 & 24(8) & 36(1.1) & 23 (0.7)& 1 \\
Generation 4 & 24 & 12(3) & & & \\
\hline
Total & 414 & 126(41) & & 478(4.6) & 32 \\
\hline
\end{tabular}
\caption{Destination of 1.3~GeV total energy carried by an 
average pion produced in hadronic shower development in lead. 
Energies are in MeV.~\label{wigm}}
\end{center}
\end{table}

The upgraded MIPP spectrometer can measure nuclear multi-particle
hadronic production using 6 beam species ($\pi^\pm$, $K^\pm$) and
$p^\pm$ in the momentum range $\approx$ 1~GeV/c-120~GeV/c. 
The TPC can measure the protons from
nuclear breakup that travel forward in the laboratory and the plastic
ball detector will detect the evaporation neutrons and protons emitted
backwards in the laboratory. We can measure 30 nuclei in 30 days of
running and obtain data of unprecedented quality and statistics on
nuclei commonly encountered in particle physics detectors.

We propose as a first priority (``A-List'') 
to measure particle production on the nuclei
\\
H2,D2,Li,Be,B,C,N2,O2,Mg,Al,Si,P,S,Ar,K,Ca,
Hg,Fe,Ni,Cu,Zn,Nb,Ag,Sn,W,Pt,Au,Pb,Bi,U
\\
and as a second priority (``B-list'')
the nuclei\\
Na,Ti,V, Cr,Mn,Mo,I, Cd, Cs, Ba

These data can be used to validate the Monte Carlos to unprecedented
accuracy or may even by usable directly as a library of events in a
fast Monte Carlo~\cite{rajalib}.

It is worth pointing out that the MIPP upgrade proposal represents a
unique opportunity to obtain such a dataset. Comparable experiments
such as HARP do not possess kaon or antiproton beams and do not have
the range in beam momentum (3-15~GeV/c primary momentum). HARP took
data The proposed NA49 upgrade does not posses the data-taking rate
($\approx$20~Hz compared to 3000~Hz in MIPP) nor the particle id
capabilities of MIPP (no forward RICH detector), though, being an SPS
experiment, it has higher beam momenta (positive beams only, maximum
beam momentum 158~GeV/c). Nevertheless, these two detectors will
provide valuable data in the near future on particle production.

Lastly, there is a misunderstanding among some that putting test
hadronic calorimeter modules in the test beam and comparing the
predictions of simulation programs such as GEANT4 and FLUKA to the
observed data can help tune the hadronic models in the simulation
programs. This myth was debunked at the Hadronic Shower Simulation
Workshop, when the Geant4 group collectively answered a question by
stating that ``We only change our models based on microscopic
data''~\cite{g4gr}. Upon being asked what ``microscopic'' meant, they
answered, thin target nuclear data. It is difficult to unfold the various 
nuclear and readout effects in calorimeter data to change the models.
In other words, the only way to
 improve the simulation models is by experiments such as MIPP, HARP and 
NA49 that measure hadro-production using thin targets.

\subsection{Tagged Neutral beams and ILC Detector R\&D}

\begin{figure}[htb!]
\centering
\includegraphics[width=\textwidth]{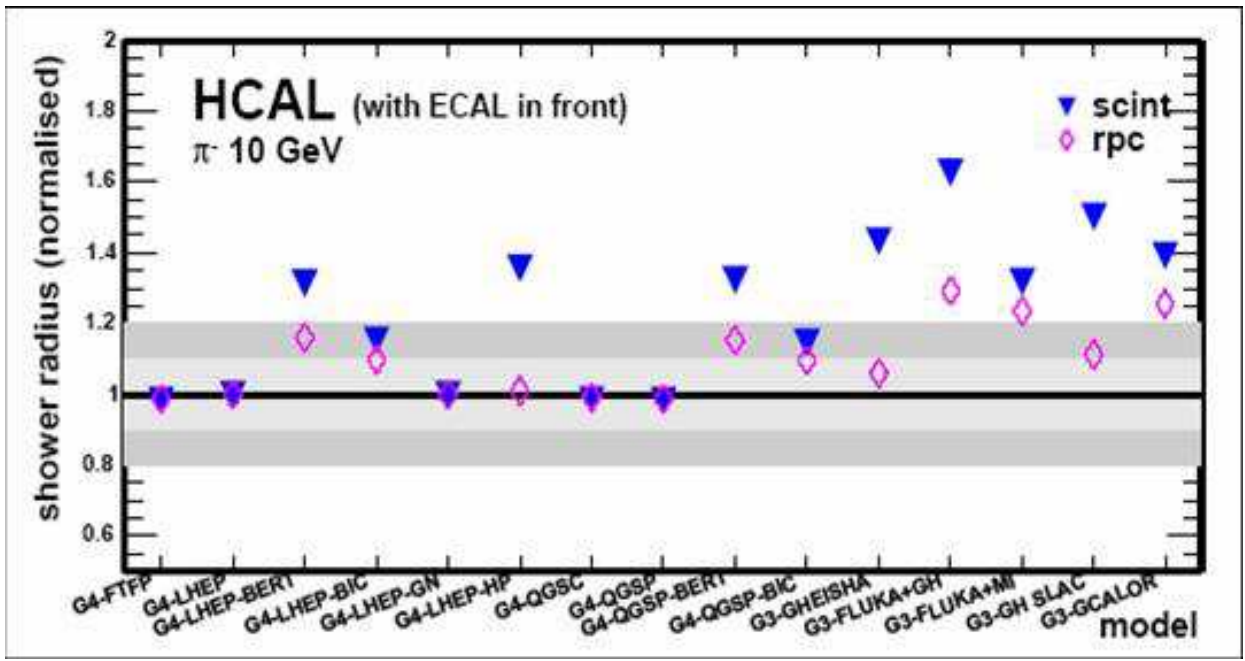}
 \caption{The width of a 10~GeV/c $\pi^-$ energy deposit in
 scintillator and RPC readout calorimeters as simulated by a host of
 simulation programs available in Geant4 (G4-) and Geant3 (G3-). The
 widths are normalized to the minimum width obtained.  }
\label{cwid}
\end{figure}

Three out of the four ILC detector concepts (SiD, LDC, GLD) are
optimized around the particle flow algorithm (PFA), which proposes to
measure the energy of jets in an event by using both the magnetic field
and the calorimeter. The charged particles are measured using the
excellent momentum resolution of the tracker  and the neutral
particles are measured using the calorimeter. The required fractional
energy resolution of a jet is $\sigma_E/E=0.3/\sqrt(E)$, E in GeV. This
hard-to-achieve performance is driven by the ILC design requirement to
be able to separate the processes $W\rightarrow jet+jet$ and
$Z\rightarrow jet+jet$. In order to measure the neutral particle
energy using the calorimeter, one needs to separate the charged
particle hits and the neutral particle hits in the calorimeter. This
dictates a highly segmented calorimeter. In order to test the design,
one needs to simulate the widths of the showers of both the charged
and neutral particles in the calorimeter.
Figure~\ref{cwid} shows the simulation of the width a 10~GeV $\pi^-$
particle entering two ILC calorimeters, one using RPC readout and the
other using scintillator readout~\cite{cwid-guy} for a variety of
hadronic shower simulators available in Geant4  and Geant3. 
The widths are normalized to the narrowest width obtained. There
is a variation in the widths of $40\%$ in the simulations. This calls
for  a data-based approach both for charged and neutral hadronic
responses. The charged response can be obtained in a regular test beam
such as would be available at Fermilab in the M-test area. The upgraded 
MIPP spectrometer offers
a unique opportunity to measure the neutral particle response to three
neutral species, the neutron, the $K^0_L$ and the anti-neutron.

The basic idea is to use the diffractive reactions
\begin{eqnarray}
 pp\rightarrow n\pi^+p \\
 K^+p\rightarrow\bar{K^0}\pi^+p ; \bar{K^0}\rightarrow K^0_L\\
 K^-p\rightarrow{K^0}\pi^-p ; {K^0}\rightarrow K^0_L\\
 \bar{p}p\rightarrow \bar{n}\pi^-p 
\end{eqnarray}
where the beam of protons, $K^\pm$ or $\bar{p}$ fragments
diffractively to produce the neutral beam. The charged particles in
the reaction are measured in the MIPP spectrometer. The beam momentum
is known to $\approx$ 2\%. So the momentum of the tagged neutral
particle can be inferred by constrained fitting (3-C fit) to better
than $2\%$, event by event. The tagged neutral particle goes along the
beam direction and ends up in a test calorimeter placed in lieu of the
present MIPP calorimeter.

This technique demands that the target is a proton and will only work
on a liquid hydrogen cryogenic target (that MIPP possesses). The
plastic ball recoil detector will act as an additional veto against
neutral target fragments such as slow $\pi^{0'}s$.

\begin{figure}[tbh]
\begin{center}
\includegraphics[width=\textwidth]{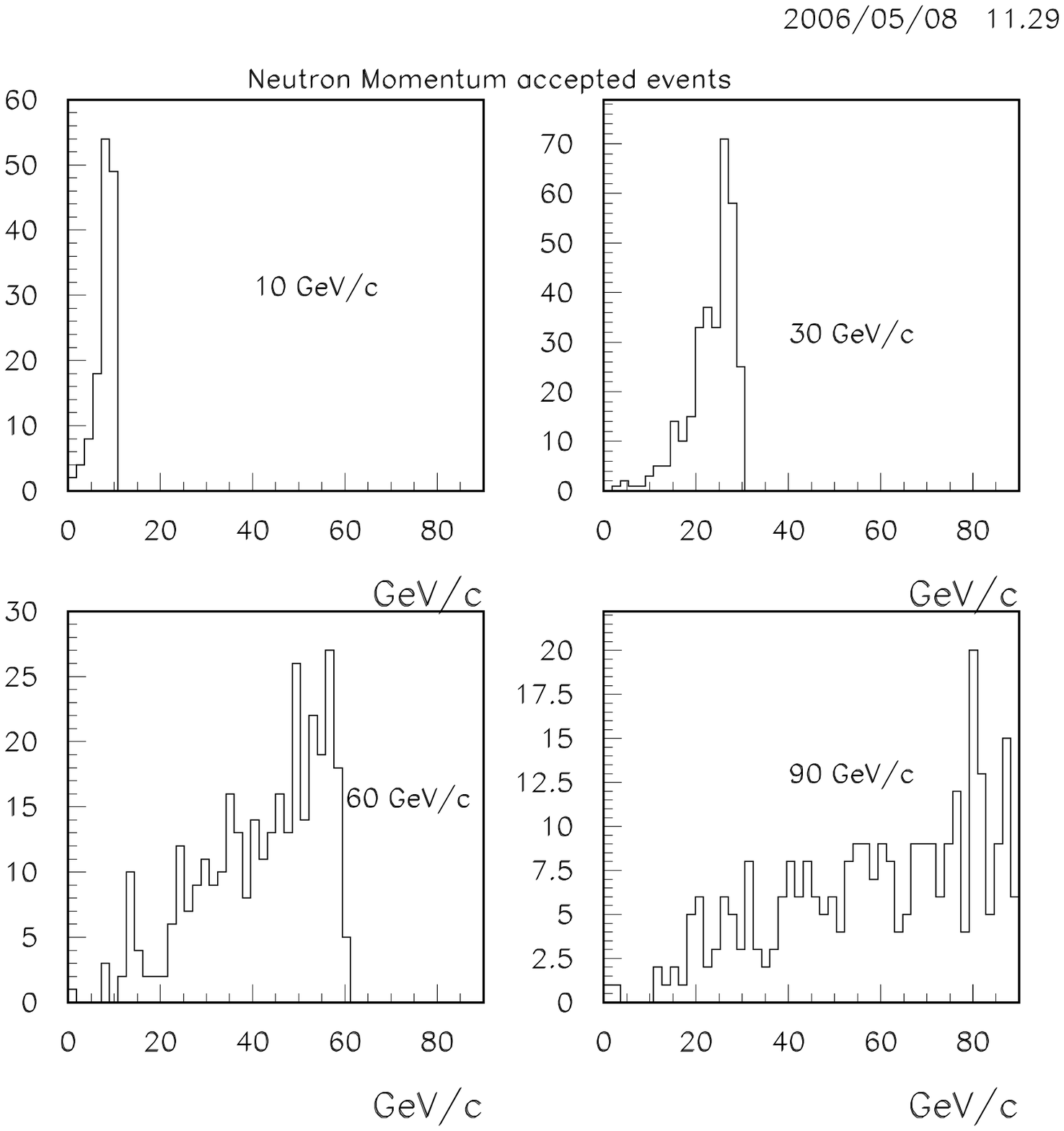}
\caption{Momentum spectrum  of accepted neutrons for incident proton momenta of
10~GeV/c, 30~GeV/c, 60~GeV/c and 90~GeV/c 
for the process $pp\rightarrow pn\pi^+$.~\label{pneut}}
\end{center}
\end{figure}

The momentum spectrum of the neutral beam is controllable by changing the beam
momentum. The method is outlined in detail in  MIPP note 130~\cite{raja130}.
The diffractive processes are simulated using the program DPMJET and the event rates estimated for a calorimeter placed in the MIPP calorimeter position. 
With the MIPP upgrade, it should be typically possible to obtain 
$~\approx$~50,000 tagged neutrons, 
$\approx$~9,000 tagged $K^0_L$, and 
$\approx$~11,000 tagged $\bar n$  per day in the calorimeter with the
 beam momentum set to 20~GeV/c. Table~\ref{tagt} shows the expected 
number of events /day as a function of beam momentum and beam species.
\begin{table}
\caption{Expected number of tagged neutrons, $K^0_L$, and anti-neutrons per day with an upgraded MIPP spectrometer.\label{tagt}}
\begin{tabular}{|c|c|c|c|c|}
\hline
Beam Momentum & Proton beam & $K^+$ beam & $K^-$ beam & ${\bar p}$ beam \\
\hline
GeV/c         & n/day    & $K^0_L$/day   & $K^0_L$/day & ${\bar n}$/day \\
\hline
10 &  20532 & 4400 & 4425 & 6650\\
\hline
20  & 52581 & 9000 & 9400 & 11450\\
\hline
30 &  66511 & 12375 & 14175 & 13500\\
\hline
60 &  47069 & 15750 & 14125 & 13550\\
\hline 
90 &  37600 &- &- & \\
\hline
\end{tabular}
\end{table}
Figure~\ref{pneut} shows the momentum spectrum of tagged neutrons
accepted in the calorimeter as a function of the beam momentum. Other
similar plots are available in MIPP note 130~\cite{raja130}.
\section{Non-perturbative QCD physics}
The upgraded MIPP detector will provide high statistcs data using hydrogen and
nuclear targets that will permit the investigation of non-preturbative
QCD with unprecedented statistics. Here we list a number of such
topics that can be addressed by the upgraded spectrometer.
Indeed, a large fraction of the liquid hydrogen running can be done 
symbiotically with the tagged neutral beam running.
\subsection{Further testing of a Scaling Law of Hadronic Fragmentation}
\label{scaling}
The scaling law in question~\cite{scale} states that the ratio of a
semi-inclusive cross section to an inclusive cross section involving
the same particles is a function only of the missing mass squared
($M^2$) of the system and not of the other two Mandelstam variables
$s$ and $t$, the center of mass energy squared and the momentum
transfer squared, respectively.
Stated mathematically, the ratio
\begin{equation}
\frac{ f_{subset}(a+b \rightarrow c+X)}
{f(a+b\rightarrow c+X)}\equiv \frac{f_{subset}(M^2,s,t)}{f(M^2,s,t)}
=\beta_{subset}(M^2)
\end{equation}
{\em i.e.}, a ratio of two functions of three variables is only a function of
one of them.  The physics behind
the scaling law may be understood by considering
inclusive cross sections as the analytic continuations of crossed
three body interactions, which factorize into a production term that
results in the formation of a short-lived fireball of mass $M^2$, which
subsequently decays into the subset in question. The formation is
governed by $s$ and $t$. The decay term is only a function of
$M^2$. It should be noted that the physics in question falls outside
the scope of perturbative QCD and as such the scaling law is not
currently derivable from QCD considerations.

In the MIPP data already taken, we managed to acquire $\approx$ 5.65 million 
events on liquid hydrogen at beam momenta 20~GeV/c, 60~GeV/c and 85~GeV/c.
These are currently being analyzed and will form the basis of testing the
proposed scaling law as a function of both $s$ and $t$.

With the upgrade, we can extend the test of 
the scaling relations with two particle 
inclusive final states, which will require higher statistics due to the 
larger number of variables to test against.

\subsection{Antiproton Interactions in MIPP}

The FAIR project has been approved by the German government and will
provide a facility for research into anti-proton and ion interactions
at GSI Darmstadt. The start of construction is planned for 2007 with the
first experiments being set for 2012 and the completion of the project
is scheduled for 2014. The cost of the project is $\approx$ 1~billion
euros. PANDA~\cite{panda} is one of the flagship experiments at FAIR and 
stands for Proton ANtiproton DArmstadt). 

The GSI-KVI group are interested in measuring cross sections of
antiprotons on hydrogen and other nuclear targets in MIPP  to
help them design the PANDA  detector.  MIPP
has antiproton beams with momenta as  high as $\approx$ 60~GeV/c and
as low as $\approx$ 3~GeV/c. The excellent particle identification
capabilities of MIPP will enable a systematic study of anti-proton
interactions in particular the annihilation cross section.

The PANDA experiment proposes to measure
$p\bar p$ interactions in the charmonium range and higher. They are
also interested in the magnitude of open charm production in $p \bar
p$ interactions of which little is known in this energy range. The
presence of pixel planes in MIPP might facilitate a measurement of
these rare processes.

\subsection{High Multiplicity Events in MIPP and the question of bosonic condensation}
We propose to investigate the production of high multiplicity events
in $pp$ interactions where excesses may exist due to the
occurence of Bose-Einstein interference~\cite{sissakian} 
in multi-pion production. This study can be done 
on the large sample of $pp$ interactions we will collect 
using the liquid hydrogen target.

     The goal is to investigate collective behavior of particles in
       multiple hadron production in pp and pA
      interactions at the beam energy Elab=30 - 120 GeV. We will study
      the domain of very high multiplicity (VHM) for 
      $z > 4$, where $z=n/<n>$.
      At large multiplicities, 
      near the threshold of reaction, all particles will 
      have small relative momentum with respective to each other. 
      In a thermalized cold and dense hadronic gas 
      a number of collective effects listed below may show up as a 
      consequence of multiboson interference.  
\begin{itemize}
     \item A large increase of partial cross
      section $\sigma$(n) of n identical particles production is
      expected, compared to the  commonly accepted extrapolation.  
      \item   Formation of jets consisting of identical particles may occur
      (pionic laser).  
\item  Large fluctuation of charged
      n($\pi^+$, $\pi^-$) and neutral n($\pi^0$) components, onset of
      Centauros or chiral condensate effects may occur.  
\item  Increase of the rate of the direct photons as the result of the
      bremsstrahlung in partonic cascade and annihilation $\pi^+$ +
      $\pi^- \rightarrow n\gamma$ in dense and cold pionic gas or
      condensate is expected. The creation of a multipion bound state is
      possible which in the course of its formation emits soft photons.  
\item  In
      the domain of high multiplicity, the major part of the center of
      mass energy is materialized leading to high density of hadronic
      system. At this condition a phase transition to cold 
      Quark Gluon Plasma may occur.  
\item The momenta of produced particles in the center of mass system
       in the VHM domain
      are small. Then the multi-particle Bose-Einstein correlation
      may be readily seen. The latter can lead to the ``hadron laser'' 
      effect, and the enhancement of soft gamma-quanta production.  
\item 
      We expect a uniform energy distribution over produced particles
      due to thermalization in this regime.
 


\end{itemize}

\begin{figure}[htb!]
\centering
\includegraphics[width=0.7\textwidth]{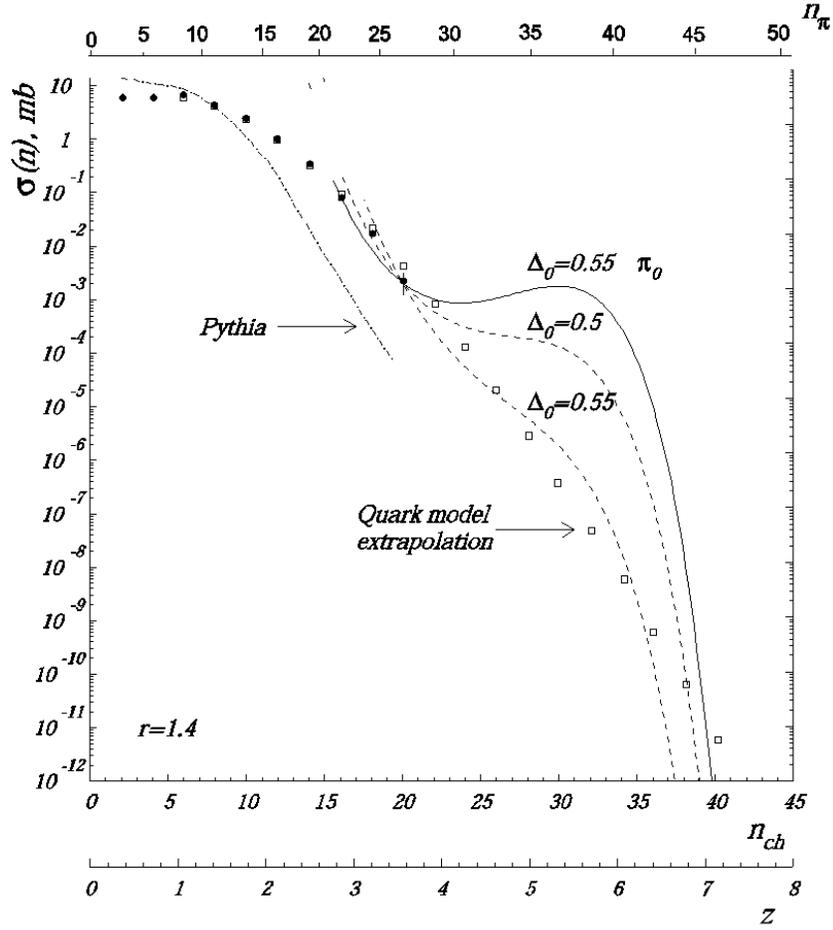}
 \caption{Multiplicity distribution in $pp$ interactions at 70~GeV
 (solid points). the curves marked by the parameter $\Delta$ are model
 calculations accounting for identical pions interference.}
\label{highmult}
\end{figure}

      The process of energy dissipation in hadron
      interactions poses a complicated problem for theory. For instance,
      the event generator Pythia gives the $pp$ 
      cross section $\sigma$(z), at $z > 2$,  two order of magnitude lower
      than the experimental value, see Figure~\ref{highmult}. Hence 
      further experimental and theoretical investigations are crucial
      to solve this problem. It may be closely connected to the vacuum
      structure of QCD and the confinement phenomenon.  
      MIPP can investigate the properties of multiboson systems in
      the domain of low temperature and high density where pions may
      be in a state of boson condensation. The estimates shows that the
      temperature of the hadronic system becomes lower than 25 MeV at a
      multiplicity of $z=5$. 
      At such a temperature, the pionic gas may be in a
      condensate state. MIPP offers a unique
      opportunity to investigate  the above-mentioned problems.  Using the
      partial cross section extrapolation (Figure~\ref{highmult}) 
      one can estimate
      the counting rate in the VHM domain. As an example, for  z=4, 
      the partial cross section is $\sigma\approx$
      0.2 $\mu b$.  In a 10 day run on the hydrogen target, we will collect 50~      Million events, in which there will be $\approx$~300 events with 
      $z\geq 4$.  
      According to theoretical estimates,  the
      multiparticle enhancement effects could become prevalent in the
      multiplicity domain z $\geq$ 4, 
      resulting in a greatly increased rate in this region.

\subsection {Baryon Spectroscopy with the upgraded MIPP}

Partial-wave analyses of $\pi N$ scattering data have yielded some of
the most reliable information about nonstrange baryon resonances.
These analyses provide information about resonance masses and total
decay widths (or their pole positions) and $\pi N$ branching fractions.
In order to determine resonance couplings to other channels,
it is necessary to study inelastic $\pi N$ scattering reactions,
such as $\pi^- p \to \eta n$, $\pi^- p \to \pi^+ \pi^- n$, and
$\pi^- p \to K^0 \Lambda$, to give only a few of many possible
examples.  Important information is also provided by meson photoproduction
experiments, such as $\gamma p \to \pi^0 p$, $\gamma p \to K^+ \Lambda$,
and $\gamma p \to \pi^+ \pi^- p$.  These hadronic and 
electromagnetic reactions are all linked
by unitarity of the S-matrix, and modern coupled-channel analyses
attempt to describe data from both hadronic and electromagnetic
channels within a single
consistent framework.  

The data obtained from $\pi N$ scattering and meson photoproduction
experiments provide crucial information about QCD in the nonperturbative
regime.  One of the important issues concerns how many internal degrees
of freedom are really needed to describe baryon resonances.
Essentially all of the known baryon resonances can be described
as quark-diquark states, whereas quark models predict a much richer
spectrum involving three dynamical quark degrees of freedom.
That is, quark models predict many more states than have been
observed experimentally.  These states are commonly known 
as ``missing resonances''. There are two likely solutions to this puzzle:
(1) the missing states simply do not exist; or (2) the missing
states have not been seen because they couple weakly to the $\pi N$
channel.  

MIPP data with  pion beams less than 5~GeV/c on liquid hydrogen will permit
coupled channel partial wave analyses to be performed on a variety of channels
such as
\begin{eqnarray*}
\pi^- p \to \pi^- \pi^0 p \\
\pi^+ p \to \pi^+ \pi^0 p \\
\pi^- p \to \pi^+ \pi^- n \\
\pi^+ p \to \pi^+ \pi^+ n 
\end{eqnarray*}
where the missing neutral is detected by the missing mass.
\subsection {Missing Cascades}
As discussed in the nonstrange (S=0) baryon spectroscopy part of this 
proposal, discovery of the excited states of the nucleon, the N*'s and 
the $\Delta$*'s, has come from partial wave analyses of these states 
being formed from pion-nucleon scattering.  Likewise, strange (S=-1) 
baryon spectroscopy, the $\Lambda$*'s and $\Sigma$*'s, has relied 
primarily on direct formation of these states via $K^-$p scattering, 
e.g. K$^-$p $\rightarrow (\Lambda^* $ or $ \Sigma^*) \rightarrow $ {\it decay 
products}.  However, discovery of the excited states of S=-2 baryons, 
the $\Xi$*'s (cascade hyperons), has been obtained primarily from 
production mechanisms.  Production of these states via the $K^-p
\rightarrow K^+ \Xi$* reaction is proposed here for the MIPP upgrade 
program.

Again, a concise review of the status of our knowledge of $\Xi$ 
resonances is found in the {\it Review of Particle Physics} \cite{rpp04}.  Quoting, 
``Not much is known about $\Xi$ resonances.  This is because (1) they 
can only be produced as a part of a final state, so the analysis is 
more complicated than if direct formation were possible, (2) the 
production cross sections are small (typically a few $\mu$b) and (3) 
the final states are topologically complicated and difficult to study 
with electronic techniques.  Thus early information about $\Xi$ 
resonances came entirely from bubble chamber experiments, where the 
numbers of events are small, and only in the 1980's did electronic 
experiments make any significant contributions.  However, nothing of 
significance has been added since our 1988 edition.''

By SU(3) flavor symmetry, the spectrum of $\Xi$* states has a 
one-to-one correspondence with N* and $\Delta$* states.  In the 
conventional quark model, the N*'s are radial and rotational 
excitations of {\it udd} and {\it uud} configurations and the $\Xi$*'s are 
excitations of {\it uss} or {\it dss} combinations.  Thus, the same question of 
``hidden'' or ``missing'' resonances appears.  Only 11 $\Xi$'s are listed 
in Ref. \cite{rpp04} (including the ground state), while 44 are predicted.  
These states are much narrower than the N*'s (tens of MeVs rather than 
hundreds), making them easier to identify and distinguish.  Hence, the 
study of the spectrum of doubly strange hyperons provides advantages in 
understanding the spectroscopy of all hadrons in particular and 
nonperturbative QCD in general.

\begin{figure}[htb!]
\begin{minipage}{35pc}
\begin{center}
\includegraphics[width=35pc]{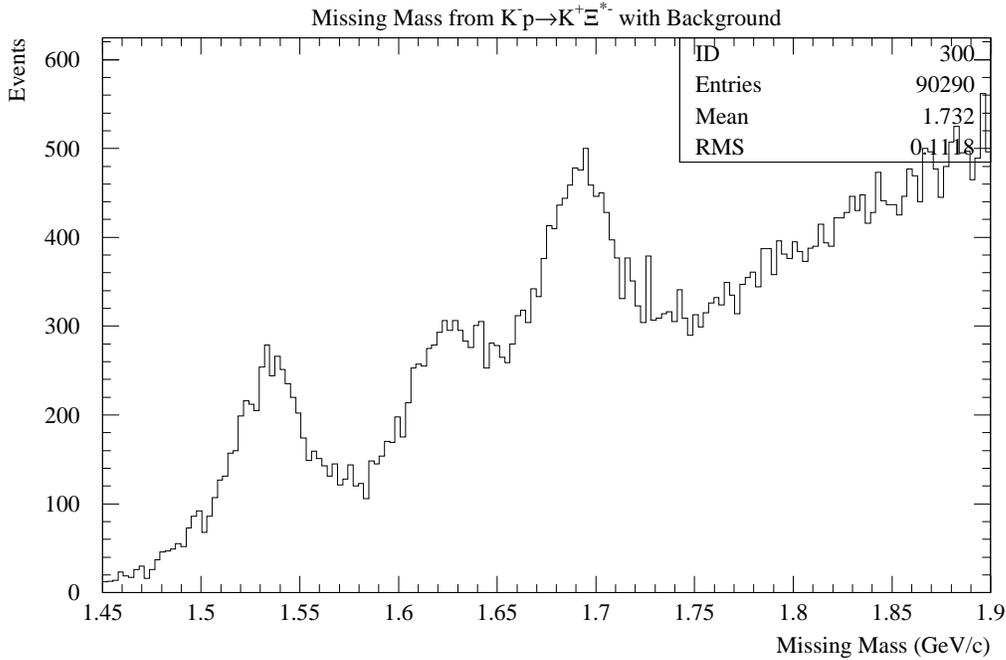}
\caption{\label{fig:low_xi_stars_with_bg} Monte Carlo simulation of
$K^-p \rightarrow K^+ \Xi$*$^-$.  The simulation includes a total of
100000 events with an assumed 5\% signal at each of the three
resonances $\Xi$(1530)$^-$, $\Xi$(1620)$^-$, and $\Xi$(1690)$^-$ from
$K^-p \rightarrow K^+ \Xi$*$^-$.  The assumed background is 34\% $K^-p
\rightarrow K^+ K^- \Lambda$, 34\% $K^-p \rightarrow K^+ \Xi^- \pi^+
\pi^-$, and 17\% $K^-p \rightarrow K^+ \Xi^- \pi^0$.  The lab momentum
is 3.0 GeV/$c$.  Masses and widths were taken from Ref. \cite{rpp04}
estimates.  A 0.5\% beam momentum uncertainty was assumed, and the
momentum resolution of the final state $K^+$ was estimated by tracking
it through the MIPP Monte Carlo, counting the number of pads digitized and 
applying multiple-scattering and digitization errors.}

\end{center}
\end{minipage} 
\end{figure}

A Monte Carlo simulation, see Fig.~\ref{fig:low_xi_stars_with_bg},
indicates that the MIPP beam momentum resolution is a critical factor
in resolving these states.  Simulated here are the lowest three
$\Xi$'s listed by the PDG
\cite{rpp04}, assuming their values for the masses and widths.  The 
middle state in Fig.~\ref{fig:low_xi_stars_with_bg}, the $\Xi$(1620),
is a one star resonance, meaning that ``evidence of existence is
poor''.  If it exists, it presents a particular problem for quark
models because of its low excitation (only 300 MeV above the ground
state).  In contrast, the first excited N* state is the Roper
resonance at 1440 MeV, 500 MeV above the ground state.

The missing cascade problem can be investigated in MIPP using low 
energy $K^\pm$ beams.

\section{Hardware details of the MIPP Upgrade}

This section describes in detail the proposed upgrades and
repairs of the Jolly Green Giant magnet coils, the
TPC Front End electronics and upgrades and additions to the
rest of the data acquisition and detector systems.
We also discuss modifications to the MIPP beamline.

\subsection{Jolly Green Giant repair}

The Jolly Green Giant magnet (JGG) provides the magnetic field for charge
and momentum measurement of particle tracks in the TPC. The aperture of
the JGG magnet is large enough to fit the TPC. The magnetic field of 0.7~T
is (except for distortions) vertical and parallel to the electric drift
field inside the TPC.

\subsubsection{Present state of the JGG}

The JGG magnet has two pairs of water cooled copper coils with a total of
1024 coil turns. Turns are insulated from each other with sheets of G10 and
epoxy. The coils have been power cycled many times and have been operated
for a long time over the past four decades. A failure in one of the coils was
repaired before the first MIPP run. The magnet was then  used in MIPP
for three years. During this time we have had electrical turn-to-turn coil
shorts and water leaks in the coils four times. Three times the magnet was
restored to an operational state. The failing coils were bypassed with external
jumpers. The operating current was then increased to obtain the same
magnetic field as before each failure. The last failure close to the end of
the first run completely destroyed the bottom two coils.

\subsubsection{New coil design}

\begin{figure}[htb!]
\includegraphics[width=5in]{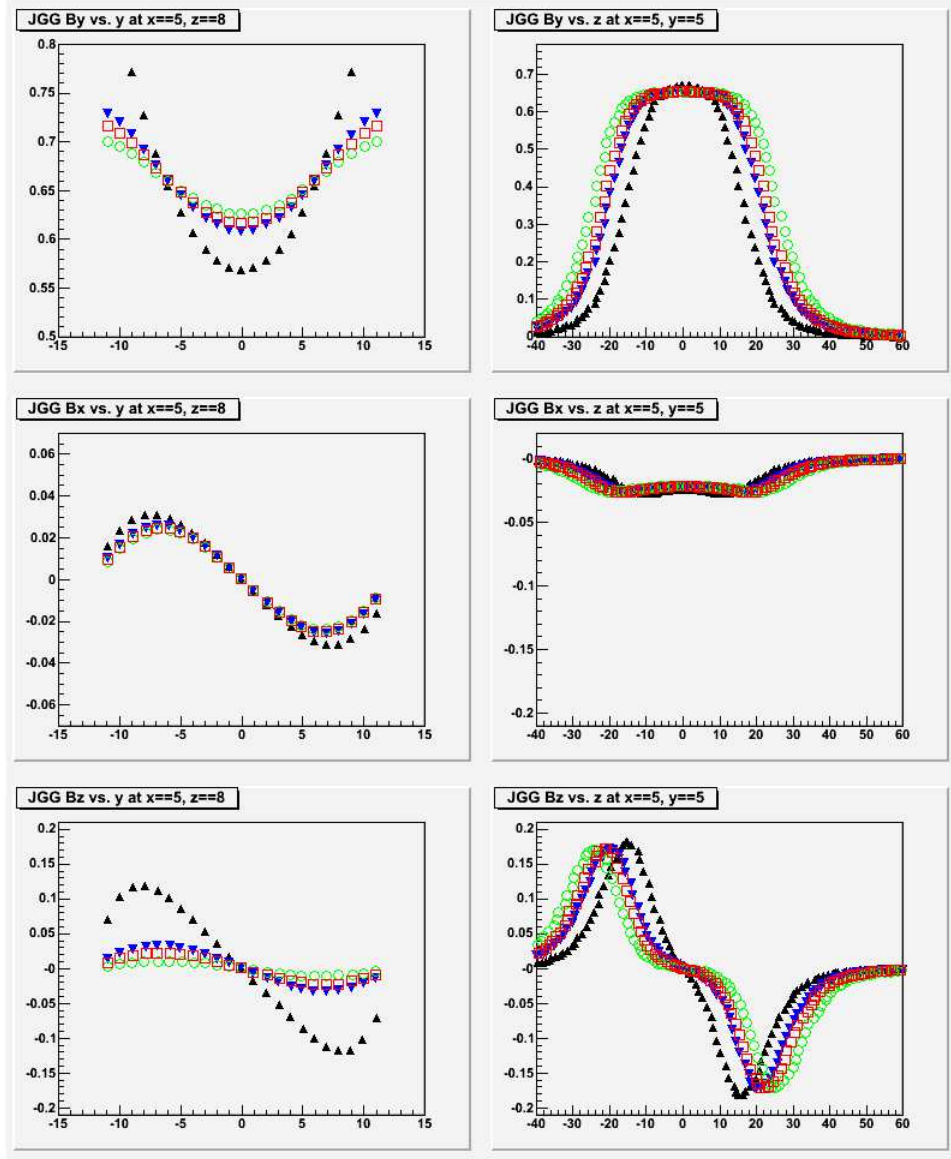}
\caption{\label{fig-JGG-field} Components of magnetic field in the JGG
in tesla plotted against grid positions (two inch per grid
space). Black triangles show the current coils. Blue inverted
triangles show the model for nine inch extended coils. Squares and
circles show extensions of 12 inches and 18 inches, respectively. The
field shapes due to the 9 inch version are considerably better than
the original field in that the extent of the bending field $B_y$ is
greater and the the other components $B_z$ and $B_x$ are better
behaved.}
\end{figure}

The need to replace the JGG coils opened up the possibility for
improvements. The magnetic field of the JGG is not very uniform. The
region of interest is the rectangular active drift volume of the TPC
centered in the magnet aperture. It extends 164~cm along the beam direction,
104~cm horizontally perpendicular to the beam, and 90~cm vertically. Within
this volume the magnetic field components perpendicular to the electric
field of the TPC reach up to 20\% of the magnetic field component parallel
to the electric field (see figure \ref{fig-JGG-field}).
The perpendicular components introduce distortions
in the TPC track data of up to 5~cm\cite{MIPPNote134}. These distortions can
be corrected but residuals of up to 1.4~mm remain. This impacts vertex
reconstruction and momentum determination. A more homogeneous magnetic field
results in better data.

The width of the coils is given by the shape of the magnet yoke. It is hard
to change. The pole pieces are 60 inches wide. This 1.46 times the width of
the TPC drift volume.  The length of the coils can be expanded more
easily with new coils expanding symmetrically upstream and downstream
of the yoke. Currently the pole pieces extend only 48 inches along the
beam. This is only 74\% of the length of the TPC drift volume. The
effect of larger coils on field uniformity was modelled for several sizes
of extensions in spring and summer of 2006\cite{MIPPNote134}. An extension
of the coils by 9 inches on each end gains a significantly more uniform
magnetic field and does not interfere with the placement of the detectors
downstream of the JGG. The further gains for extensions larger than 9 inches
are smaller.
With the new coils the pole pieces will have the same size along the beam
direction as the TPC drift volume. The resulting $\mbox{E}\times\mbox{B}$
distortions will be less than 3~cm throughout the drift volume of interest,
half of the distortion with the old coils. After correcting for the
new distortions remaining residuals will be at most 0.5~mm. This is
more than twice as good as the results with the old coil.

Besides this work on the coil geometry a lot of work was done on the
detailed design of new coils. For cost reduction the new coils will be
made from aluminum rather than copper. The coil conductor will have a larger
cross section. Two new coils will have 360 turns. The heating
calculation for the final design has been performed and found to be 
satisfactory. 
The impact on the magnet power supply
and power bus was evaluated. The new coil specifications are listed in
MIPP Note 137\cite{MIPPNote137}.

\subsubsection{Coil replacement}

The orders for aluminum for new coils and for the coil fabrication have
been placed in summer 2006.
The old coils have to be removed. Detectors in and upstream of the
JGG will need to be moved out of the way. The new coils will be larger
than the old coils. The pole pieces have to be lengthened in the beam 
direction to fill the resulting gap.

\subsubsection{Ziptracking the new magnetic field}

The magnetic field of the JGG with new coils has to be mapped. The Ziptrack
system was used to map the magnetic fields of both MIPP analysis
magnets before our commissioning run. 
We propose to upgrade the Ziptrack system with new Hall
probes because the cables and connectors on the old Ziptrack Hall probes
have become unreliable over time. Replacing cables and connectors on
the existing Hall probes is not cost effective because the Hall probes
would need to be recalibrated.

The PC currently used to control the Ziptrack and collect the data needs
to be upgraded and the Ziptrack software needs to be adapted to the new
hardware.

\subsection{The TPC Front End Electronics Upgrade}

We discuss below the details of the scheme to speed up the front-end
electronics of the MIPP TPC. This is done by acquiring 1100 ALTRO/PASA
chips originally designed for the ALICE experiment at the LHC. The
production run for the ALICE chips is scheduled in the coming
months. The STAR experiment at Brookhaven is ordering $\approx$ 10,000
ALTRO/PASA chips for upgrading the electronics of its TPC.  MIPP plans
to acquire 1100 ALTRO/PASA chips in the same chip production run
thereby reducing the cost (by a factor of $\approx$ 5) by sharing the
overhead with the STAR collaboration.

\subsubsection{Brief Description of the MIPP TPC}

\begin{figure}[tbh!]
\includegraphics[width=\textwidth]{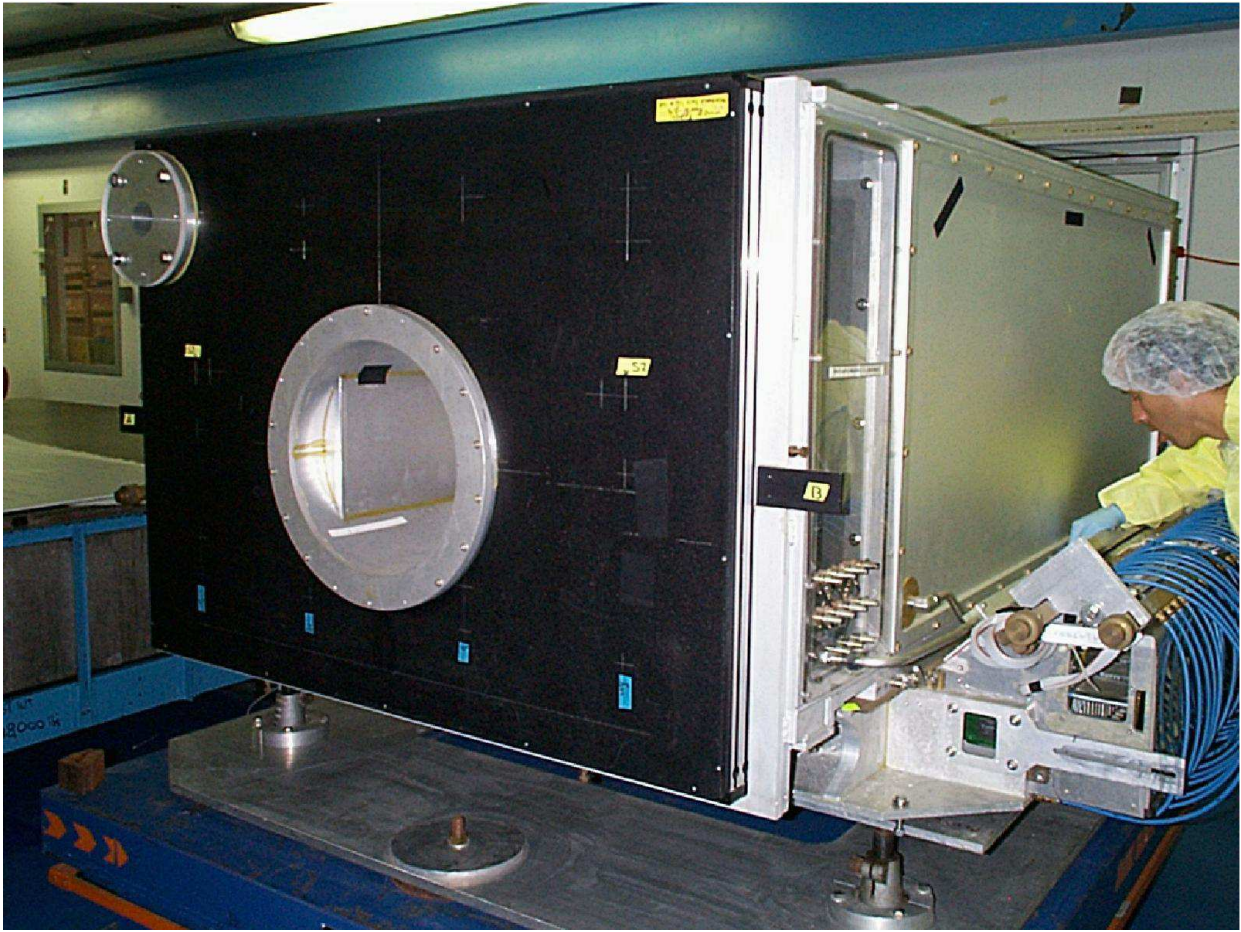}
\caption{The MIPP TPC.}
\label{tpc}
\end{figure}

The MIPP TPC \cite{Rai90} was originally designed and used at LBNL
in the EOS (E987) experiment and later at BNL (E895).  The TPC encompasses
an active gaseous volume of 96~cm wide by 150~cm long by 75~cm high (the drift
direction), corresponding to a maximum collection time of $16\,\mu{\rm sec}$.
To minimize space charge build up, the TPC incorporates a gating grid 
(currently limited to a maximum pulse rate of 3~kHz) which 
is pulsed to allow only ionization related in time.
Because of limitations in the readout 
electronics described below the trigger rate is presently limited to 
about 30~Hz.  Figure~\ref{tpc} shows  the MIPP TPC as viewed from the 
upstream end.

The information from the 15,360 channels in the  TPC is used to determine 
with high
precision, in three dimensions, charged particle tracks emerging from
the target station mounted on the front aperture of the detector.
This chamber has the ability to independently record over 3,800,000
individual data points for a single interaction event and forms the
basis of the precision momenta and dE/dx measurements for each
particle trajectory.  The original device that was refurbished for
MIPP was designed with a readout system that limited the total data
acquisition rate to a maximum of 60~Hz.  Redesign and updating of the
TPC front end electronics, replacing the aging 20 year old components
with new high density components, is projected to allow a 100 fold
increase in the maximum readout rate of the detector to a theoretic
limit of 3~kHz.

Currently the readout of the TPC is limited by the multiplexed serial
readout system which operates on non-zero suppressed data samples for
each given event.  In this manner the maximum readout speed is limited
to 60~Hz, due to the high channel count readout and slow (1~MHz)
multiplexing/digitization system.  The observed occupancy however, for
a typical interaction event in the TPC is only on the order of
5\% of the total channel count. This results in the possibility
of greatly increasing the readout capabilities of the detector by
performing the initial data filtering on board the front end
electronics and reducing by at least an order of magnitude the data
through-put that is currently required for a single system read.  The
readout can be further enhanced by improving the digitization time
required for each pad row and increasing the over all parallelization
of the readout system.

The design goal of the proposed electronics upgrade is to bring the
speed of the readout system to  3~kHz
for normal operation of the system.  Operation of the system at 3~kHz
requires that sustained readout of the chamber be accomplished in less
than 0.3~ms.  Non-uniformities in event rate induced by beam structure,
restricts this rate in such a manner that the operational time for
full event readout should not exceed 0.2~ms during burst operation for
sustained high speed data acquisition.

The average zero suppressed data size for events as measured during
the MIPP physics run was determined to be on the order of 115~kilobytes
for a multi-track interaction event.  The raw data rate when combined
with transaction overhead results in the requirement that the output data
pathway be designed to accommodate a single spill burst data rate of
575~megabytes/s.  The proposed upgrade addresses this throughput via a minimum
5-way parallelization of the output data-way, resulting in a requirement
of only 115megabytes/s per primary data pathway which is compatible with
commercial data bus implementations.

The upgrade of the TPC front end cards (FECs) to meet these requirements
has been studied using a pair of custom
designed ASICs that have been engineered, tested and produced for A
Large Ion Collider Experiment (ALICE) collaboration at the LHC for use
in their more than 570,000 channel time projection chamber.  The
system incorporates two separate chips, ``PASA'' an analog
preamp/shaper and ``ALTRO'' a fast ADC/filter which provides event
buffering, baseline corrections, signal filtering and zero
suppression.  The two chips are integrated in a standardized front end
card with a dedicated data bus that is synchronized to the main data
acquisition system via a series of readout control units (RCUs).  The
system has also been adopted by STAR collaboration at BNL, 
the BONUS at Jefferson Lab, as well as by the TOTEM
experiment at CERN.

\subsubsection{ALICE ASICs}
\label{sect:alice_asics}

\begin{figure}[tbh!]
\begin{center}
\begin{minipage}{15pc}
\includegraphics[width=15pc]{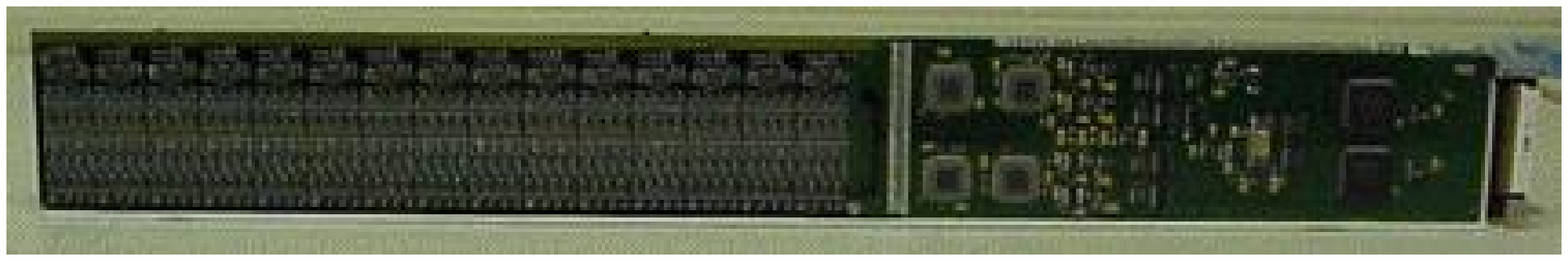}
\end{minipage}
\begin{minipage}{15pc}
\includegraphics[width=15pc]{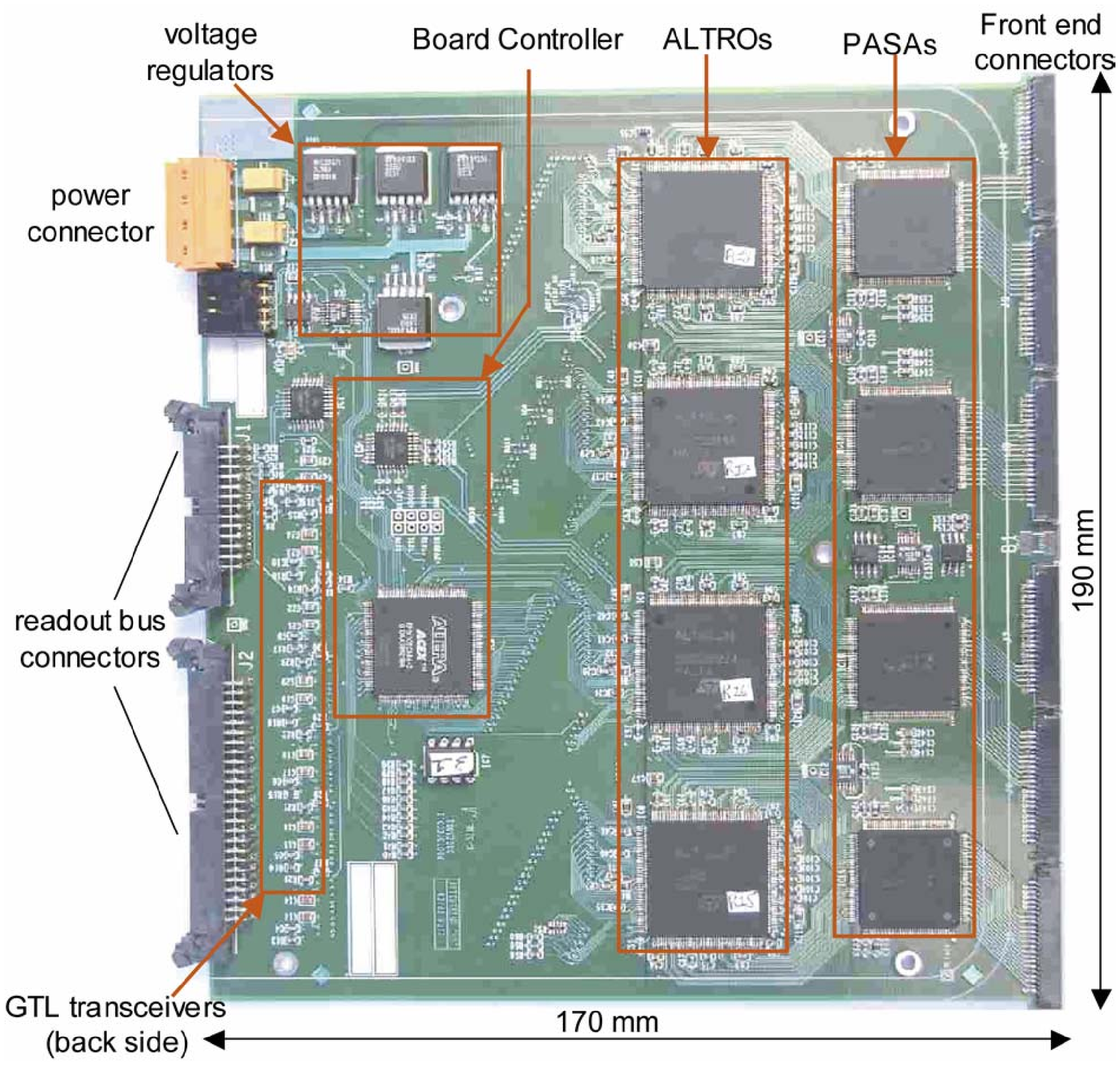}
\end{minipage}
\caption{TPC front end electronics boards for MIPP and ALICE.  The
difference in physical form factor requires a redesign of the ALICE
board to match the MIPP cold plates and connectors. Note that the two boards 
are not to scale and the MIPP TPC stick is roughly as long as half the 
TPC width.}
\label{tpc_stick_alice}
\end{center}
\end{figure}

To accommodate the readout of 570,136 charge collection pads, each
sampling  a maximum of 1000 samples per event, the ALICE
collaboration designed and engineered two custom ASICs to operate in
the high rate environment of the LHC heavy ion program.  The ALICE
readout design, as would be incorporated into the upgrade of the MIPP
TPC would replace both the analog and digital portions of the current
front end electronics cards.  Each of the existing 128 analog/digital
electronic ``sticks'' would be
removed and replaced in a one to one manner with an ALICE FEC
( both of which are shown
in Fig.~\ref{tpc_stick_alice}), redesigned to match the physical
dimensions of the aluminum cold plate upon which the current
electronics are mounted.  Additionally the cards would be fitted to
use zero insertion force (ZIF) socket compatible with the current TPC
chamber connections and interlocks.  This redesigned FEC follows in
all other respects the electrical design and characteristics of the
current CERN board layouts.
%
%

\begin{figure}[tbh!]
\begin{center}
\includegraphics[width=\textwidth]{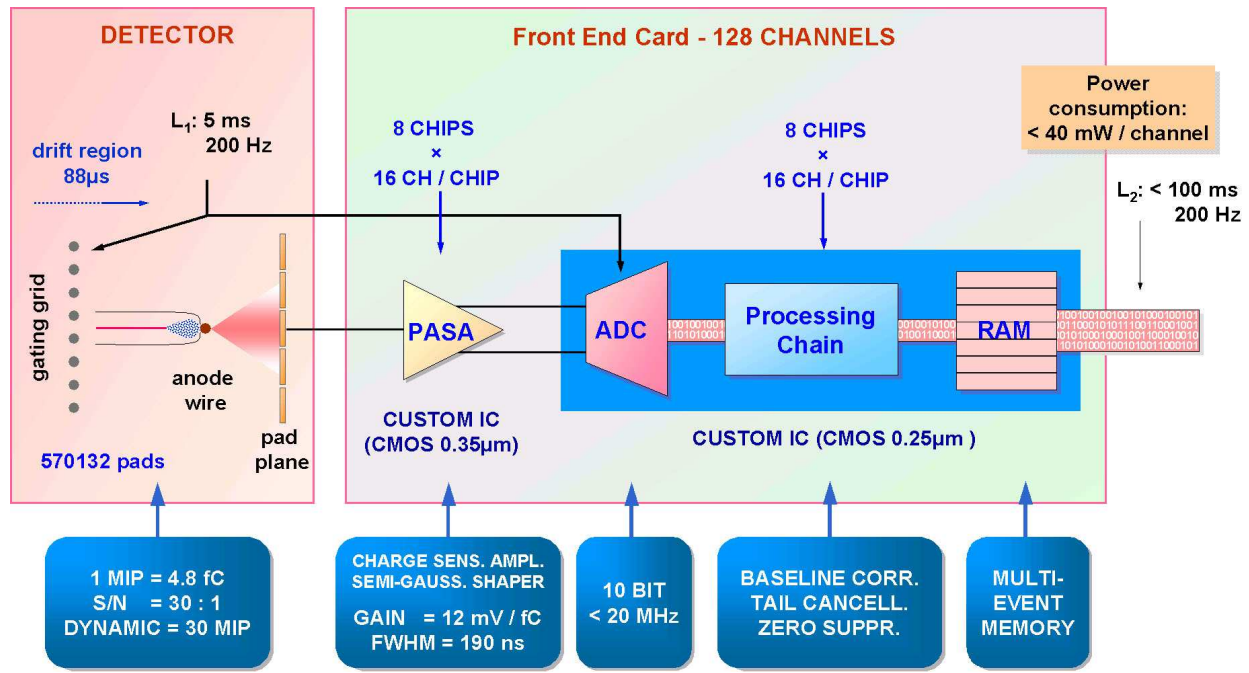}
\caption{ALICE front end card and readout system block diagram\cite{ieee_musa}.}
\label{alice_block_diagram}
\end{center}
\end{figure}
\begin{figure}[tbh!]
\begin{center}
\includegraphics[width=0.6\textwidth]{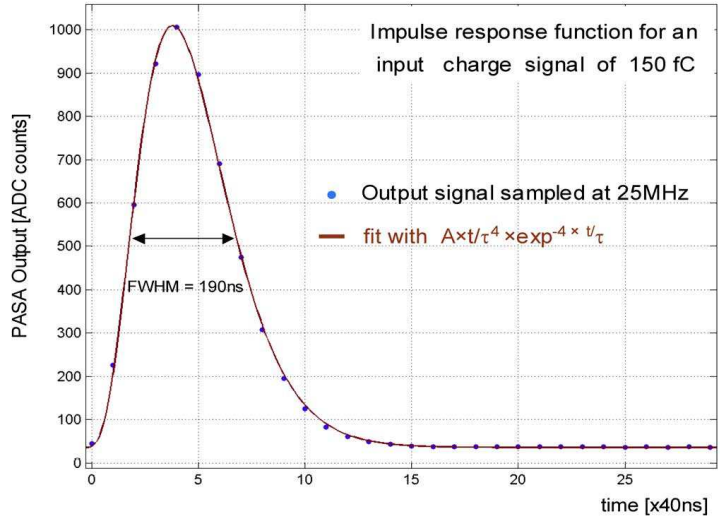}
\caption{PASA output function for an initial test charge of 150~fC\cite{ieee_bosch}.}
\label{pasa_waveform}
\end{center}
\end{figure}
\begin{figure}[tbh!]
\begin{center}
\includegraphics[width=0.6\textwidth]{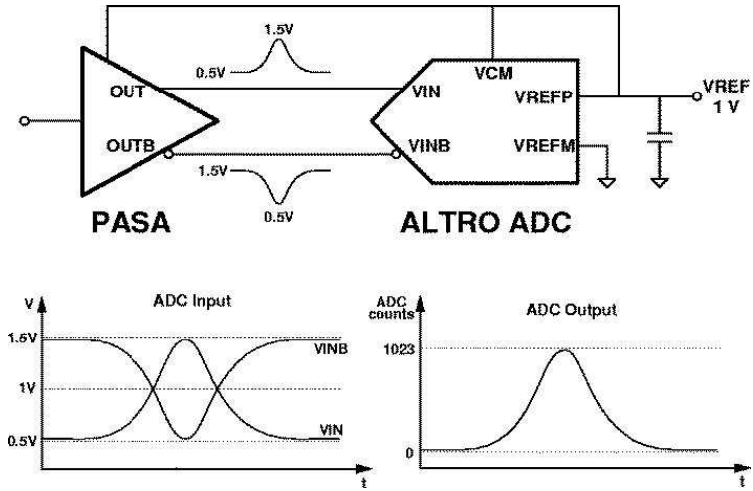}
\caption{PASA to ALTRO digitization logic\cite{campagnolo_2003}.}
\label{pasa_altro_logic}
\end{center}
\end{figure}
\begin{figure}[tbh!]
\begin{center}
\begin{minipage}{15pc}
\includegraphics[width=15pc]{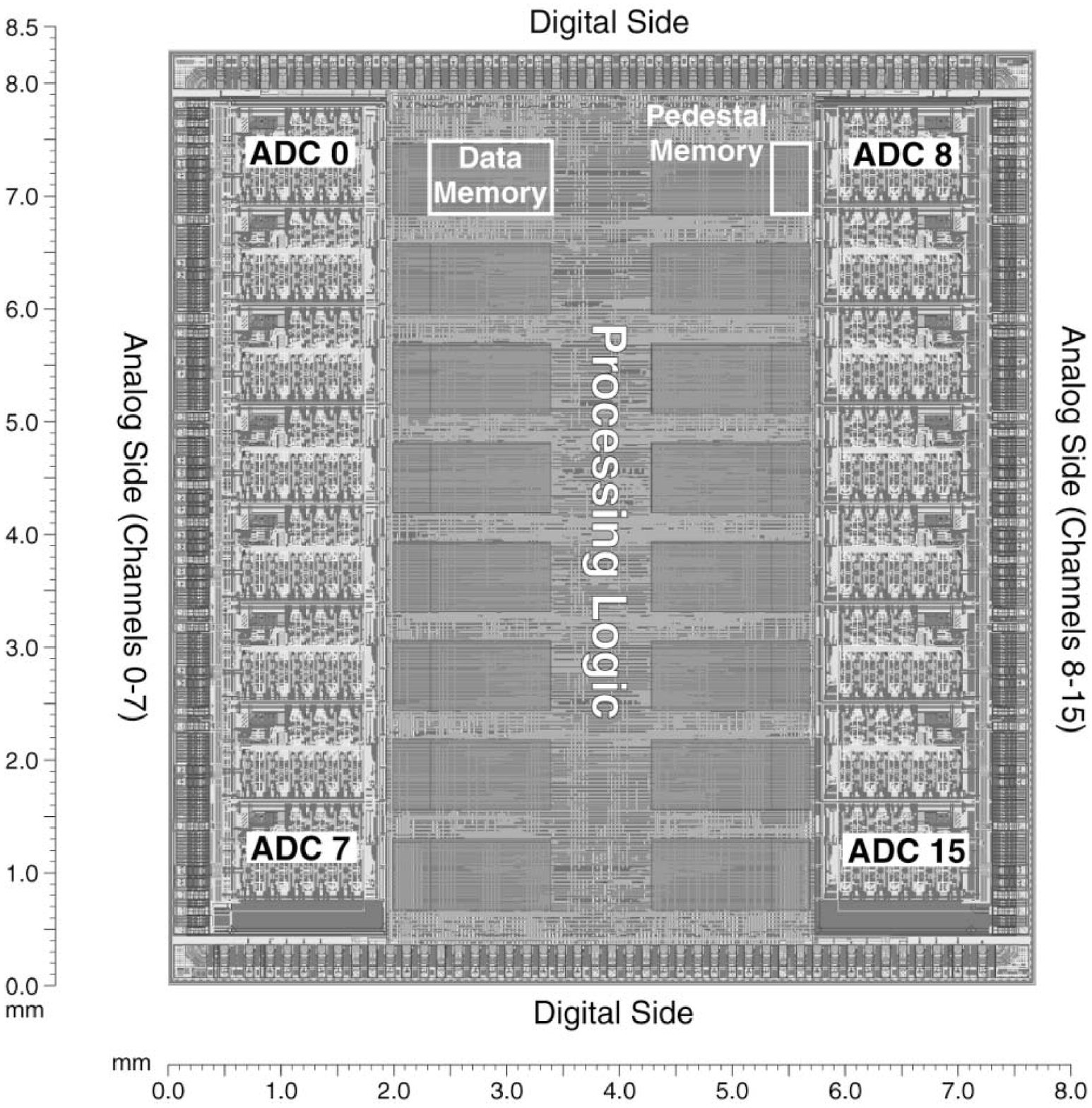}
\label{altro_asic_photo}
\end{minipage}
\begin{minipage}{15pc}
\includegraphics[width=15pc]{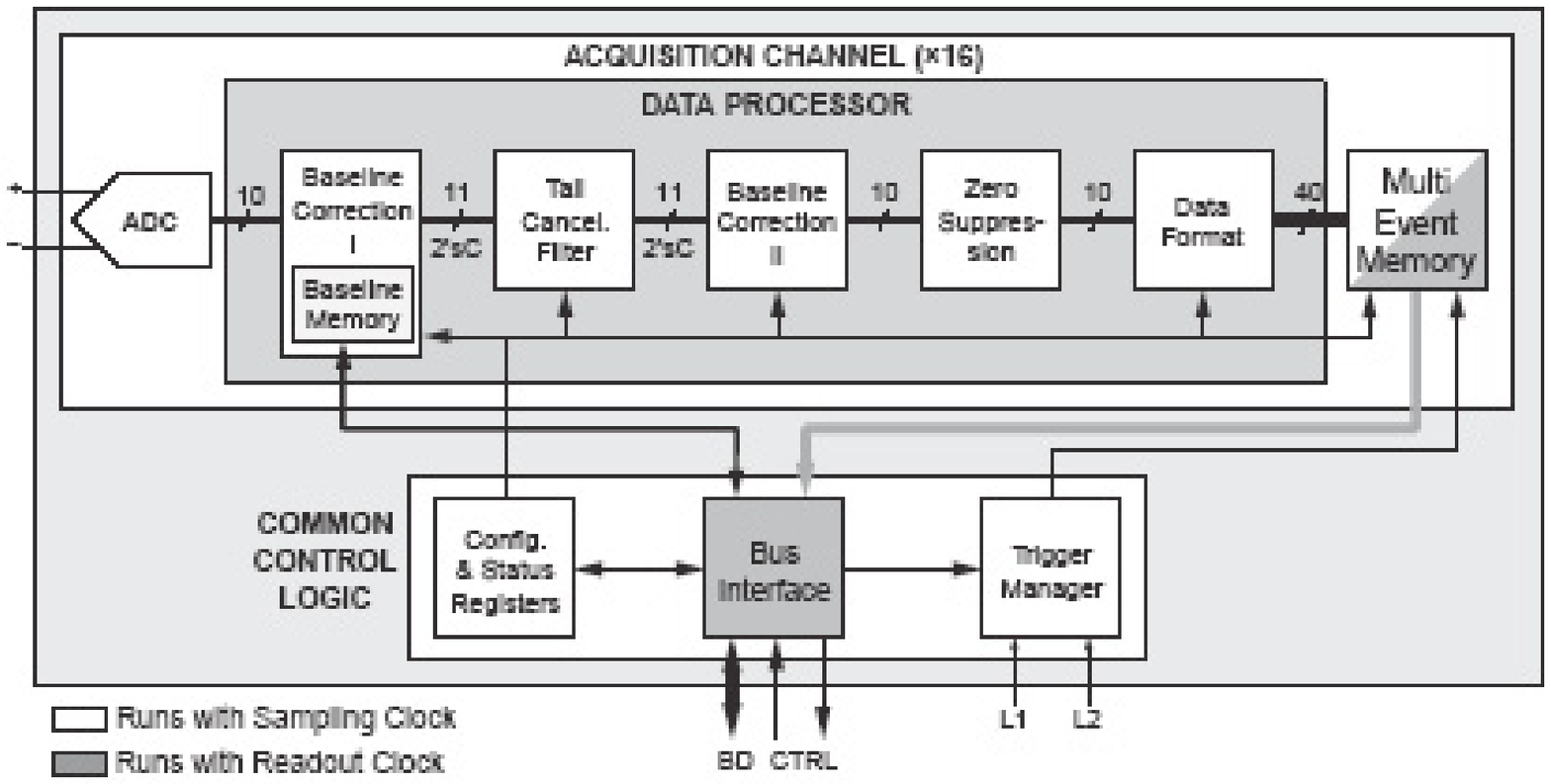}
\label{altro_asic_blockdiagram}
\end{minipage}
\caption{The ALTRO chip developed at CERN
services the readout of 16 channels by integrating a fast ADC, signal
processing and event buffering in a single package with interface to a
high speed data bus and programing lines.}
\label{altro_asci_diagrams}
\end{center}
\end{figure}

The ALICE system as shown in Fig.~\ref{alice_block_diagram} is
divided into two stages.  The raw signals from the detector pad rows
are first fed into a custom designed integrated circuit referred to as
``PASA'' which serves as the preamp and pulse shaper for each
channel\cite{mota_2000}.  The raw charge collected from the sample
window is reshaped into a sharply peaked output distribution of width
$\mathcal{O}(190~ns)$, as shown in Fig.~\ref{pasa_waveform}, which
is matched to the input requirements of the ALTRO chip for accurate
digitization. Each PASA chip services 16 readout pads and is matched
to the ADC inputs of the ALTRO chips shown schematically in
Fig.~\ref{pasa_altro_logic} for a single digitization channel.
The ALTRO ASIC as shown in Fig.~\ref{altro_asci_diagrams} is a 16
way parallel system including on each channel a 10~MHz ADC, digital
signal processor, and memory buffer.  The operation of the chip is
compatible with normal fixed-target data acquisition operation.
Although the signals are sampled continuously the data is processed
(pedestal subtraction, shaping and sparsification) only upon receipt
of a Level 1 trigger signal.  The processed data is stored in the
memory buffer upon receipt of a Level 2 accept signal or otherwise
discarded. The chips are controlled over a 40~bit wide ``ALTRO BUS''
developed at CERN \cite{subiela_2002}\cite{ieee_musa} for
communication with with a Readout Control Unit (RCU).

\begin{figure}[tbh!]
\begin{center}
\includegraphics[width=0.8\textwidth]{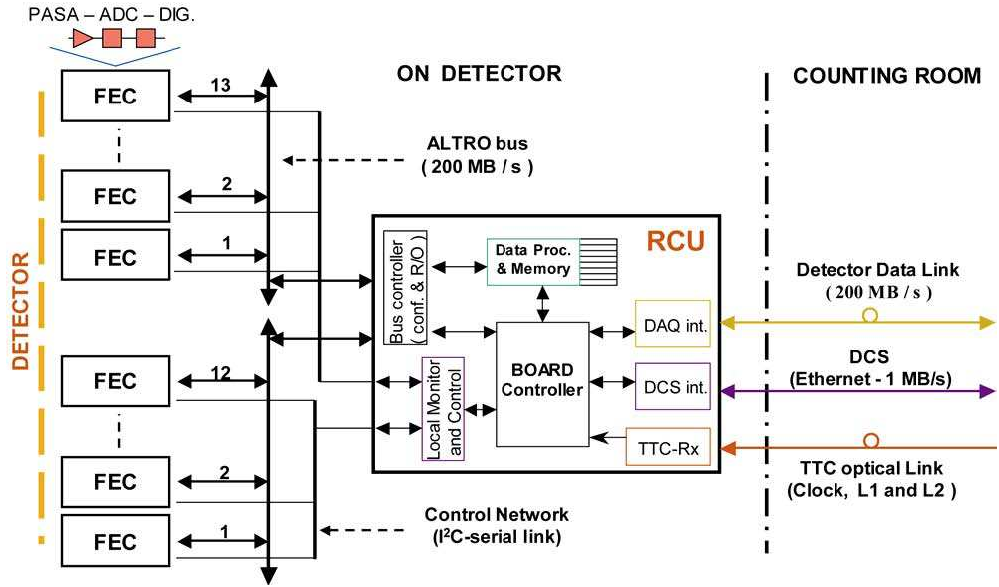}
\caption{Readout control unit interface between front end cards and main data acquisition system\cite{bosch_2002}.}
\label{rcu_readout_overview}
\end{center}
\end{figure}

\begin{figure}[tbh!]
\begin{center}
\includegraphics[width=0.8\textwidth]{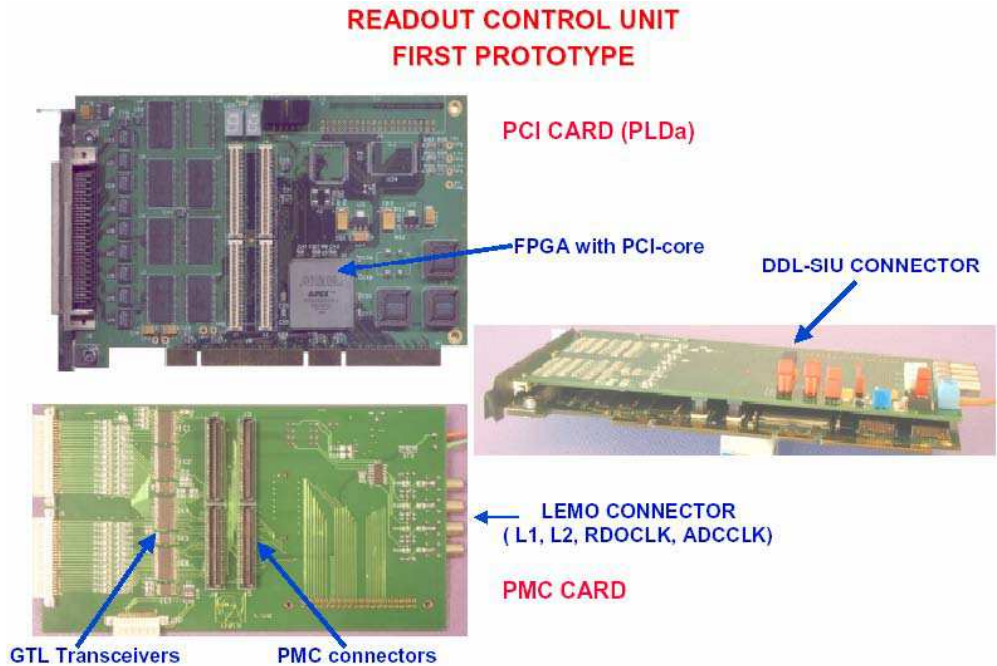}
\caption{Readout control unit PCI interface prototype\cite{bosch_2002}.}
\label{rcu_pci_prototype}
\end{center}
\end{figure}

\begin{figure}[tbh!]
\begin{center}
\includegraphics[width=0.6\textwidth]{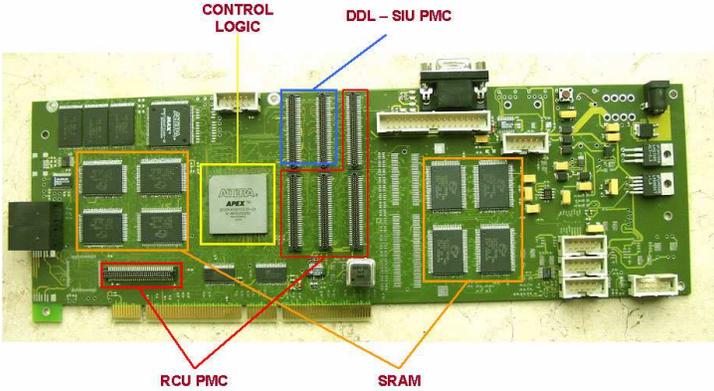}
\caption{Readout control unit alternate interface prototype\cite{bosch_2002}.}
\label{rcu_prototype2}
\end{center}
\end{figure}

\begin{figure}[tbh!]
\begin{center}
\includegraphics[width=0.6\textwidth]{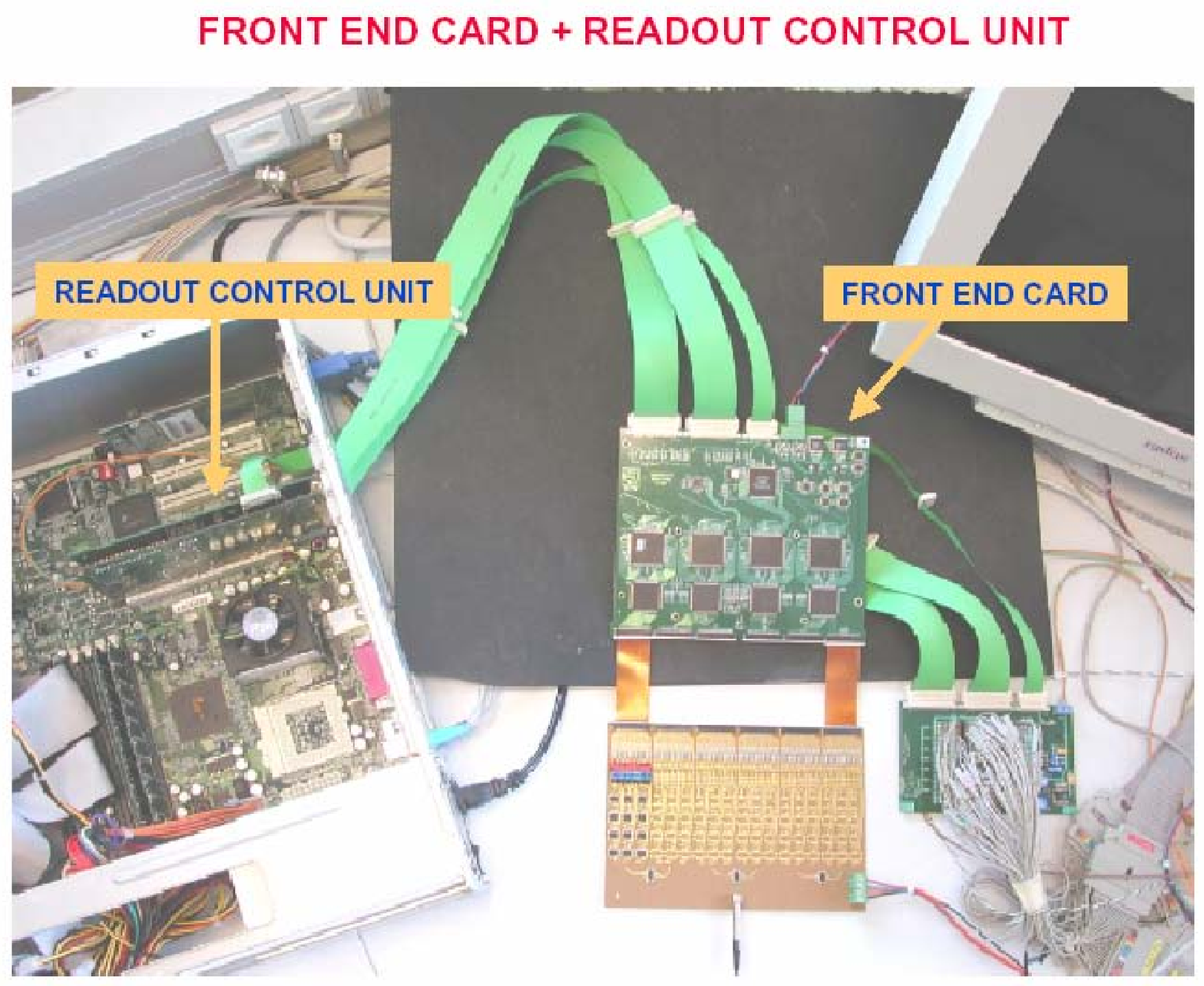}
\caption{Readout control unit test station\cite{bosch_2002}.}
\label{rcu_teststand}
\end{center}
\end{figure}

The RCU system, as shown in Fig.~\ref{rcu_readout_overview},
serves as the interface between the experimental acquisition and
control software and the front end cards.  Each RCU is designed to
interface with a set of 12 front end cards providing a single high
speed data pathway for the zero suppressed event data.  The protocol
of the output data path is customizable and has been shown to operate
both with the high speed serial link protocols developed at CERN, as
well as the more common USB protocol and peripheral connection
interface (PCI) to bridge the data directly into single board computer
memories for further processing.  Prototypes of the various RCU
interface cards as shown in Fig.~\ref{rcu_pci_prototype} and
Fig.~\ref{rcu_prototype2} and have been built and tested both by
the ALICE and BONUS collaborations using standard PC based test stands
like the one pictured in Fig.~\ref{rcu_teststand}. The current
prototyped RCUs would need no further adaptations to interface
directly with the current VME signal board computer system used in the
MIPP system.

Implementation of the ALICE front end electronics in the MIPP TPC
requires that several additional modifications be made to match the
operational needs of the existing hardware.  The time window for event
scanning and digitization will be reduced from 1000 samples per event
to 250 samples to match the drift time over the active volume of the
detector.  The reduced number of samples then allows for additional
segmenting of the ALTRO event buffer in such a manner that the FEC
cards will be able to fully buffer 8 events at a once.  To ensure that
the heat load generated by the new front end boards is compatible with
the existing cold plates and water cooling system, provisions have
been made to operate the ALTRO bus at 20~MHz instead of the 40~MHz
design frequency.  These modifications are projected to both increase
stability and retain the utility of as much of the existing equipment
as is possible.
The component requirements for the full system upgrade of the time
projection chamber are listed in
Table~\ref{table:tpc_chip_requirements}.  The projected yield for the
PASA and ALTRO wafers based upon the previous production run is
estimated at 82\%.  When yield is included, it is estimated that
1200 raw dies would be required to obtain enough components to
instrument the detector.
\begin{table}
\begin{center}
\begin{tabular}{|c|c|c|c|}
\hline
Component & Channels & no. Per FEC & Total Required \\
\hline
Front End Circuit Board & 120 & 1 & 128 \\ 
ZIF Sockets & & 1 & 128 \\
Preamp/Shaper (PASA) & 16 & 8 & 960 \\ 
ADC/Filter/Memory (ALTRO) & 16 & 8 & 960 \\ 
Readout Control Units (RCU) &  & 1:16 & 8 \\ 
Single Board VME PCs & & & 1 \\ 
Gigabit Network Switch & & & 1 \\
\hline
\end{tabular}
\caption[]{Component requirements for upgrade of the MIPP time projection chamber for operation at 3~kHz.}
\label{table:tpc_chip_requirements}
\end{center}
\end{table}

The cost per channel for the ALTRO electronics solutions, dependent
upon chip yield, is estimated at \$10 per channel based upon the
electronics costs for instrumenting the BONUS TPC at Jefferson lab.
The total cost of the front end electronics modifications is estimated
at a direct cost of \$180,000 without contingency.  Additional
cost is incurred in the procurement of single board VME style
computers for event filtering and synchronization.  The single board
processor is estimated to cost \$4800 dependent upon final
specifications and memory buffering requirements.  The total
direct cost of equipment for upgrading the MIPP time projection
chamber is estimated at \$190k without contingency.

The contract signed by CERN and Fermilab would deliver the 
1100 ALTRO/PASA chips needed by this upgrade scheme to be 
delivered after they are tested and verified to work.
 Faulty chips would be replaced by CERN.

\begin{figure}[tbh!]
\begin{center}
\includegraphics[width=\textwidth]{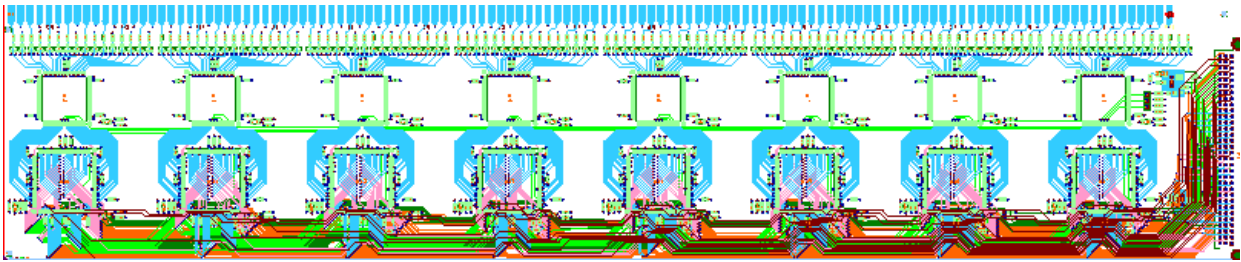}
\caption{Fermilab layout of the TPC stick using Altro/Pasa chips}
\label{fnal_stick}
\end{center}
\end{figure}

Fermilab has started the layout work for the TPC front end boards
using the Altro/Pasa chips. Figure~\ref{fnal_stick} shows the layout
of a complete TPC stick using Altro/Pasa chips.

\begin{figure}[tbh!]
\begin{center}
\includegraphics[width=0.6\textwidth]{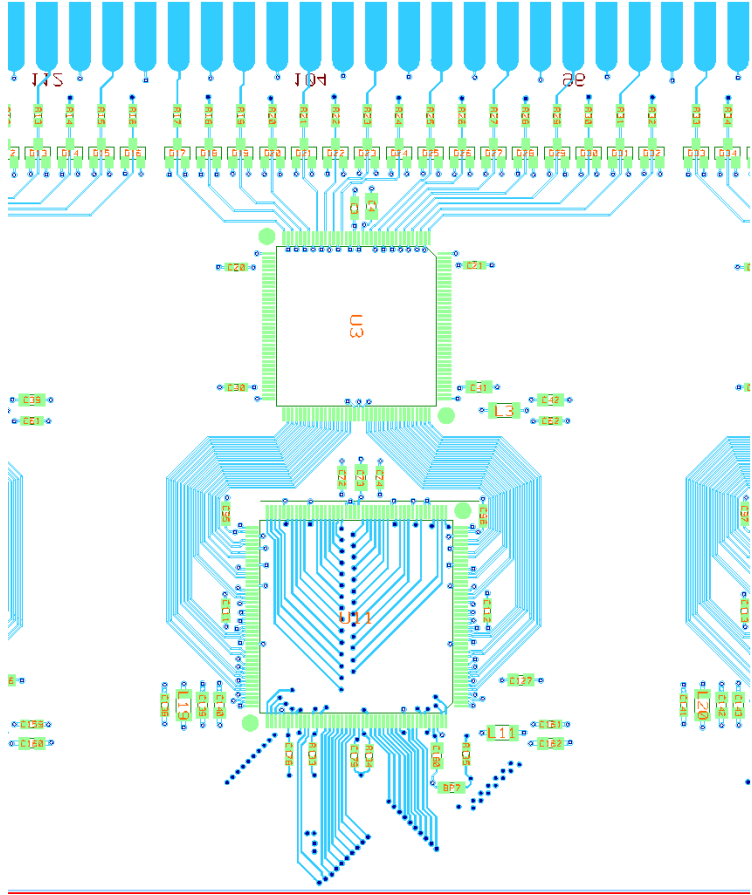}
\caption{Detail of the Fermilab layout of the TPC stick using Altro/Pasa chips
for one Altro/Pasa unit}
\label{fnal_stick_detail}
\end{center}
\end{figure}

Figure~\ref{fnal_stick_detail} shows the detail of the layout for one Altro/Pasa
unit.

\subsection{MIPP trigger system upgrades}

MIPP trigger consists of two parts. The raw beam trigger is formed by
coincidence between scintillator counters (TBD and T01) placed at the
entrance of the MIPP hall and just before our interaction target
respectively. Particle identification is performed on the beam using
the information from the beam \v Cerenkov system for higher energy
beam and by the time of flight counters for lower energy beam. We were thus 
able to trigger on 6 species of beam particles ($\pi^\pm, K^\pm$ and
$p^\pm$). The interaction trigger in MIPP consisted of information
from a scintillation counter (SCINT) placed after the target combined
with information from the first drift chamber downstream of this. This
interaction trigger has performed well during our physics run but its
purity/efficiency suffered for lower multiplicity events due to the
inability to tell apart single and multiple tracks in SCINT because of
Landau tails in the pulse height distributions. It also suffered from
occasional oscillatory behavior in the drift chamber. These
inefficiencies will be corrected during offline analysis of the
present run to obtain the correct multiplicity cross sections. The SCINT 
counter also introduced  
0.5\% interaction length, which
would would result in triggers caused by interactions in the counter.

In the upgraded MIPP experiment, we propose to remedy these defects by a new
interaction trigger based upon the fPix Silicon pixel detectors developed for
the BTeV experiment and being proposed for the Phenix detector at BNL.

\subsubsection{Interaction Trigger}

There will be one Silicon pixel plane upstream of the target and two
planes downstream of the target. Each layer will consist of an array
of six by eight fPix chips. The pixels on each chip are 
400~$\mu m$ by 50~$\mu m$, with the finer segmentation in the y
vertical direction. The fast chip hit signals from each fPix chip will all be
independent so that it can be input to a trigger processor, but the
readout of each individual chip will be coupled across the rows so
that each silicon detector plane provides six rows to read out. The
fast chip hit signal will be latched on the rising edge to be 50~ns 
wide. The  fast chip hit signal will be inactive for 130
to 200~ns. 
With our planned 300~kHz beam rate with  the beam being spread
over four chips, this will not impose any substantial dead time
limitations. Approximately 86\% of  the beam will hit
four chips upstream of the target. The two planes downstream
of the target will be used to form track-pointing to make sure that
the interactions come from the target and not from the 300~$\mu m$
silicon or other material.

Several different categories of triggers will be used. The first type
are non-interacting beam tracks that go without interacting in our 1\%
target and hits are observed in a ``bull's eye'' in the downstream
pixel counters where we expect beam. These non-interacting beam
triggers are highly pre-scaled and used for monitoring the experiment.
An interaction trigger is formed when we  do not have
a beam track in an expected ``bull's-eye'' region. This bull's eye
region is defined in a dynamic way for each beam track depending on
its point of impact in the first silicon plane. The other trigger
planned is based on total multiplicity of hits in the two pixel planes
downstream of the target. The interaction trigger logic on the fPix
signals will be performed in a custom interaction trigger board.

\begin{figure}[htb!]
\begin{center}
\includegraphics[width=\textwidth]{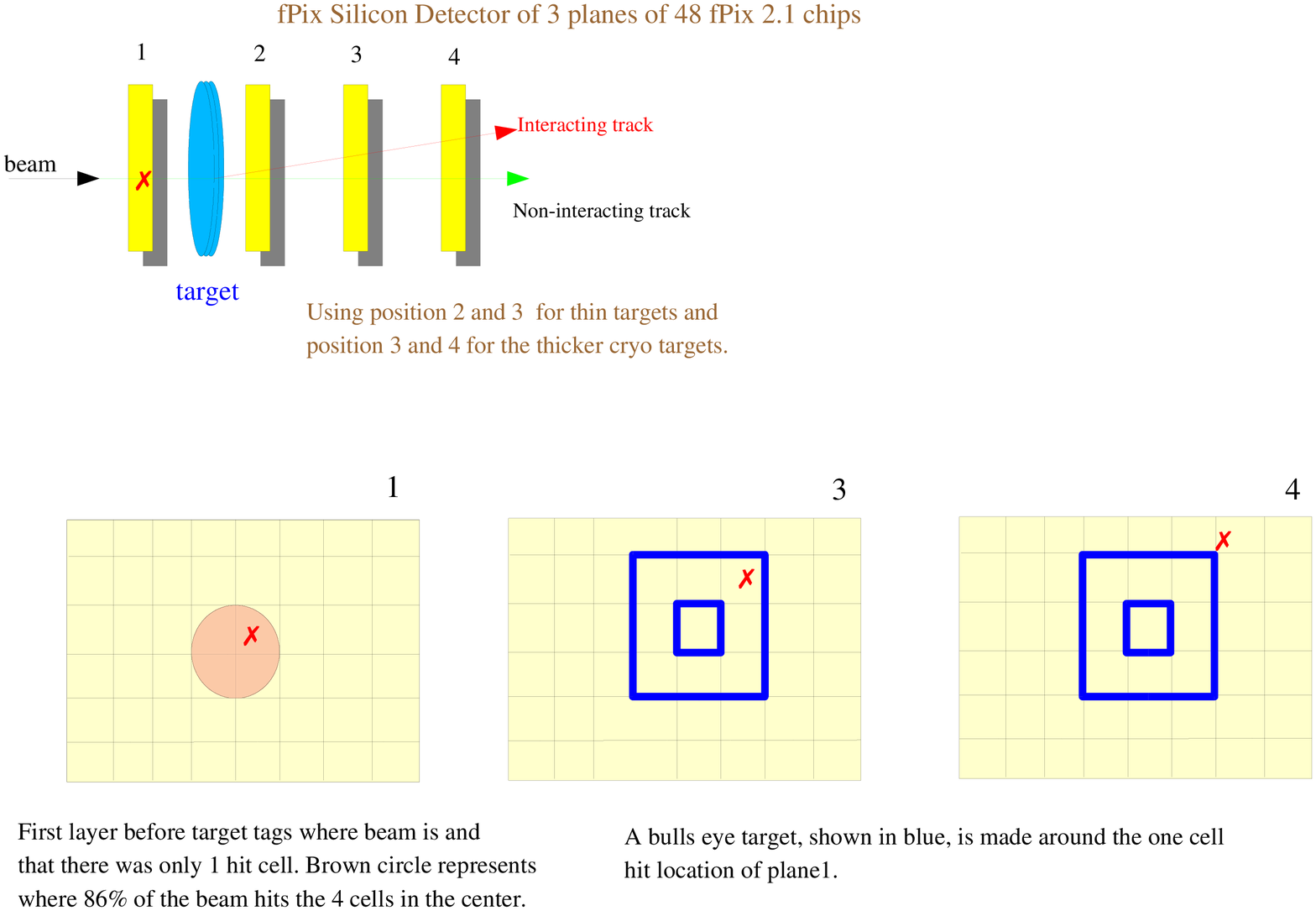}
\end{center}
\caption{Schematic of the new silicon based trigger system.}
\label{trigger1}
\end{figure}

The trigger schematic is shown in
Figure~\ref{trigger1}.

As in the original MIPP trigger system, we will keep 
the ability to pre-scale triggers so that we take 80\% of events
that have an interaction trigger with an equal amount of pions, kaons
and proton interactions. The other 20\% will be un-interacted
beam triggers, to monitor the efficiency and
performance of detector, trigger efficiency and dead time.
It should be pointed out that this trigger scheme will have greater efficiency
in triggering on low multiplicity events and so is of importance for the ILC 
tagged neutral beam events which have two charged tracks and one missing 
neutral.

\subsubsection{MIPP Trigger Electronics}
 
The MIPP trigger electronics will remain mostly as currently configured
 with some reorganization of the NIM and CAMAC crates. The main goal will be to
streamline the trigger system and remove some problematic modules
which have dead channels from the two years of usage during the 
previous run. This work consists mostly of 
getting partially dead modules replaced by PREP, reorganizing
the current crates and incorporate the new veto wall and fPix chip hit
signals as the new interaction trigger.

As in any cross section measurements we plan to take $\approx$ 10\% 
of all triggers as empty targets. For the experimental operation with thin
nuclear targets this will be done by taking a spill with one target,
(12,000 events in a 4 second spill every 2 minutes)
and the target wheel will automatically be
advanced to the next target. The wheel consist of 8 slots one of
which will be empty. By moving the target automatically between spills, we
plan to make the  most effective use of the beam and have empty target
runs that reflect the beam quality backgrounds.

Table~\ref{trupg} gives a summary of the equipment needed for the
trigger upgrade.

\begin{table}
\begin{tabular}{|c|c|}
\hline
Equipment & Cost \\
\hline
3 Planes of fPix detectors & \$87,100 \\
fPix mounting, cooling, LV, bias and PCI boards & \$50,000 \\
Interaction Trigger logic board  & \$8,800\\
\hline
\end{tabular}
\caption{Equipment summary for trigger upgrade\label{trupg}.}
\end{table}

\subsubsection{Trigger Tasks}
The following tasks have to be performed to build the trigger system.

\begin{itemize}
\item PPD tasks\\
a)Engineering support to mount fPix interaction trigger planes
on a stand and position to 100 micrometers around the target

b) Trigger board that receives the 144 fast chip hit signals generates
a fixed width gate based upon the rising edge of the signal. Time is
about 3 months to design (4 weeks), build (2 weeks) and program (1.5
months) a FPGA board including testing.

\item CD tasks\\
 a) Engineering support for the fPix chips readout into DAQ and
using the fast chip hit signals for interaction triggering.\\  
b) PREP to replace partially dead modules and provide new modules for the new
elements in the trigger.

\item MIPP tasks \\
a) Programming of the Physics triggers in FPGA trigger board. \\ 
b) 4 months of a physicist to streamline the trigger
system, replace partially dead modules and incorporate the new veto
wall and fPix interaction trigger into the DAQ trigger system. Two
weeks of beam time will be needed to understand and fine tune this
system.
\end{itemize}

\subsection{Upgrade for Chamber Readout Electronics}

Currently the drift chambers are read out through pre-amplifiers and
discriminators inherited from the E690 experiment that use CAMAC TDCs.

The large MWPC's are read out through RMH electronics that have not been
supported for several years.

This system was maintainable through the first MIPP physics run
and produced good data from three beam chambers and six tracking
chambers downstream of the experimental target.
However, there are several issues that need to be resolved.

We managed to keep the RMH system in working order up to now using spares.
However, it is not possible to guarantee that this readout system
will work through another run.

The readout of the CAMAC TDCs currently uses the CES~CBD8210. This
system is obsolete and not maintainable any more. CAMAC hardware
in general is becoming less well supported. Currently the TDCs are
read out without the use of DMA. The current readout system could
perhaps be upgraded to transfer data in DMA mode. A new solution for
the CAMAC readout and the change to DMA transfer would likely be
sufficient to obtain a readout time as needed for operation at 3000~Hz.
However, a new readout will provide much more flexibility at only
incrementally higher cost.

The existing system uses a large number of high current 
low voltage power supplies.
Each drift chamber uses two 5~V and one 10~V supply. All of these supplies
are aging and several have failed during the past run resulting in
minor downtimes.

The current system dissipates a large amount of heat in the area
of the experiment between the two analysis magnets. This has caused
significant problems with air conditioning in MC7. The air
conditioning itself is an old system. With the present heat load
the air conditioning system in MC7 would need to be upgraded.

In preparation for the first MIPP physics, new
electronics for the RICH detector readout were built. The RICH readout uses
front end boards that read out the RICH PMT's and send data to
VME readout boards. This solution has been working well and
essentially the same readout board will be used for the new TPC
electronics. This readout can be adapted for all chambers.

\begin{figure}[ht]
\includegraphics[width=\textwidth]{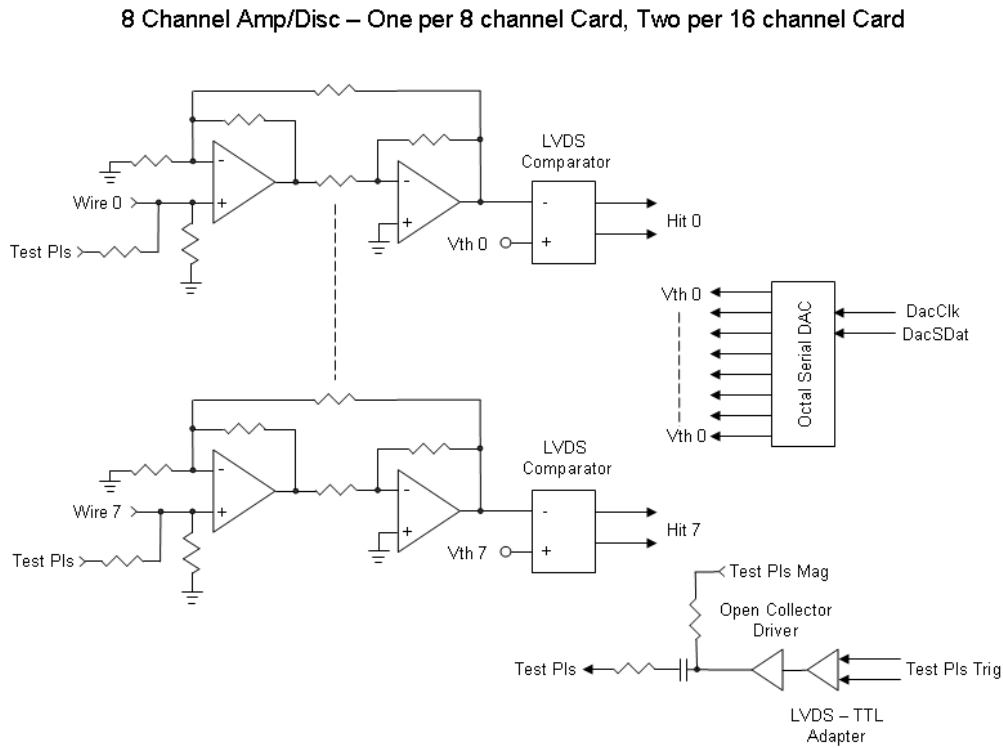}
\caption{\label{fig-Sten4} Chamber pre-amp card schematic.}
\end{figure}
\begin{figure}[ht]
\includegraphics[width=\textwidth]{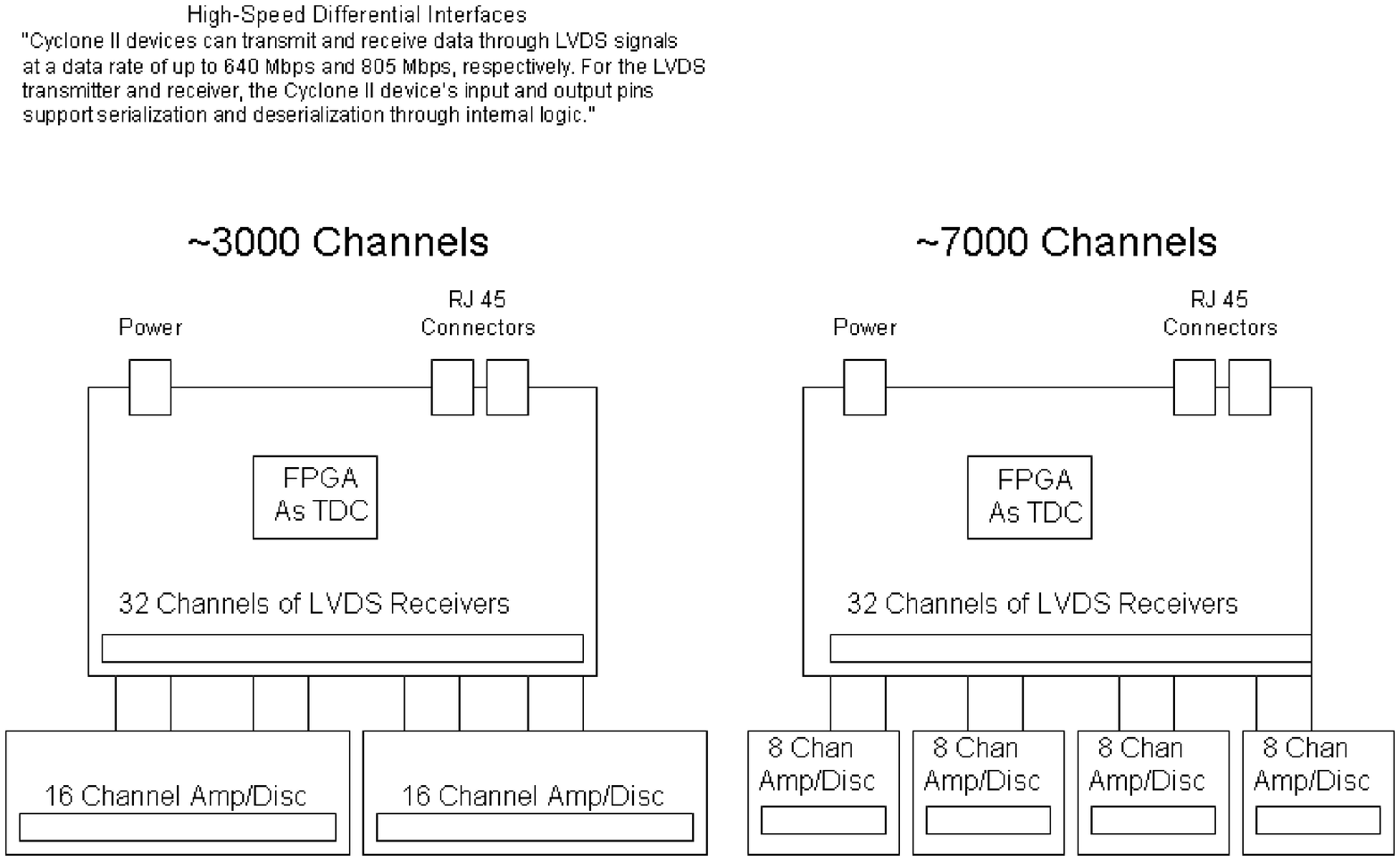}
\caption{\label{fig-Sten3} Chamber pre-amp and discriminator cards. The 16 channel pre-amp cards will be used with the MWPCs. The 8 channel cards are for drift and beam chambers.}
\end{figure}

For the new readout, we propose to build new pre-amp cards with
8 and 16 channels per card for the drift chambers and MWPC's,
respectively (see figure \ref{fig-Sten4}). These cards will be mounted on the chambers in the
same way that preamp cards are mounted presently. Several of these daughter
cards will connect to 32 channel TDC front end cards as shown schematically
in figure \ref{fig-Sten3}. These cards
will be based on the RICH front end cards. The RICH cards provide
a latch for each channel whereas the new chamber cards will use TDCs.
Each card will hold enough memory to buffer data for an entire
4 second slow spill.

\begin{figure}[ht]
\includegraphics[width=\textwidth]{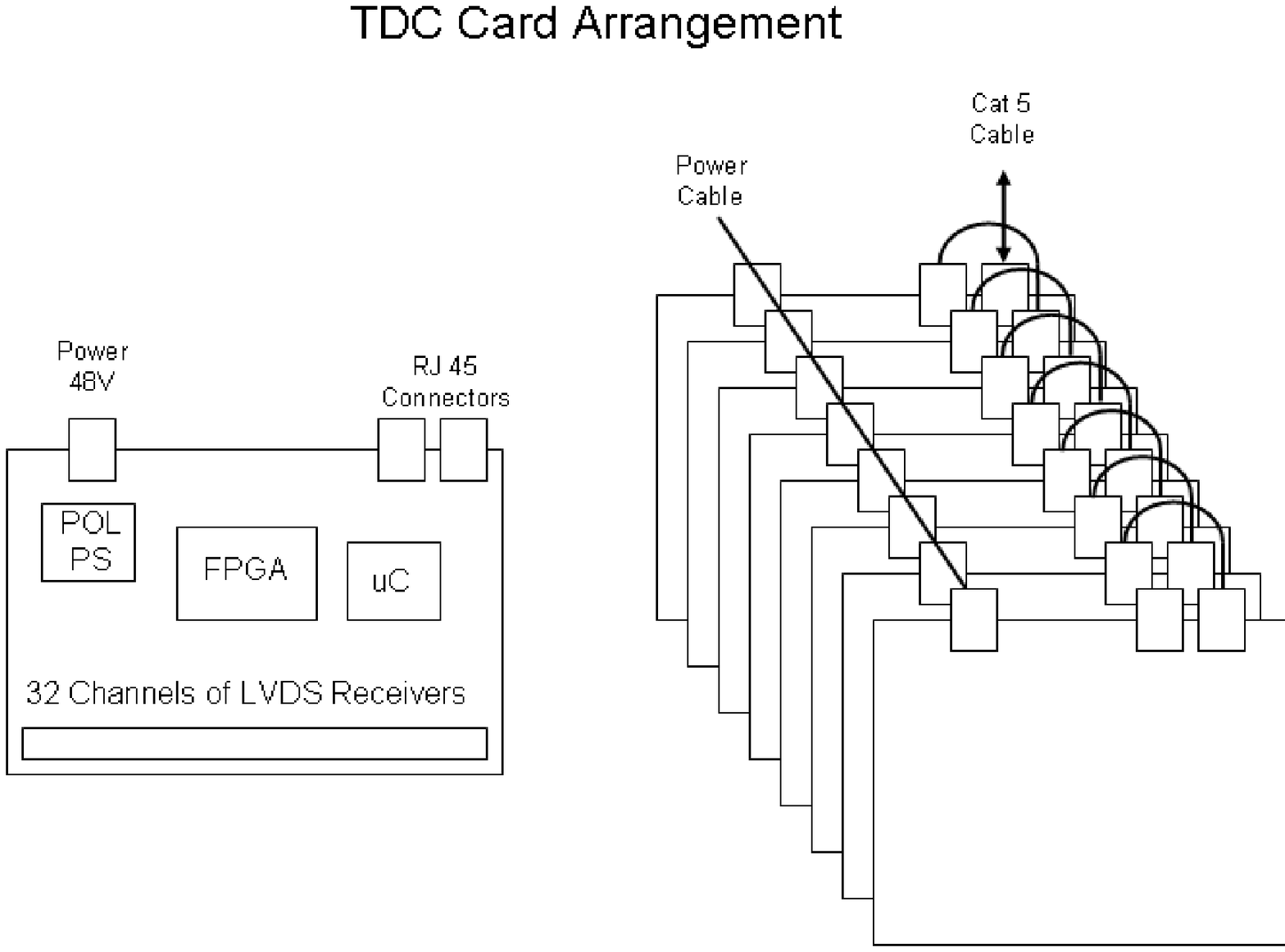}
\caption{\label{fig-Sten2} Chamber TDC cards schematic.}
\end{figure}
\begin{figure}[ht]
\includegraphics[width=\textwidth]{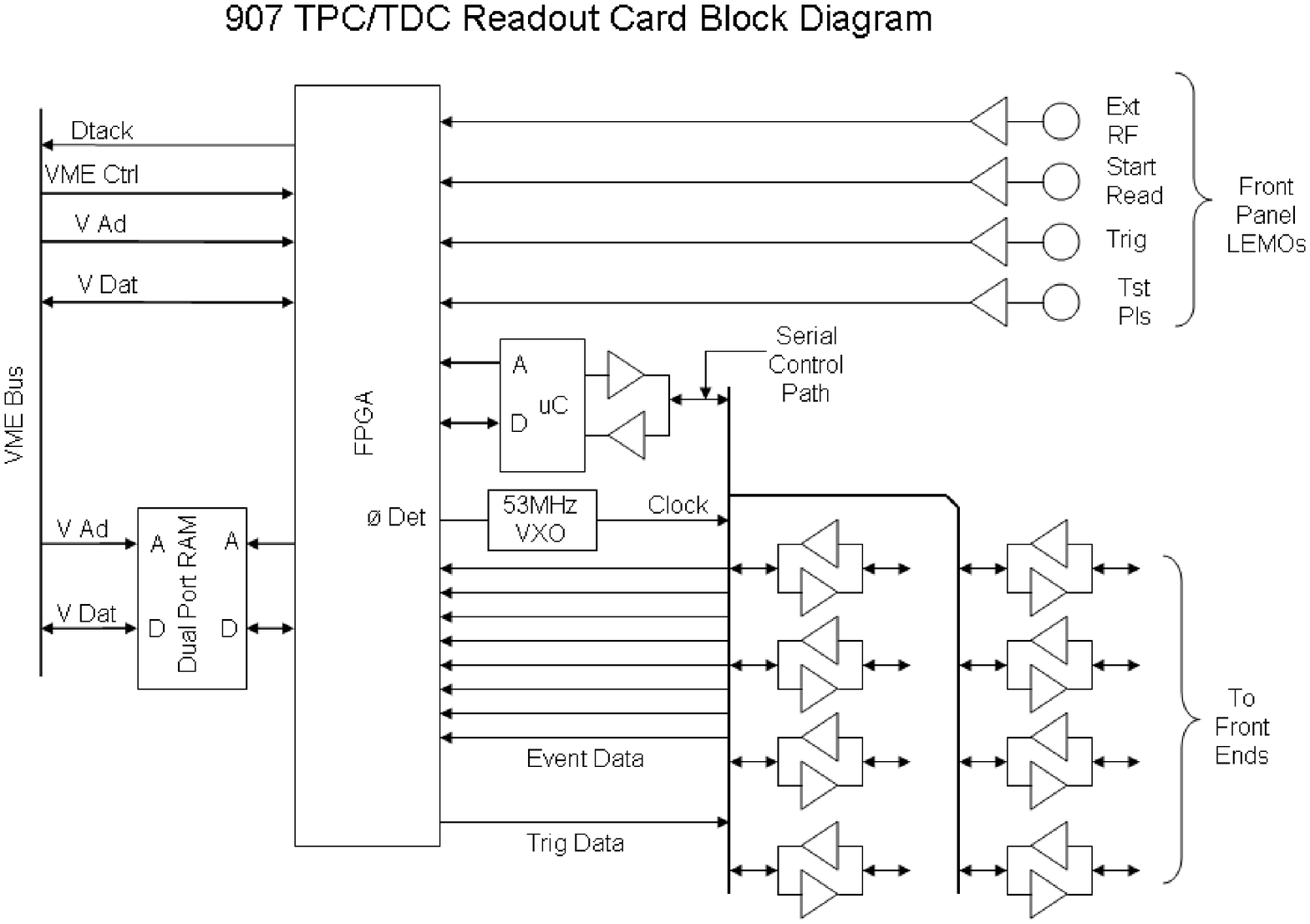}
\caption{\label{fig-Sten1} VME readout card schematic.}
\end{figure}

The Chamber front end cards will be daisy-chained (see Figure~\ref{fig-Sten2})
onto a total of
five VME readout cards in the same way that multiple RICH front end
cards are read out.

The total cost for the new chamber readout electronics is estimated
at \$121k for material and \$29k for labor. The labor
cost is low due to the simple design of the new readout based
closely on existing electronics designed by the same engineers.
The largest material costs are \$56k for 1100 daughter cards and
\$49k+\$7.5k for 325 front end TDC cards and mounting
structures/mechanical protection.

A detailed list of tasks is included in the Gantt chart.

\subsection{Time of Flight, T0, and threshold \v Cerenkov Readout}

The current Time of Flight (ToF) readout uses LeCroy 2229 CAMAC TDCs.
The 2229 modules
have a long conversion time when used in full range mode as is desired
by the MIPP experiment. The Time of Flight signals are also read out with
CAMAC 4300 ADCs needed for slew corrections on the TDC signals. These ADCs
are FERA modules, but are currently read out through the CAMAC backplane.

Long delay cables are used in order to receive a trigger for the
common TDC start and ADC gate. These ribbon cable delays are sensitive
to the environmental temperature. This causes fluctuations in the
timing response that are large compared to the timing resolution
needed for ToF particle identification. Although much work is being
done to correct for these fluctuations in offline analysis, a change
in the readout to eliminate these temperature dependent variations
will make the ToF system more robust and significantly simplify the
offline analysis, reducing both time and man-power needed for data
analysis.

The threshold \v Cerenkov (CKOV) detector is read out through ADCs and
multi hit TDCs. It has 96 channels. The current readout uses CAMAC hardware.

The T0 detectors in the beamline provide the beam definition for the
trigger and the timing mark of the interaction in the target. They are
also used at low beam momentum for beam particle identification. A
total of three scintillators in the beam line is read out with 12
photomultiplier tubes. With proposed modifications to the trigger we
would need to add delay to the signals of the T0 system. This would
degrade timing resolution.

If the CAMAC readout was to be maintained we would need to obtain new
readout for four CAMAC crates for the ToF readout and for two CAMAC
crates for the CKOV readout. The currently used CES CBD 8210 module is
no longer maintainable. The new solution for CAMAC readout is the
Hytec1365 module.  It provides readout for one CAMAC crate, but does
not allow to put multiple crates onto a CAMAC branch. Thus the
experiment would need to purchase six of the Hytec1365 modules to read
out the ToF and CKOV. These modules by themselves would cost
\$30k. The total upgraded readout system for these detectors would be
more expensive.

Instead we propose to build a new readout for these detectors. Again
the back-end will be provided by the VME readout cards that are also
used by the TPC and Chamber readouts. The front end boards will be
similar to the front end cards used in the RICH and proposed for the
Chamber readout with the difference that ToF and CKOV need ADC readout
and the ToF needs high resolution TDC readout.

Front end boards with the TripT chip (also used by the \MINERVA{}
experiment) and a high resolution TDC chip (the TDC-GPX chip from ACAM
GmbH, also used by LHC-b) will provide 30~ps timing resolution for the
ToF (better than the 2229 currently used) and multihit
capability needed for the CKOV. The TDC-GPX chip can be operated in
different modes. The mode with 30~ps resolution and 2 channels per chip
fits the needs of the ToF while coarser resolution with 8 channels per
chip reduces cost for the CKOV. The new electronics will be able to
buffer hits. The delay cable on the ToF can be eliminated.

Design of the front end cards for the ToF and CKOV is expected to not
pose significant challenges. The TripT chip and the analog part of the
circuit will essentially be copied from a \MINERVA{} design and the
TDC-GPX chip is entirely digital.

The total cost of the new readout for the ToF and CKOV is estimated at
\$34k. The production cost is small because the total number of
channels is small (106 ToF + 12 T0 + 96 CKOV + spares). The design
cost is small because the system uses components that are common to
other systems.

\subsection{Calorimeter Readout Upgrade}

The MIPP EM calorimeter consists of 640 proportional tube channels.
Every event has 50\% of these channels above pedestal mean. Currently
there are 160 ADC channels reading out four times for each event into
a data buffer. This takes 800~$\mu s$ per event.

The readout time needs to be reduced to significantly below 300~$\mu s$
in the upgrade. We propose to install a system of four crates of
FERA ADCs to read out the EM calorimeter.
Two crates of FERA ADCs each will be read out on the front
bus by two FERA drivers that sparsify the data. The data above pedestal
is then read through the CAMAC back plane using two Hytec CAMAC controllers
with buffer memory that send the data to a PC through ethernet.

Setting the pedestal subtraction to the pedestal mean
plus 2 or 3 sigma of its width will reduce the number of channels
with data further. With 80 words of data per CAMAC crate per
event and DMA CAMAC data transfer we expect to achieve 40~$\mu s$
readout time per event.

The cost for this upgrade is driven by the cost for the two Hytec1365
CAMAC readout controllers.

\subsection{Upgrades to the Online Data Acquisition System}

The MIPP  DAQ system used in the 2004 to 2006 runs was originally
designed and coded in 2003. It utilized components that were readily
available at the time such as the CES CBD8210 CAMAC branch drivers,
six VME power PCs and assorted other electronics. The DAQ system
itself used the Fermilab CD event builder of that time (r2dm) which
during the run constrained our system to remain at RedHat Linux 7.3
operating system (due to ACE/ITC). The CBD8210 CAMAC branch driver can
no longer be maintained nor are the old power PCs capable of being
upgraded to the newer operating systems. Hence the DAQ system for the
upgraded MIPP experiment is in need of a major overhaul. This is
important not only for the larger physics data set expected in the
upgraded MIPP experiment which is 100 times more demanding at 3 kHz
than the old system running at 30 Hz, but also in sight of the long
term use of the MIPP experiment when eventually the calorimeters of
the MIPP experiment will be replaced by ILC test calorimeters
for tagged neutral beam studies. 
So having a DAQ system whose electronics and 
operating systems can be maintained through the year 2010 is 
of importance.

From May to September 2006 a  group of MIPP physicists and the
Fermilab Computing Division met weekly to determine the route that
this MIPP DAQ upgrade should proceed. It also involved discussions
with the PPD electronic engineers who are building new electronics for
the TPC and wire chamber readout. The new DAQ can be broken up into
four groups: CAMAC, VME, fPix and computers. In addition there are
requirements that we wish to address for local data storage, long term
data storage for Physics analysis, network infrastructure and long
term data analysis CPU needs.

The DAQ hardware upgrades consist of replacing the CAMAC Crate
controllers and old Power PC's with
a newer version, and interfacing the new equipment being built. Most
of the work on the DAQ upgrade is in software code development for the CAMAC
and the newer PowerPC's. We will continue to use the DAQ user interface
and run control code from the existing MIPP DAQ system. The experiment
will continue to maintain these parts of the code and the code specific
to the readout of all detector systems. We are asking the Computing Division
for support on three specific tasks only. These tasks are the ethernet
communication of the Hytec module, operating system support on the PCs,
and upgrade of the event builder.

\subsubsection{The CAMAC systems}

It is economical to keep some of the electronics in the CAMAC system
which includes ADC, TDC, and assorted trigger electronics modules. The
two CBD8210 CAMAC branch drivers control 12 CAMAC crates in the old
DAQ system. The CBD8210 module cannot be maintained further and six
options were available to us for replacement. We concluded that the
Hytec CAMAC crate controller 1365v4 is the CAMAC crate controller of
choice. It has several advantages: it will be maintained through 2010,
it comes with enough on-board memory to buffer an entire spill for a
fully loaded CAMAC crate, has a data transfer rate that matches our
needs, has an event time stamp feature for each event and the upgraded
v5 features with a firmware upgrade will have pedestal subtraction
sparsification features that we requested. Each 1365 crate controller
is read out through a fast 100 Mbit ethernet link to the central DAQ
computers independently.

The wire chamber readout in the upgraded MIPP experiment will be
changed to new electronics that does not involve CAMAC. The Plastic
Ball will be added as a recoil detector in the upgrade. It currently
has a CAMAC based readout that will be upgraded to FERA ADC or a
dedicated system using the BTeV qie chips. The new readout will
contain a total of four to seven CAMAC crates read out through the
Hytec controllers.

\subsubsection{The VME systems}

Currently MIPP has the RICH PMT readout in a VME crate shared with the
power PC and the old CBD8210 CAMAC branch driver. The RICH PMT readout
board will remain and additional VME modules designed by the PPD
electronic shop for TPC and wire chambers will be added. Currently the
six power PC's (four for the old TPC electronics and two for the CAMAC
and RICH readout) will be replaced by two new Power PC 5500's. There
is no need to upgrade the old RICH readout boards except to read them
out on an event by event basis during the spill with the new Power PC
5500.  The new TPC and wire chamber readout will have event buffers
for the spill so that at the end of each spill the sparsified data
will be sent by the data transfer process of VME modules to the
the DAQ central computer with the Power PC's processing the data.

\subsubsection{fPix readout}

A new tracking silicon detector system is planned for the target
region. These will be used as part of the new interaction trigger
system which is described in the trigger section, but its readout will
need a dedicated PC computer with an extended PCI bus that can handle
18 PCI cards, one for each row of fPix chips of the 3 silicon
planes. During the spill the dedicated fPix computer will constantly
readout each event into a local memory or a disk file. This will then be
transferred at the end of the spill to the central DAQ computer.

\subsubsection{DAQ system Computer}

The central DAQ computer will receive and assemble event data and copy
it to tape storage.  The computer must be fast enough to be able to
readout the two Power PC 5500 with TPC and wire chamber data, the four
CAMAC crate interfaces, and the fPix computer. This will be done
through an upgraded event builder.

The expected data rate is 1~Gbyte per spill. This must be transfered
and handled by the computers before the next spill, i.e. within 56
seconds.  The nominal rate is that MIPP will get a 4.0 second spill
every two minutes, but during some fraction of data taking the system
must be able to handle one spill every minute.  All processors, (Power
PC 5500's and Hytec 1365v5's) have  Gigabit and 100 Megabit fast ethernet
ports and should be able to handle the highest rate expected in only a few
seconds to get the data to the main DAQ computer and enable out-of-spill 
calibration and monitoring data acquisition. The main DAQ
computer should be able to process the spill through the event builder
in the remaining 56 seconds and store it onto a mirrored disk
system. The local disk system needs to have a capacity of 1 TB to
provide local storage for $\sim 1$ day of data. The local storage will
allow continued data taking in case of network problems between the
experiment and the tape backup at FCC.

This computer also has to provide access to the data for online
monitoring. This is presently done through a cross mounted disk using
NFS. Another model would be for the DAQ software to send a data stream
to the monitoring PC through the network, thus avoiding double disk
access for monitoring.

A separate PC will run the graphical human interface to set up run
conditions, start and stop runs, perform pedestal runs, etc. For the
main DAQ computer a dual CPU system with a 3 GHz processor with a high
speed bus can provide sufficient computing power to support event
building and data transfer to Feynman and the monitoring PC.

\subsubsection{Monitoring Computer}

In order to reliably monitor data quality, two approaches are
used. Summary histograms with event statistics provide data on
dead/hot channels and similar statistics for all events in the last
spill which will be displayed and compared to reference plots for
presentation to the shift operator on five display monitors. The
computer will automatically do a comparison of these distributions and
audio alarms will be provided when there is an obvious problem. Also a
fraction of the events have to be fully reconstructed and displayed
constantly on a  monitor for full event data validation. This
track reconstruction is CPU intensive but we would expect that 10
events out of the $\sim$10000 taken each spill need to be fully reconstructed
to monitor physics quality.

This set of tasks is best handled by a dual CPU computer that receives
a copy of the data stream. The system also will need six monitors 
that the  computer can display the data on
at the end of each spill.

\subsubsection{Data transfer to ENSTORE, 
Data storage in Feynman Computer center and Offline analysis needs} 

\begin{figure}[htb!]
\begin{center}
\includegraphics[width=\textwidth]{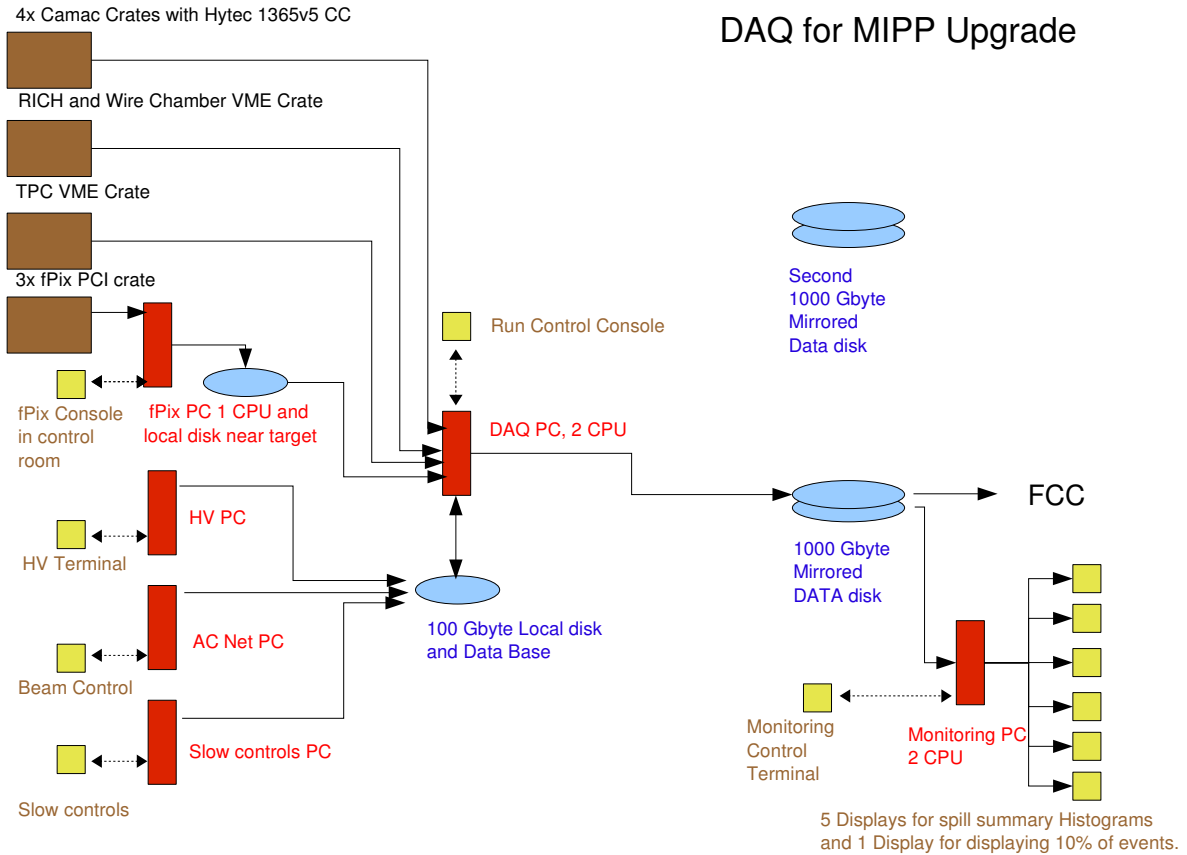}
\caption{Schematic of the Upgraded MIPP DAQ system.}
\label{daq}
\end{center}
\end{figure}

Our average event size is 100~kBytes.  The 3~kHz event rate expected
in the upgraded MIPP experiment results in 1.2~Gbytes of data per
spill. With 58\% accelerator uptime we will receive 420~spills per day
and record 500~Gbytes of data per day. We expect to transfer this data
to the Feynman Computing Center ENSTORE tape robot continuously during
the day. A 6~Mbytes/second transfer rate through the site
network and into ENSTORE is needed. The CDF and D\O\ experiments
currently obtain 30 to 60 Mbytes/second. Thus the existing ENSTORE
system should be sufficient.  
 For offline storage and CPU needs, please see the
discussion in the section entitled ``Proposed Run Plan''.  See
table~\ref{tdaq} for a summary of the DAQ equipment costs.  See
Figure~\ref{daq} for a schematic of the proposed MIPP data acquisition
system.

\begin{table}
\begin{tabular}{|c|c|c|}
\hline
Equipment &  breakdown & Cost \\
\hline
2 Power PC 5500   &      2x\$4800                                        &    \$ 9600 \\
5 Hytec 1365v5    &      4x\$4634 + \$2316 site license + spare \$4634   &    \$33486 \\
2 3 GHz PC        &      2x\$2759                                        &    \$ 5518 \\
2 1 Tbyte disk    &      2x\$1021                                        &     \$2042 \\
6 write only displays &  6x\$189\ + 6x\$36 driver boards                  &    \$1350 \\
1 PC for fPix     &      \$1778                                          &     \$1778 \\
2 13 slot PCI extender & \$1999                                          &     \$3998 \\
\hline
Total             &                                                      &    \$57472 \\
\hline
\end{tabular}
\caption{Equipment summary for DAQ upgrade equipment\label{tdaq}.}
\end{table}

\subsubsection{DAQ Task Summary}
The MIPP/CD DAQ group has identified the following tasks as being
central to the upgrade. In their initial discussion, the following was 
felt to be an optimal division of tasks between the experiment and the 
computing division.

\begin{itemize}

\item (CD task)  Update event builder for the new kernel and
operating system (Estimate 6 weeks of 1 person full time.)  
\item (CD task) 
Readout package for the fast ethernet of the new Hytec 1365v5 CAMAC
crate controllers (2 Weeks)  
\item (CD task) Get the Linux kernel
operating on the two new Power PC 5500 (Estimate 
8 weeks of 1 person full time in two blocks 
of 4 weeks, with a spare 2 weeks at the
end)

\item (MIPP task)  Provide all the CAMAC commands to read out all the
CAMAC modules. (Estimate 1 month)
\item (MIPP task)  Modify the DAQ for the new TPC(12 weeks)
 and wire chamber readout modules (4 weeks), CKOV, ToF  (Estimate 4 weeks)
\item (MIPP task) Calorimeter FERA ADC readout (5 weeks)
\item (MIPP task) Modify the Event Monitor(2 weeks)

\item (PREP task) Test the electronics modules being re-used (4 Weeks)
\item (PREP task) Maintain all associated CAMAC, NIM and VME  modules.

\item (PPD task)  Maintain custom electronics built for RICH, TPC,
                  wire chambers, Interaction Trigger and TOF-TDC.

\end{itemize}
We foresee a $\approx$~6~month commissioning run during which work will 
continue to integrate the new pieces of software together. See section on 
``Proposed Run Plan'' for a discussion of these issues. 

\subsection{Cryogenic Target Upgrade}

\subsubsection{Modifications to the Cryo-target}

The cryogenic target and refrigeration system built for MIPP  worked well
during the commissioning and physics run. MIPP has acquired over 5 million 
events on liquid hydrogen. There are, however, 
some design changes we would like to perform
The target volume is contained in a vacuum space connected to the cryogenic
system through a transfer line. The beam has to travel through part of
this transfer line to get to the target. The inner diameter of the
transfer line, based on the E690 design is currently 
small. A very small fraction of the beam
particles in the tail 
hit the transfer line (with the exact fraction depending on the
beam tune) and causes triggers. 
This causes a large background none-the-less because the
experimental target is 1\% of an interaction length thin, while beam
particles encounter a much larger fraction of an interaction length if
they hit the transfer line. The existing transfer line also is not
compatible with the Plastic Ball detector that we plan to install as
a recoil detector just upstream of the TPC to cover the hemisphere
upstream of the target. In addition to the new transfer line the mounting
structure of the refrigerator on the JGG magnet yoke will have to be
modified to be compatible with the extended JGG coils and the Plastic
Ball.
]
These modifications will cost \$51k in material and labor.

\subsubsection{Modifications to operate with cryogenic Nitrogen}

The cryogenic target system will need to be modified to allow for
cryogenic Nitrogen operation. Liquid nitrogen is heavier than
liquid hydrogen or deuterium and it is warmer. Thus a new target
flask will have to be designed and fabricated and the target control
system will have to be expanded to allow control at liquid nitrogen
temperature. The cost for material and labor for this upgrade will
be \$33k. This target can also be filled with liquid Argon.

\subsubsection{Cryocooler}

One component of the cryogenic target system is the cryocooler. It
has been working up to now. However, it is old and we do not have
a spare. We have identified the cryocooler as a single point of
failure that could result in significant downtime to the experiment.
A new spare cryocooler will cost \$40k. Additional parts and labor
to make the new cryocooler compatible with the existing target will
cost \$20k.

\subsection{Gas System and Slow Control Upgrades}

The detectors in the MIPP experiment use a variety of gases and gas
mixtures. These are provided by an extensive gas system. Overall
this gas system has been working well. However, there are some
aspects to the system that we propose to improve further.

\subsubsection{Methylal bath system}

The drift chambers operate on a mixture of Argon, Isobutane, and
a small fraction of Methylal. The Methylal is added by flowing
a fraction of the mix of the other two gases through a bubbler in
a liquid Methylal bath.

The Methylal refrigerator on the drift chamber gas mixing system
requires frequent maintenance due to water ice building up. It also
needs to be filled frequently due to its small reservoir size. We
propose to upgrade the Methylal refrigerator to a system that fills
automatically from a larger reservoir and  works with less
maintenance. This will make the system more reliable and also
reduce contact of shift personal with the Methylal, thus increasing
safety. We estimate the cost for this task at \$12k.

\subsubsection{RICH Vessel Fill Automation}

The RICH vessel is filled with CO2 at slightly above atmospheric
pressure. It was designed to be a gas tight vessel to keep constant
gas density in the radiator volume. The vessel has a small leak
that could not be found. This is not a problem because the changes
in gas density can be computed from temperature and pressure of the
gas, both of which get monitored electronically. However, the system
needs to be maintained above atmospheric pressure to avoid oxygen
contamination of the CO2. This requires manual intervention weekly
to bi-weekly depending on the weather. In order to minimize the
pressure fluctuations and hence the corrections to the data we
propose to install an automated CO2 fill system. This system will
cost \$8k in material and labor to implement.

\subsubsection{Other Gas Upgrade}

The TPC is operating with P10 gas, a mixture of 10\% methane in
Argon. We developed a mixing system for this gas mixture. When we
had problems with breaking wires in the TPC we switched to a backup
system of premixed P10 cylinders. The existing
system was designed as a backup to the mixing system rather than a
primary supply system. A cylinder lasts only two days. The bottle
pressure is not monitored electronically. The overhead for bottle
changes and gas orders was significant. The system allows for the
introduction of air contamination into the P10 during bottle changes
that could result in broken wires.
We propose to upgrade the P10 system to be supplied from a semitrailer.

The beam \v Cerenkov detectors (BCkov) use vacuum pumps to run below
atmospheric pressure. The vacuum pumps can fail and in worst case
introduce oil into the BCkov vessels. With upgraded vacuum instrumentation
any such failure can be detected early. This will increase system
reliability and significantly reduce the chance of a failure that would
result in a long downtime.

\subsubsection{Slow Controls Upgrades}

The temperature in the experimental Hall is currently being monitored
in several places. Due to the sensitivity of some of the detectors
to temperature we propose to add more temperature probes throughout
the experimental hall and especially in the region around the time of
flight detector.

The threshold \v Cerenkov detector pressure sensors need to be replaced
before the next physics run. These and other sensors need to be
recalibrated.

Control and monitoring of new components for the gas system and cryogenic
target require additional slow control infrastructure. These upgrades
will cost \$34k.

\subsection{Photomultiplier Tubes for RICH and \v Cerenkov detectors}

The RICH detector images \v Cerenkov rings with an array of 96 columns of 32
half-inch diameter photomultiplier tubes. Currently 15 columns are
instrumented with 12 stage Hamamatsu R560 PMT's and 52.5 columns utilize
10 stage FEU60 PMT's. The remaining 912 positions in the RICH are not
instrumented due to the loss of PMT's in a fire during commissioning of
the MIPP experiment in Spring of 2004. The remaining  PMT's were distributed
over the RICH array. With the present configuration, the RICH performed well
during the MIPP physics run in 2005/2006. However, rings with a small diameter
from particles near the \v Cerenkov threshold will be fit with better precision
and accuracy if the readout is instrumented completely. The 12 stage PMT's
provide a better sensitivity for single photons than the 10 stage PMT's.

The threshold \v Cerenkov detector provides particle identification for
particles with a momentum close to 10~GeV/c. It contains 96 cells, each read
out with a two inch diameter PMT. Seven of these 96 PMT's need to be replaced
to obtain a fully working \v Cerenkov detector. 

\subsection{The Plastic Ball Recoil Detector}

The Plastic Ball detector was originally built at GSI, Darmstadt to
study high multiplicity heavy ion collision events for 
the BEVALAC experiment at LBNL. It then moved to  CERN to be in the WA80
experiment in the early 80's~\cite{pball}. After the completion of the
CERN-SPS heavy-ion experiment, WA98, the plastic ball moved to KVI,
Gr\"oningen, Netherlands. KVI have informed us that it is available
for use at Fermilab, having finished its physics run at KVI. The GSI,
Darmstadt group is interested in measuring antiproton cross sections
using the upgraded MIPP spectrometer as this will enable them to
better design the PANDA~\cite{panda} detector.

\begin{figure}[tbh!]
\begin{center}]
\includegraphics[width=0.5\textwidth]{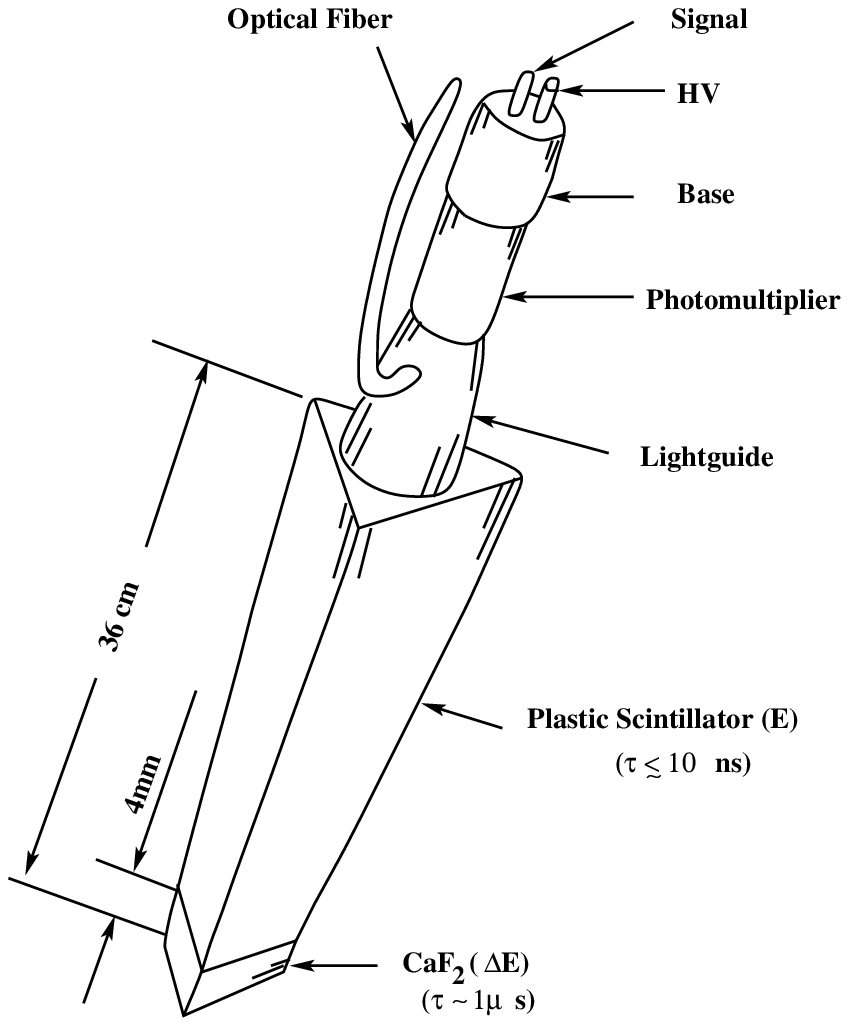}
\end{center}
\caption{Schematic of a single plastic ball phoswich module. 
The detector consists of a scintillator sandwich of a $CaF_2$ module 
and a Plastic scintillator, readout through the same photomultiplier.}
\label{phos}
\end{figure}

The fully configured plastic ball consists of 815 ``phoswich''
modules, one of which is shown schematically in Figure~\ref{phos}. The
module consists of a $CaF_2$ scinitillator sandwiched to a plastic
scintillator, both of which are readout with a single photomultiplier tube.
The photomultiplier signal is measured using two gates with different
widths. Light from $CaF_2$ is slow (1250~ns gate) and that from the
plastic scintillator is fast (250~ns gate). The fast gate measures the
plastic scintillator light only and the slow gate measures both the
plastic scintillator and the $CaF_2$ light. The charged particles  
deposit energy in both layers where as the neutral particles such as 
neutrons deposit energy in the hydrogen rich plastic scintillator.

\begin{figure}[tbh!]
\begin{center}
\includegraphics[width=\textwidth]{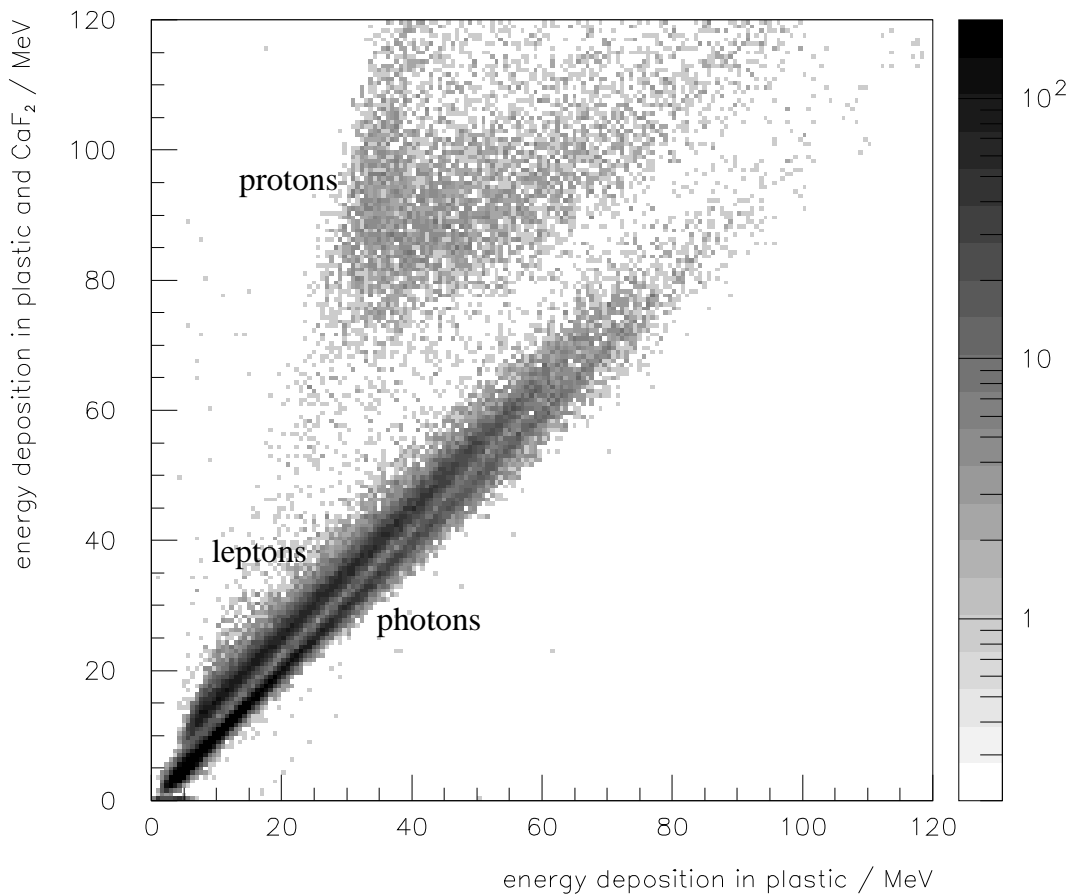}
\end{center}
\caption{On the ordinate is plotted the sum of the energies 
deposited in $CaF_2$ and the plastic scintillator and on the 
abscissa is plotted the plastic scintillator energy only. 
The protons are clearly separated from the photons.}
\label{edep}
\end{figure}

Figure~\ref{edep} shows the identification of particles using the
plastic ball by comparing the two signals. One can see the separation
between the proton signal which is seen in both scintillator layers
where as the neutral signal is seen only in the plastic scintillator. 
The neutrons and photons are distinguished further by time-of-flight.

\begin{figure}[tbh!]
\begin{center}
\includegraphics[width=0.5\textwidth]{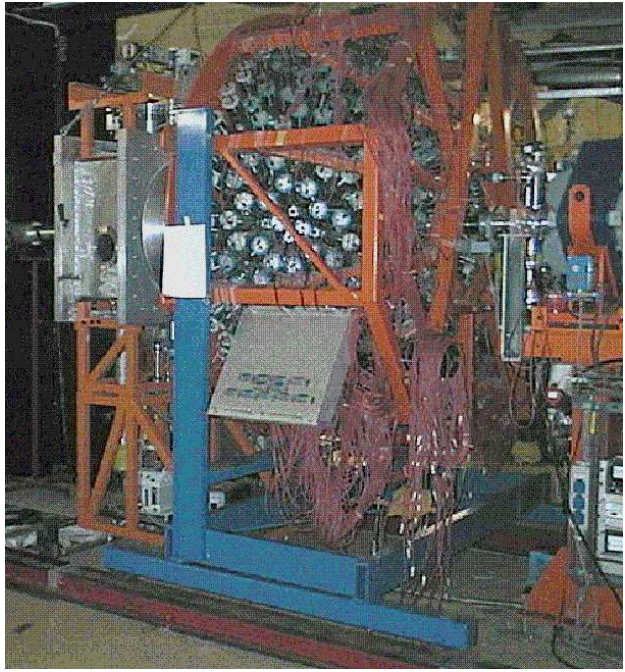}
\end{center}
\caption{Photograph of the fully configured plastic ball at KVI.}
\label{plast}
\end{figure}

\begin{figure}[tbh!]
\begin{center}
\includegraphics[width=0.5\textwidth]{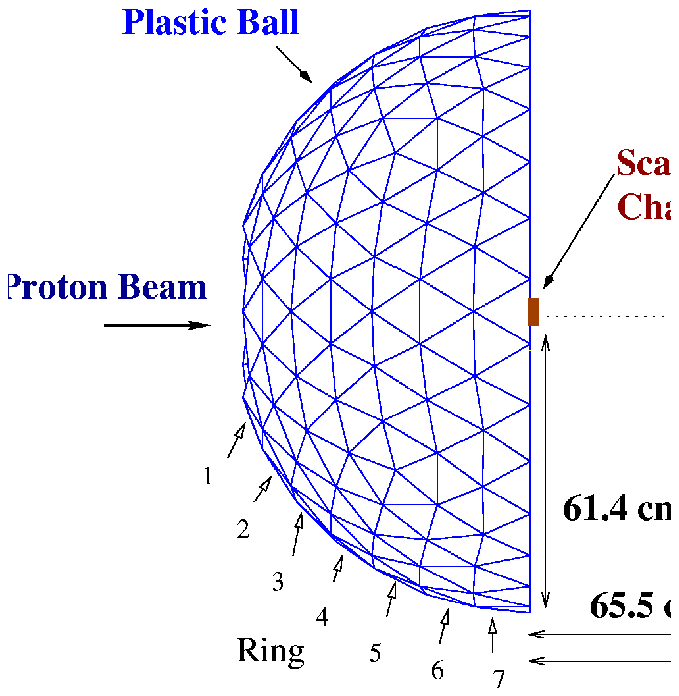}
\end{center}
\caption{This hemispherical arrangement of 340 phoswich modules 
would be similar to the way we would use it in MIPP,
 upstream of the TPC.}
\label{plast_hemis}
\end{figure}

Figure~\ref{plast} shows the full plastic ball detector as configured
at KVI.  Figure~\ref{plast_hemis} shows a schematic of 340 phoswich
modules arranged as a hemisphere. This is similar to the configuration
we would use in MIPP, with the hemisphere covering the target upstream
of the TPC.
\subsubsection{Transportation to Fermilab and Integration into MIPP}

The intention of the GSI physicists to take part in the MIPP upgrade
has only been communicated to us recently and we have not had time to
design the mounting of the Plastic Ball hemisphere into MIPP. We have
estimated a cost figure to do this, after talking to Fermilab
engineers. GSI is in the process of assigning manpower and resources
to take part in MIPP upgrade and the responsibilities will become
clear at that point. The Plastic Ball detector has to be mounted in a
removable fashion such that the TPC may be repaired. During NuMI
target running, the Plastic Ball has to be removed from the beamline
as well, since the NuMI charged tracks produced at the upstream end of
the long target need to enter the TPC without encountering the Plastic
Ball.

\subsection{Modifications to the beam line}
We propose to add low-current  power supplies (Lambda- Genesis) as in
M-test to run our beamline in the low momentum mode. During low
momentum running, we will switch from the regular power supplies to
these new ones for all the magnets in the secondary beamline. In
addition, we plan to add Hall probes to each magnet to monitor the
field to avoid hysteresis effects due to iron yoke residual fields to
ensure that each magnet is set to the correct field appropriate for
the particular beam momentum setting.

\subsection{Beam Veto upgrade}

In the first MIPP physics run some of the triggered events,
especially at low beam momentum, have beam halo and spray
particles hitting the experiment in coincidence with events of interest.
The beam background particles can be separated from the events
of interest when they are significantly out of time with the triggered
event. But additional tracks within several 100~ns cannot always
be separated from the primary event vertex in the TPC. ADC gates of
other detectors in the experiment are ~400~ns long. The energy deposit
of background inside this window cannot be distinguished from the
event signals.

The experiment is currently using a beam veto counter to inhibit the
trigger when spray is present. However, this counter covers
only an area around the beam of approximately $1~\mbox{ft}^2$.
In the upgrade the experiment will read out at a rate of approximately
100 times the rate of the first run. Thus the beam intensity has
to increase by a similar factor. The beam backgrounds will also increase.

We propose to expand the beam veto counter into a veto wall that
will cover the area of the TPC and the proposed recoil detector. This
is an area of $48 \times 60~\mbox{in}^2$. The signals from spray particles
need to be collected in time for the trigger formation.
An array of six 10 inch wide fast plastic scintillator paddles with
light attenuation $>$100~cm will satisfy these requirements.
The Rexon RP420 scintillator is a good inexpensive choice.
For reliable light detection, each paddle will be read out with
photomultiplier tubes on each end.
The 12 signals will be latched and processed by the upgraded trigger
electronics to form the trigger inhibit signal. 
High Voltage for the PMT's will be provided through the existing LeCroy
1440 HV system.

\subsection{Rewinding the TPC}

The TPC contains three planes of wires. Charged particles ionize the P10
gas inside the TPC drift volume. Ions drift up to the solid cathode at
the top of the drift volume. The cathode is at -10~kV. Electrons
drift down to the ground plane of wires. Just above the ground plane is
a plane of wires used to gate the TPC. This gating grid is transparent
to the drifting charges when all of its wires are at the surrounding
potential $V_0$. Outside the gate window half of the wires in this plane
are put at $V_0 - 150$~V and alternate with the other
half of wires at $V_0 + 150$~V so that drifting electrons terminate in the
gating grid wires. When the grid is gated, electrons drift through the
gating grid and through the ground plane into the amplification region.
Anode wires at +1250~V form the third plane of wires. They are just below
the ground plane to create a large gradient in the amplification region.
Drift electrons get multiplied in this region and terminate on the
anode wires. They create image charges on the TPC pad plane just below
the anode wire plane. These pads are read out by the TPC electronics.
The anode wire plane actually consists of anode wires alternating with
field shaping wires at ground potential.

\begin{figure}[ht]
\includegraphics[width=\textwidth]{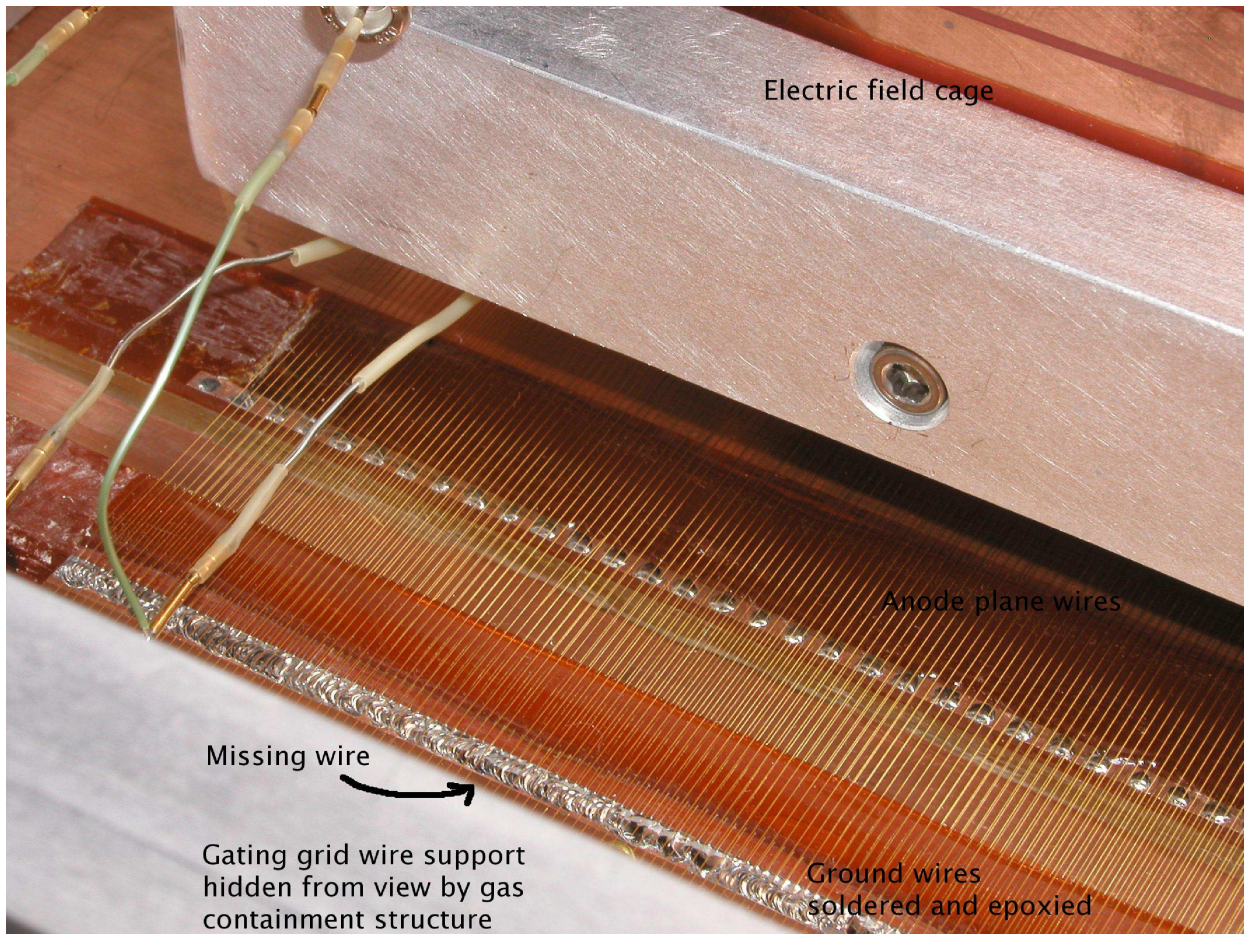}
\caption{\label{fig-TPCwire} One section of TPC wires during repair.}
\end{figure}

During commissioning for the first MIPP physics run several wires in
each of the wire planes broke. The TPC was opened multiple times and
broken wires were carefully extracted. Figure \ref{fig-TPCwire} shows
a section of wires during the repair. The wires broke due to
contaminations in the gas. All wires were exposed to this gas. Some
of the anode HV sections can not be operated at the full 1250~V. This
results in gaps in sensitivity in the TPC.
Due to the mounting
of the wire planes it is not possible to replace anode wires without
removal of the two upper wire planes.

We propose to replace the wires in all planes. Lab 6 at Fermilab has a
facility that can be used to wind wire onto construction frames and
then transfer the wires onto the TPC wire support structures. The TPC
wires are mounted in sections onto supports. These supports are not
designed to hold the wire tension unless they are mounted in the TPC
or on transfer frames. Frames and transfer hardware handling of the
TPC wire planes need to be designed and built. The TPC wires need to
be removed and the frames cleaned. Then new wires can be places on
the existing mounting.supports.

This work is similar to projects previously performed at lab 6. The
only less familiar aspect involves placement of two different kinds
of wires onto the anode frame to get alternating anode and field
shaping wires with the correct spacing. The expertise for this job
exists at Fermilab.

The cost for this project is approximately \$16k in material and labor.

\subsection{Wire Chamber repairs}

MIPP uses six large wire chambers and three smaller beam chambers.
The high voltage on one of the drift chambers (DC4) started to trip
frequently during the last weeks of the first MIPP physics run.
One of the MWPCs has a defect that causes two of the four planes to
be inoperational.

A spare MWPC is now in lab6. We refurbished it during the
last run. We propose to exchange the broken MWPC with the spare and
repair it. We also propose to diagnose the reason for the DC4 HV trips
and repair this chamber. Both types of chamber have been repaired by
MIPP physicists before. All necessary hardware exists. The cost for this
project is approximately \$5k.

\section{Cost and Schedule}

\subsection{Cost estimate of the MIPP Upgrade}

The MIPP upgrade project cost is shown in detail in table \ref{tab-Cost}.
The total cost for all tasks including non-Fermilab contributions and
labor is \$2 million.

\begin{table}[htb!]
\includegraphics[width=\textwidth]{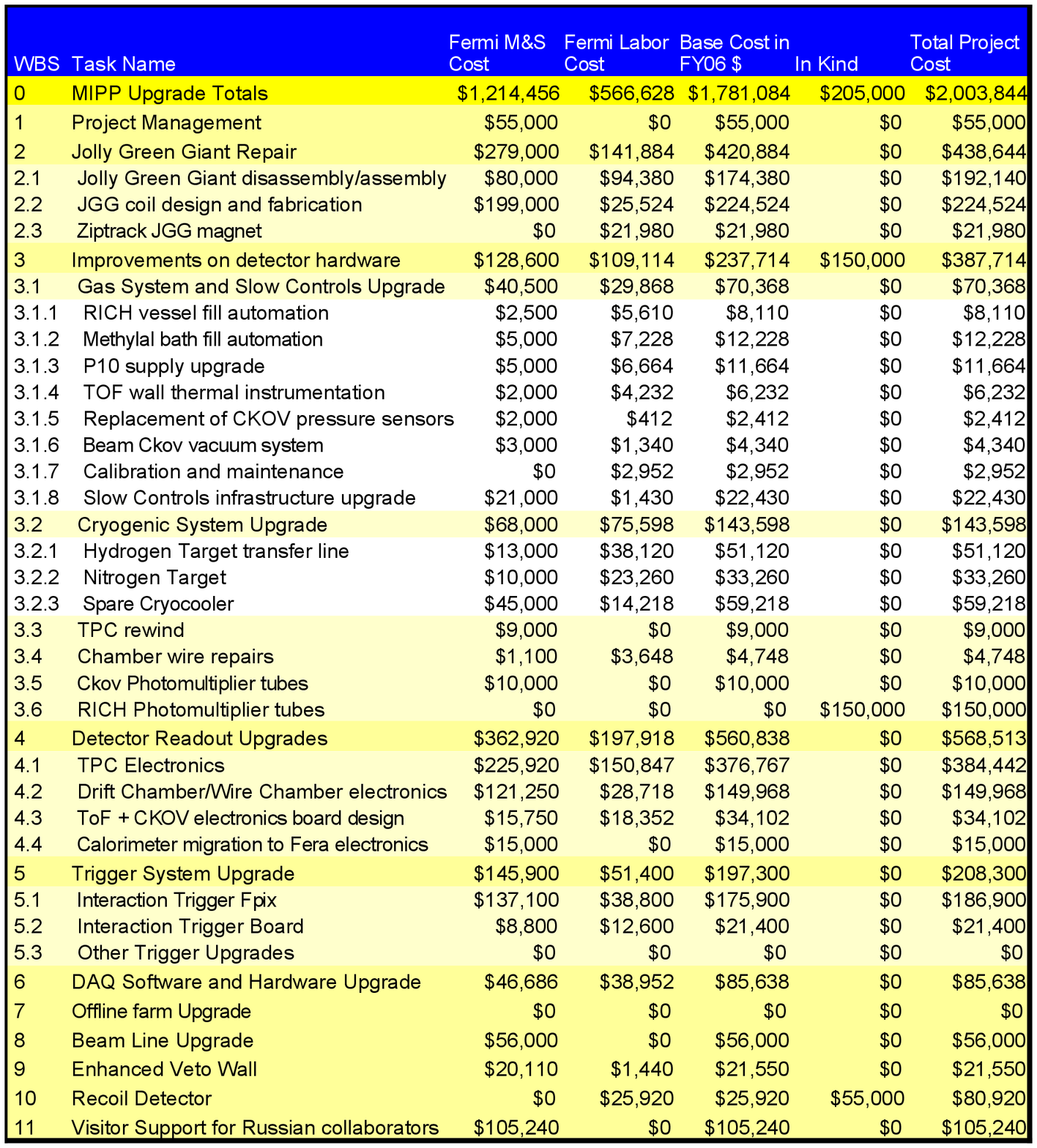}
\caption{\label{tab-Cost} Cost estimate for the MIPP upgrade project.}
\end{table}

The most essential tasks for a successful run are the JGG repair, detector 
readout upgrades, and DAQ upgrades. These add to M\&S costs of approximately 
\$715,000. Also important for the success of the project are the improvements 
to the detector hardware, especially the cryogenic target, the trigger system,
 and the enhanced beam veto wall and recoil detector.

The JGG coil replacement and TPC readout electronics upgrade are the two 
tasks with the largest costs. These costs and the schedules for these tasks
 have been estimated in great detail. The orders for new JGG coils and for 
the chips for the TPC front end electronics have been placed.

\subsection{Schedule}

\begin{figure}[htb!]
\includegraphics[width=\textwidth]{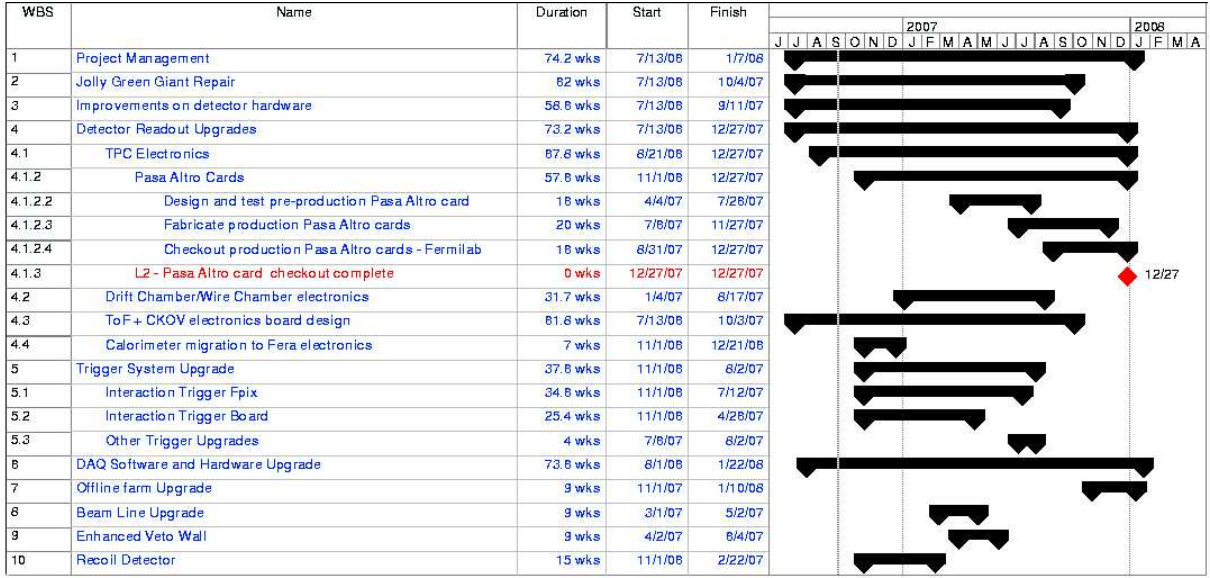}
\caption{\label{fig-Gantt} Schedule for the MIPP upgrade project.}
\end{figure}

The schedule for all upgrade tasks is shown in the Gantt chart in
figure
\ref{fig-Gantt}. The critical tasks in the timeline are the development, 
fabrication, and testing of TPC readout electronics and the integration 
into the upgraded DAQ system. Replacement of the JGG coils on schedule is
 also critical because many other tasks cannot proceed until the JGG coils 
are in place and the Ziptrack is finished and removed from MC7. The new TPC 
electronics will be operational at the end of January 2008. All other tasks can
 be finished on or before this date.

The project contains more than 450 tasks. The WBS is shown at summary level 
in the figure in this document. The full Gantt chart and cost and resource 
usage plots are available in a separate document.

\section{Proposed Run Plan}

We feel we will need a commissioning run lasting $\approx$ 6~months to
debug and integrate the various components in the experiment.  With
the detector running at 3~kHz, we can generate a large quantity of
data very quickly.  After the commissioning, we propose a three stage
run plan designed to use the data obtained in an optimal fashion.
During the first phase we plan to acquire 110~million events as
detailed in table~\ref{phase1}. We propose to include 12~ nuclei that
are most commonly encountered in calorimetry in this list.  This data
sample is a factor of 6 larger than what we have acquired so far in
MIPP and would require $\approx$ 11~Terabyte of storage space to hold
the raw data and approximately 3-5 times that to store the
reconstructed data, with tracks and particle ID. If one were to keep
the offline turn-around time to be similar to what we have currently
in MIPP, we would like a farm size of $\approx$ 600 nodes of today's
node speed. In 2 years time, the node speeds will  have
become faster.

\begin{table}[tbh]
\begin{tabular}{|c|c|c|c|}
\hline
Target & Number of Events & Running Time & Physics Need \\
       & (Millions)       &  (Days)      & Group \\
\hline
NuMI Low Energy target & 10 & 2 & MINOS, \MINERVA \\
NuMI Medium Energy Target & 10 & 2& \MINERVA{}, \NOVA \\
Liquid Hydrogen & 20 & 4 & QCD, PANDA, DUBNA \\
Liquid Nitrogen & 10 & 2 & ICE CUBE \\
12 Nuclei & & & Nuclear Physics \& \\
$D_2$,Be,C,Al,Si,Hg,Fe,Ni,Cu,Zn,W,Pb & 60 & 12 & Hadronic Showers \\
Total Events & 110 & 22 & \\
Raw Storage & 11~TBytes & & \\
Processed Storage & 55~TBytes & & \\
\hline
\end{tabular}
\caption{Phase 1 Run Plan.}
\label{phase1}
\end{table}
\begin{table}[tbh]
\begin{tabular}{|c|c|c|c|}
\hline
Target & Number of Events & Running Time & Physics Need \\
       & (Millions)       &  (Days)      & Group \\
\hline
18 Nuclei & & & Nuclear Physics\\
Li,B,$O_2$,Mg,P,S,Ar,K,Ca & & & Nuclear Physics \&\\
Ni,Nb,Ag,Sn,Pt,Au,Pb,Bi,U & 90 & 18 & Hadronic Showers \\
10 Nuclei B-list & & & Nuclear Physics \& \\
Na,Ti,V, Cr,Mn,Mo,I, Cd, Cs, Ba & 50 & 10 & Hadronic Showers \\
Total Events & 140 & 28 & \\
Raw Storage & 14~TBytes & & \\
Processed Storage & 70~TBytes & & \\
\hline
\end{tabular}
\caption{Phase 2 Run Plan.}
\label{phase2}
\end{table}

During phase 2, we plan to complete the remaining 18 nuclei of the
A-List, as detailed in the section on hadronic shower simulation and then
proceed with the B-list if there is need. The second phase of running
is detailed in table~\ref{phase2}.

Before we go to phase 3, it is possible that additional running time is 
requested for the missing baryons search depending on the outcome of the 
analyses performed on the hydrogen data taken in phase 1.

During phase 3, we plan to go into the tagged neutral mode, where we
run the liquid hydrogen target and allow the ILC calorimetry to run
simultaneously in place of the MIPP calorimeter. This mode is shown in 
table~\ref{phase3}. The running times for
this will be dictated by the schedule of the 
ILC calorimetry test modules available then.
\begin{table}[tbh]
\begin{tabular}{|c|c|c|c|}
\hline
Target & Number of Events & Running Time & Physics Need \\
       & (Millions)       &  (Days)      & Group \\
\hline
Liquid $H_2$ & 5 million events/day & As needed & ILC Tagged neutral beams \\
\hline
\end{tabular}
\caption{Phase 3 Run Plan.}
\label{phase3}
\end{table}
It should be pointed out that the running times in Tables~\ref{phase1}
and \ref{phase2} are the actual amount of beam on live times. They do
not include the time taken to setup the various target conditions,
target cool-down times (in case of cryo-targets) or calibration runs
(such as magnetic field off runs, target empty runs). 
These end effects could become quite  significant when 
computing the actual duration of a particular phase.

The upgrade factor of 100 in DAQ speed can be thought of as permitting
the acquisition of 10 times the data we acquired so far in 10 times
less time. This results in considerable savings in machine time and
manpower to keep the experiment operational. The quality of data
acquired will be considerably improved over what we had in the first
run, due to improvements in triggering, recoil detector and Jolly
Green Giant field quality.

\section{Conclusions}

We have proposed a cost-effective upgrade solution to the MIPP DAQ
that will make MIPP a powerful spectrometer with hitherto
unprecedented particle identification and acceptance. Such a
spectrometer will be capable of improving our understanding of
hadronic shower simulations significantly and will help a large number
of experiments which have non-perturbative QCD processes as a signal
or a background understand their systematics better. Particularly
helped will be Fermilab neutrino experiments (MINOS, \MINERVA{} and \NOVA)
as well as the cosmic ray experiments such as ICE CUBE. MIPP will also
acquire much wanted $\bar p p$ data to help the PANDA collaboration
and also test various hypotheses in non-perturbative QCD (Inclusive
scaling relations, missing baryon resonances and multiplicity
enhancements at high multiplicity). In addition, it can also serve as a
source of tagged neutral beams to help the international linear
collider detector community to benchmark the particle flow algorithm.


\begin{thebibliography}{99}
\bibitem{e690}
The multi-cell \v Cerenkov detector was initially built for 
Brookhaven Experiment E766 and later in Fermilab
experiment E690  and then used in several other Brookhaven experiments. In
MIPP, we fill the detector with the gas $C_4F_{10}$ which has the
appropriate refractive index at atmospheric pressure.
\bibitem{ecal}The electromagnetic calorimeter was fabricated
by MIPP and uses lead as the radiator and an array of proportional
tubes with 2.54 cm wire spacing as the readout.  It has ~10 radiation
lengths and has 10 longitudinal segments.
\bibitem{hcal} The hadron calorimeter is recycled from the
HyperCP collaboration and uses scintillator fibers embedded in iron as
readout. It has 9.7 interaction lengths and has four longitudinal
segmentations each of which is segmented in two transversely.
\end{thebibliography}
\end{document}